\documentclass[lettersize,journal]{IEEEtran}
\IEEEoverridecommandlockouts
\usepackage{cite}
\usepackage{amsmath,amssymb,amsfonts}
\usepackage[ruled,linesnumbered,lined]{algorithm2e}
\usepackage{graphicx}
\usepackage{textcomp}
\usepackage{xcolor}
\usepackage{bm}
\usepackage{stfloats}
\usepackage{url}
\usepackage{verbatim}
\usepackage{multirow}
\usepackage{comment}
\usepackage[caption=false,font=normalsize,labelfont=sf,textfont=sf]{subfig}
\usepackage{hhline}
\usepackage{diagbox}
\usepackage{pdfrender}
\newcommand*{\boldcheckmark}{%
  \textpdfrender{
    TextRenderingMode=FillStroke,
    LineWidth=.5pt, 
  }{\checkmark}%
}

\begin{document}
\title{Multiple Access Techniques for Intelligent and Multi-Functional 6G: Tutorial, Survey, and Outlook}

\author{Bruno Clerckx,~\IEEEmembership{Fellow,~IEEE,} Yijie Mao,~\IEEEmembership{Member,~IEEE,} Zhaohui Yang,~\IEEEmembership{Member,~IEEE,} Mingzhe Chen,~\IEEEmembership{Member,~IEEE,} Ahmed Alkhateeb,~\IEEEmembership{Member,~IEEE,} Liang Liu,~\IEEEmembership{Senior Member,~IEEE,}  Min Qiu,~\IEEEmembership{Member,~IEEE,}  Jinhong Yuan,~\IEEEmembership{Fellow,~IEEE,} Vincent W.S. Wong,~\IEEEmembership{Fellow,~IEEE,} and  Juan Montojo,~\IEEEmembership{Senior Member,~IEEE}

\vspace{4mm}
\par \textit{(invited paper)
\vspace{-2mm}
}
\thanks{Bruno Clerckx is with the Department of Electrical and Electronic Engineering, Imperial College London, London SW7 2AZ, U.K (email: b.clerckx@imperial.ac.uk).
\par Yijie Mao is with the School of Information Science and Technology, ShanghaiTech University, Shanghai 201210, China (e-mail:
maoyj@shanghaitech.edu.cn).
\par Zhaohui Yang is with College of Information Science and Electronic Engineering, Zhejiang
University, Hangzhou 310027, China, (e-mail: yang\_zhaohui@zju.edu.cn).
\par Mingzhe Chen is with the Department of Electrical and Computer Engineering and Institute
for Data Science and Computing, University of Miami, Coral Gables, FL, 33146 USA
(Email: mingzhe.chen@miami.edu).
\par Ahmed Alkhateeb is with the School of Electrical, Computer and Energy
Engineering, Arizona State University, Tempe, AZ 85287 USA (e-mail:
alkhateeb@asu.edu).
\par Liang Liu is with the Department of Electronic and Information Engineering, The Hong Kong Polytechnic University, Hong Kong, SAR, China (e-mail:
liang-eie.liu@polyu.edu.hk).
\par Min Qiu and Jinhong Yuan are with the School of Electrical Engineering and
Telecommunications, University of New South Wales, Sydney, NSW 2052,
Australia (e-mail: min.qiu@unsw.edu.au; j.yuan@unsw.edu.au).
\par Vincent W.S. Wong are with the Department of Electrical and Computer Engineering, The University of British Columbia, Vancouver,
BC V6T 1Z4, Canada (e-mail: vincentw@ece.ubc.ca).
\par Juan Montojo is with Corporate R\&D, Qualcomm Inc., San Diego, CA (e-mail: juanm@qti.qualcomm.com).
}
}

\maketitle

\begin{abstract} Multiple access (MA) is a crucial part of any wireless system and refers to techniques that make use of the resource dimensions (e.g., time, frequency, power, antenna, code, message, etc) to serve multiple users/devices/machines/services, ideally in the most efficient way. Given the increasing needs of multi-functional wireless networks for integrated communications, sensing, localization, computing, coupled with the surge of machine learning / artificial intelligence (AI) in wireless networks, MA techniques are expected to experience a paradigm shift in 6G and beyond. In this paper, we provide a tutorial, survey and outlook of past, emerging and future MA techniques and pay a particular attention to how wireless network intelligence and multi-functionality will lead to a re-thinking of those techniques. The paper starts with an overview of orthogonal, physical layer multicasting, space domain, power domain, rate-splitting, code domain MAs, and MAs in other domains, and highlight the importance of researching universal multiple access to shrink instead of grow the knowledge tree of MA schemes by providing a unified understanding of MA schemes across all resource dimensions. It then jumps into rethinking MA schemes in the era of wireless network intelligence, covering AI for MA such as AI-empowered resource allocation, optimization, channel estimation, receiver designs, user behavior predictions for different MA schemes, and MA for AI such as federated learning/edge intelligence and over the air computation. We then discuss MA for network multi-functionality and the interplay between MA and integrated
sensing, localization, and communications, covering MA for joint sensing and communications, multimodal sensing-aided communications, multimodal sensing and digital twin-assisted communications, and communication-aided sensing/localization systems. We finish with studying MA for emerging intelligent applications such as semantic communications, metaverse, virtual reality, smart radio and reconfigurable intelligent surfaces, and massive connectivity and random access in Internet-of-Things, before presenting a roadmap toward 6G standardization. Throughout the text, we also point out numerous directions that are promising for future research.
\end{abstract}

\section{Introduction}

\subsection{From Communication-centric 5G to Intelligent and Multi-functional 6G}
\IEEEPARstart{N}ext generation wireless networks, such as 6G and beyond, will face mounting challenges such as higher data throughput and (spectral/ energy) efficiency, enhanced reliability, massive connectivity, global coverage across terrestrial and non-terrestrial networks, and a growing heterogeneity in the quality of service (QoS). Addressing those challenges is essential to meet the demands of further-enhanced mobile broadband (FeMBB) for augmented reality (AR) / virtual reality (VR), extremely ultra reliable and low-latency communication (eURLLC) for full automation, control, and operation in industrial environment and connected robotics, ultra massive machine type communication (umMTC) for Internet-of-Things (IoT), etc \cite{8869705}.

\par Importantly, future wireless networks will not only provide communications services but also offer a wide range of new functionalities such as sensing, intelligence, computation, localization/navigation, powering. This tendency for multi-functional networks is well exemplified by the increasing number of research areas in the past decade that have looked at various forms of wireless systems integration. Indeed, radio waves can be used for numerous applications, most commonly communications, but also power in the form of wireless energy harvesting and wireless power transfer, sensing in the form of radar, localization, etc. All of those areas have traditionally been studied separately and have led to different disciplines within Electrical and Electronic Engineering but have also evolved into vastly different industry sectors \cite{8476597}. In the past decade, the community has progressively experienced a paradigm shift in wireless network design, namely, unifying transmission and processing of many quantities and functionalities, such as information, power, sensing, localization etc so as to make the best use of the radio frequency (RF) spectrum and radiation as well as the network infrastructure for the multi-purpose of communicating, energizing, sensing, locating, computing, but also for the synergies that all those disciplines can bring to each other once properly integrated. This quest for integration, convergence, and multi-functionality has led to the new research areas of integrated sensing and communications (ISAC), integrated sensing, localization, and communications, wireless information and power transfer (WIPT), edge computing and intelligence, and integrated artificial intelligence (AI) and communications \cite{8869705,9330512}. Additionally, network intelligence, using AI and machine learning (ML), will become pervasive in the design, control and optimization of the multi-functional networks and the network itself will become the underpinning tool to enable AI applications \cite{8808168}. 

\subsection{The Crucial Role of Multiple Access}


\par Capturing multi-functionality and intelligence in future wireless network design will enable using wireless to its full potential, hence enabling trillions of future intelligent users to sense, compute, connect, energize, analyze anywhere, anytime, and on the move \cite{9502719,9675011}. One major challenge and opportunity that such intelligent multi-functional 6G and beyond network brings is that the notion of ``wireless networks" and ``users" should be understood in a much more broader context compared to 1G--5G era. With the emergence of integrated wireless communications/power/sensing/computing/AI networks, users refer to communication devices, sensing targets, devices to be charged, AI nodes, training devices, or any other form of services/functionalities that the network could provide.

\par At the core of wireless network design lies the multiple access (MA) technique whose pivotal role is to serve and process all these ``users" and decide how to allocate them resources, including time, frequency, power, space (e.g., antennas, beams), signal (e.g., messages, codes, etc), in the most efficient way. The design of future intelligent and multi-functional networks brings new challenges and opportunities for wireless network designers and in particular when it comes to MA. It is crucial to comprehend how MA techniques can address these future demands and how they need to be re-thought in light of the network multi-functionality and intelligence paradigms.

\par Time/frequency domain multiple access (TDMA/FDMA) were popular in 2G, code-division multiple access (CDMA) in 3G, orthogonal frequency division multiple access (OFDMA) coupled with space-division multiple access (SDMA) in 4G and 5G. Though SDMA-OFDMA has remained the dominant MA in the past 20 years, the past decade has also seen a wide interest in other forms of MA schemes, often classified into non-orthogonal multiple access (NOMA), making use of the power domain, the code domain, or other domains \cite{NOMA38812, MUST36859}. Classifying MA techniques has been challenging due the proliferation of new schemes in the past decade. Unfortunately, the widely used classification into non-orthogonal MA vs orthogonal MA, i.e., NOMA vs OMA, 
is over-simplistic and tends to amalgamate many different MA schemes under the non-orthogonal umbrella without contrasting them or truly understanding the essence of all those schemes \cite{10038476}. Such classification has caused unnecessary confusions and misunderstandings in the past few years \cite{9451194}. Instead of contrasting orthogonal vs non-orthogonal, \cite{10038476} suggested that a different classification should be considered in
next generation wireless networks and showed that the fundamental question behind MA design should instead be how to manage multi-user interference. Answering this question shed the light on the differences between various non-orthogonal approaches to MA designs and on a new classification of MA schemes based on how the interference is managed. Importantly, this exercise brought to light the powerful and emerging Rate-Splitting Multiple Access (RSMA) that unifies into a single MA scheme four seemingly unrelated strategies, namely OMA, power domain (PD)-NOMA, SDMA, and physical-layer multicasting \cite{bruno2019wcl}. The capability of RSMA to unify and therefore be more universal than other MA schemes makes practical implementation and operation easier. Indeed, one could claim that a single unified and general MA scheme would be easier to implement and optimize than a combination of multiple MA schemes, each optimized for specific conditions. This can be increasingly important in multi-functional 6G and beyond networks where the range and diversity of services, use cases, and deployments explode.

\subsection{Objectives and Organization}

\par This paper has three objectives: 1) a tutorial paper to educate the readers about the fundamentals of a wide range MA schemes, 2) a survey paper to give the readers access to the state-of-the-art MA schemes and literature, and 3) an outlook paper to guide the readers with new research directions. Specifically the paper contributes in the following ways. 

\par \textit{First}, we provide a tutorial and survey (in Section \ref{section_II_MA_overview}) of a wide range of MA techniques: OMA in the form of TDMA/FDMA/OFDMA, SDMA \cite{4350224}, and PD-NOMA \cite{7263349}; RSMA \cite{10038476, Mao2022Survey}; code-domain (non-orthogonal) multiple access (CD-NOMA) departing from traditional 3G CDMA \cite{Proakis08} including low-density spreading multiple access (LDS-CDMA), sparse code multiple access (SCMA), multi-user shared multiple access (MUSA), successive interference cancellation amenable multiple access (SAMA); and other multiple access schemes exploiting other domains such as interleave-division multiple access (IDMA), pattern division multiple access (PDMA), compute-forward multiple access (CFMA), lattice-partition-based multiple access (LPBMA), spatially coupled multiple access, layered-division multiplexing (LDM), index modulation multiple access (IMMA), delay-Doppler domain multiple access (DDMA), etc. The advantages and disadvantages of those MA schemes and the interplay between them are discussed, before drawing observations and conclusions on how to rethink the role and design of MA schemes. This overview departs from the conventional discussion on orthogonal versus non-orthogonal approaches found in 5G \cite{10038476} and recent tutorials \cite{7263349,10038476,9451194,Mao2022Survey,10273395,9832611,9832622,9832618,zhiguo2022NOMARISsurvey} whose focus was on RSMA and its sub-MA schemes OMA, SDMA, PD-NOMA, but not on code domain approaches nor on other domains MA. 

\par
\begin{figure}[t]
\centering
\includegraphics[width=\linewidth]{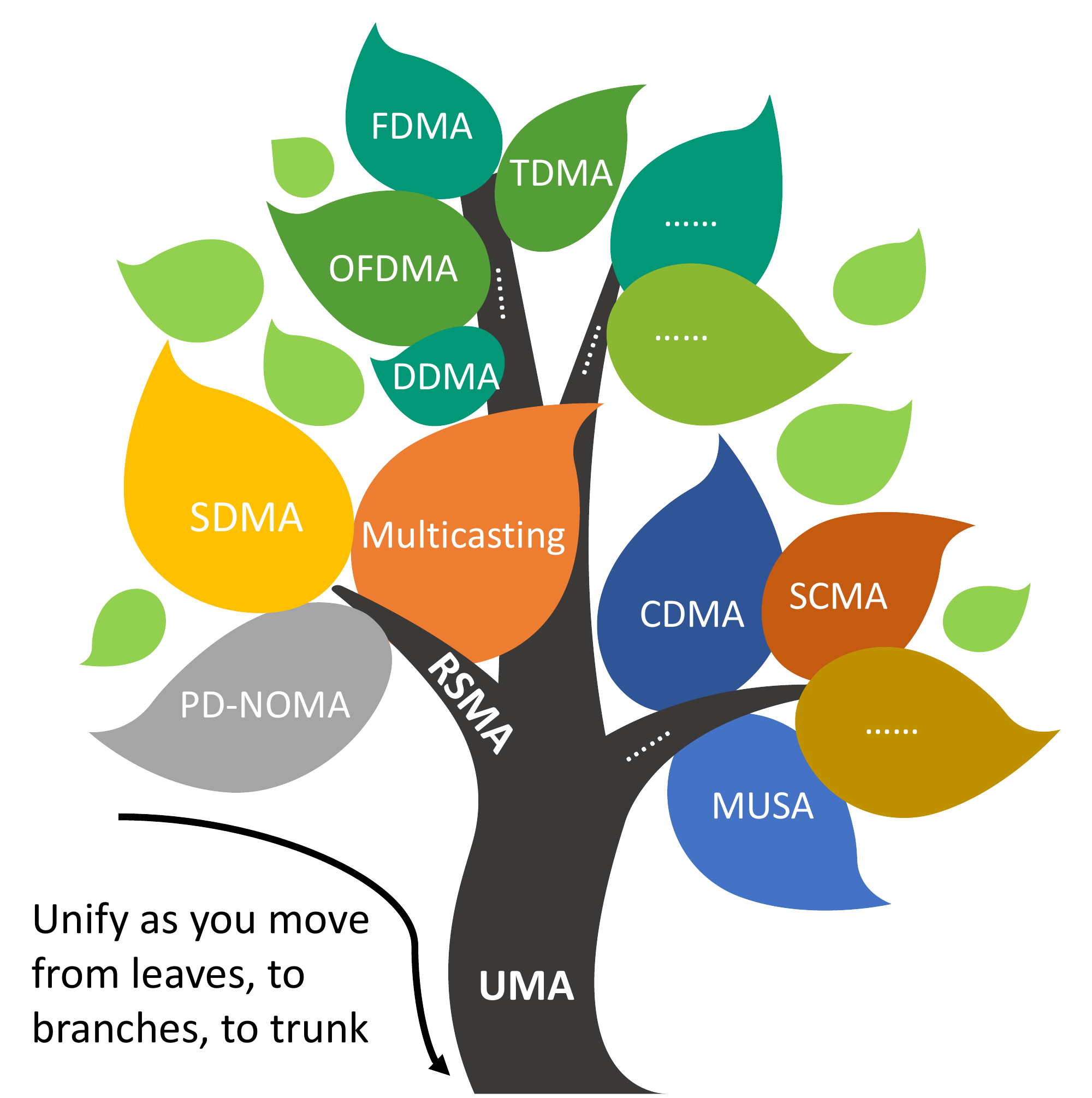}
\caption{Shrinking the knowledge tree of MA by unifying MA schemes as we move from leaves, to the branches, to the trunk. UMA, not yet found, would be the holy grail of MA scheme unification.} \label{fig:RSMA_tree}
\end{figure}

 \textit{Second}, we discuss and motivate (in Section \ref{UMA_section})  the importance of shrinking the knowledge tree of the MA literature in order to identify MA schemes that can exploit multiple dimensions and unify them. Inspired by RSMA to that shrinks the knowledge tree by providing a unified and conceptually simple understanding of a morass of results on OMA, SDMA,
PD-NOMA, physical-layer multicasting, we motivate future research toward universal multiple access (UMA). UMA should further shrink the knowledge tree of MA schemes by unifying RSMA with all other dimensions, such as code domain MAs, and ultimately provide a unified and conceptually simple understanding of the current and future morass of MA schemes as illustrated in Fig. \ref{fig:RSMA_tree}. Clearly, there is no UMA yet that optimally suits all applications and use cases, but such scheme and research directions are expected to become increasingly important in multi-functional 6G and beyond networks where the range and diversity of services, use cases,
and deployments explode. This research avenue and vision has not been discussed in prior works.

\par \textit{Third}, we discuss MA in the era of network intelligence and the interplay between AI and MA (in Sections \ref{Section_AI_for_MA} and \ref{Section_MA_for_AI}). We first identify how AI can be leveraged to enhance MA designs (in Section \ref{Section_AI_for_MA}). This includes AI-empowered resource allocation and optimization for different MA schemes, AI-empowered channel estimation for different MA schemes, AI-empowered receiver designs for advanced MA schemes, and AI for user behavior predictions in MA. We then investigate the converse, namely how to design MA schemes for AI applications (in Section \ref{Section_MA_for_AI}). We here touch upon MA for federated learning/edge intelligence and over the air computation. We conclude the discussion by identifying further research avenues on the interplay between MA and AI. Such treatment significantly departs from other tutorials \cite{10038476,9451194,Mao2022Survey,10273395,9832611,9832622,9832618}. This is the first tutorial paper providing a comprehensive review of the interplay between AI and MA, addressing both AI for MA and MA for AI. 

\par \textit{Fourth}, we discuss MA in the era of network multi-functionality and the interplay between MA and integrated sensing, localization, and communications (in Section \ref{Section_MA_for_ISAC}). We elaborate on how MA designs should be tailored for joint sensing and communications, multimodal sensing-aided communications, for multimodal sensing and digital twin-assisted communications, and for communication-aided sensing/localization systems. We identify the shortcoming of existing MA schemes in multi-functional networks and interesting research topics for future works.
This treatment differs from other tutorials such as \cite{9832622} that focuses on demonstrating RSMA superiority over SDMA and NOMA in ISAC or \cite{9737357} that does not focus on MA designs for ISAC.  

\par \textit{Fifth}, we discuss MA designs for emerging intelligence applications (in Section \ref{Section_MA_for_Intelligent_Applications}) such as semantic communications, metaverse, virtual reality, smart radio and reconfigurable intelligent surfaces, and massive connectivity and random access in Internet-of-Things, before presenting a roadmap toward 6G standardization (in Section \ref{Section_6G}). This differs from other tutorials such as \cite{10038476, Mao2022Survey, 10273395} that focus on RSMA for some of those applications.

\par Table \ref{tab:Acronyms} details the main abbreviations used throughout this work.

\begin{table*}
\caption{List of abbreviations.}
\label{tab:Acronyms}
\centering
\begin{tabular}{|l|l||l|l|}
\hline

BC    & Broadcast Channel                   & (u)mMTC   &  (ultra) massive Machine-Type Communication                                     \\
CDMA  & Code Division Multiple Access             & MU--LP   & Multi-User Linear Precoding                    \\
CFMA   & Compute-Forward Multiple Access            & MU-MIMO & Multi-User Multiple-Input Multiple-Output                               \\
CoMP  & Coordinated Multi-Point               & MUSA & Multi-User Shared Multiple Access                         \\
CSI   & Channel State  Information                & NOMA     & Non-Orthogonal Multiple Access                            \\
CSIT/R  & Channel State Information at the Transmitter/Receiver    & NOUM     & Non-Orthogonal Unicast and Multicast   \\
C-RAN & Cloud-Radio  Access  Networks              & OFDMA    & Orthogonal Frequency Division Multiple Access                     \\
 DDMA &    Delay-Doppler Domain Multiple Access  & OMA      & Orthogonal Multiple Access  \\
DoF       &   Degree-of-Freedom                                          & PDMA & Pattern Division Multiple Access                     \\
DPC       &   Dirty Paper Coding                                          & QoS      & Quality of Service                       \\
DPCRS   &  Dirty Paper Coded Rate-Splitting                           &  RF       & Radio Frequency                                          \\
(F)eMBB   &  (further-)enhanced Mobile Broadband Service                      &   RIS     &      Reconfigurable Intelligent Surfaces                                  \\
FDD     &  Frequency  Division  Duplex                                       &  RS       & Rate-Splitting     \\
FDMA     &   Frequency Division Multiple Access                   &  RSMA     & Rate-Splitting Multiple Access                                        \\
F-RAN          &  Fog-Radio Access Networks                      & SAMA & Successive Interference Cancellation Amenable Multiple Access                         \\
GRS          &  Generalized Rate-Splitting                       &   SC       & Superposition Coding   \\
HRS        &   Hierarchical Rate-Splitting                     & SCMA     & Sparse Code Multiple Access                         \\
IDMA       &   Interleave-Division Multiple Access              &  SDMA     & Space Division Multiple Access                                \\
IMMA    &    Index Modulation Multiple Access                   &  SIC      & Successive Interference Cancellation                            \\
ISAC   &  Integrated Sensing and Communications                  & SISO     & Single-Input Single-Output                     \\
IRS  &   Intelligent  Reconfigurable   Surface        & SNR      & Signal-to-Noise Ratio                               \\
LDM &  Layered-Division Multiplexing                  &  SWIPT    & Simultaneous  Wireless  Information  and  Power  Transfer                                     \\
LDS-CDMA    & Low-Density Spreading Multiple Access         &     SIMO & Single-Input Multiple-Output          \\
LPBMA    &  Lattice-Partition-Based Multiple Access                           &   TDD       &     Time  Division  Duplex        \\
MA    &   Multiple Access          &   TDMA        &       Time-Division Multiple Access                                                     \\
MAC &  Multiple Access Channel      &  UMA  & Universal Multiple Access                                                   \\
MIMO    &   Multiple-Input Multiple-Output     &  UAV      & Unmanned Aerial Vehicles                                                               \\
MISO     & Multiple-Input Single-Output   &   (e)URLLC    & (extremely) Ultra-Reliable  Low-Latency  Communicati                                                  \\
\hline
\end{tabular}
\end{table*}

\section{An Overview of Multiple Access Techniques}\label{section_II_MA_overview}


In this section, we provide an overview and discuss pros and cons of OMA, Physical Layer Multicasting, SDMA, PD-NOMA, RSMA, CD-NOMA, and other MAs exploiting other dimensions. We draw some general observations before identifying possible paths toward UMA.

\subsection{Orthogonal Multiple Access}


\par 
Orthogonal multiple access (OMA) is a fundamental MA technique that has been widely used in mobile communication systems. In OMA, the radio resources at hand are strategically divided into distinct, non-overlapping frequency bands, time slots, or codes, each meticulously allocated to an individual user. OMA adheres to a typical principle of interference management, which is to prevent multi-user interference.

\par 
There are four well-established OMA strategies specifically tailored for  1G to 4G communication systems, namely,  frequency division multiple access (FDMA), time division multiple access (TDMA), code division multiple access (CDMA), and orthogonal frequency division multiple access (OFDMA) \cite{tsefundamentalWC2005}. Specifically, FDMA divides the available spectrum into non-overlapping frequency bands, each accommodating one user. TDMA partitions time into time slots allocated to different users. CDMA utilizes orthogonal, user-specific codes to spread the modulated user symbols, serving multiple users simultaneously in the same time-frequency resources without causing multi-user interference (under ideal propagation conditions). OFDMA divides the frequency and time resources into narrow subcarriers and time slots, which are grouped into resource units and allocated to the users. 

\par 
While OMA has garnered widespread acceptance in past communication systems, its efficacy  becomes restricted in light of the explosive expansion of wireless communication worldwide.
Here we outline the advantages and disadvantages of OMA:

\begin{itemize}
    \item  \textit{Advantages:}

    \begin{enumerate}
     \item \textbf{Widespread adoption:} OMA is well-established and extensively utilized in current communication systems, making it easier for facilitating smooth and effortless network expansion.
    \item \textbf{Simplicity:} OMA simplifies transceiver design, implementation, and management.
    \item \textbf{Interference free:} OMA prevents interference between users, enabling interference-free transmissions and thereby enhancing the overall quality and reliability of communications. It excels in managing low to moderate user loads effectively. It is effective for handling low to moderate user loads.
    \end{enumerate}
\end{itemize}

\begin{itemize}
    \item  \textit{Disadvantages:}

    \begin{enumerate}
     \item \textbf{Inefficient spectrum utilization:} In OMA, there is a risk that even low-rate users, such as IoT sensors with minimal resource requirements, may occupy an entire resource block, leading to inefficient spectrum utilization.
    \item \textbf{Low capacity:} OMA allocates each orthogonal radio resource to an individual user. The system capacity is restricted by the total number of available radio resources. This limitation hinders its ability to accommodate the surging user demand experienced in modern communication systems.
    \item \textbf{High signaling overhead:} To enhance system performance of OMA, it is essential to implement well-designed user scheduling, which typically results in a significant increase in signaling overhead.
    \end{enumerate}
\end{itemize}

\subsection{Physical Layer Multicasting}
Multicasting usually refers to the transmission of a message intended to multiple users, i.e., one-to-all. Popular example are radio and television where the message of interest to multiple users is decoded by those users. However, multicasting can also be understood in a wider context where the transmitter transmits multiple unicast (i.e., one-to-one) messages, each intended to one user, by encoding them jointly into one stream to be decoded by all users. Users would then retrieve from the decoded stream the part intended to them. The encoding and transmission over the air is effectively a physical layer multicasting since all messages are encoded into one multicast stream to be decoded by all users. The difference with conventional multicasting is that only part of the multicast stream is intended to a given user, instead of the entire stream. Physical-layer multicasting contrasts with OMA where messages are encoded in independent streams and transmitted on orthogonal resources.  

\begin{itemize}
    \item  \textit{Advantages:}

    \begin{enumerate}
     \item \textbf{Coding gain:} By combining messages and encoding jointly multiple medium-size packets together into a single stream, the encoded stream is longer and the coding gain is increased, which leads to higher reliability. Such feature is particularly useful in non-terrestrial systems, such as geostationary satellite communications based on DVB-S2X technology \cite{DVBS2X}, where each spot beam of the satellite serves more than one user simultaneously by transmitting a single coded frame. Since different beams illuminate different group of users, such satellite system follows a physical layer multigroup multicast transmission \cite{9257433}. 
    \item \textbf{Low latency:} Since all messages are encoded jointly into one stream to be decoded by all, users can decode their messages simultaneously and do not have to wait for their message to be transmitted consecutively as in TDMA, therefore reducing latency in the network.
    \item \textbf{Interference free:} Multicasting prevents interference between users since there is only one stream transmitted, enabling interference-free transmission and thereby enhancing the overall quality and reliability of communications.
    \end{enumerate}
\end{itemize}

\begin{itemize}
    \item  \textit{Disadvantages:}
    \begin{enumerate}
     \item \textbf{Inefficient spectrum utilization:} In multicasting, all users, from the weakest to the strongest, need to be able to decode the stream. Consequently the transmission rate of the stream is always determined by the weakest user, therefore leading to inefficient spectrum and resource utilization. This can be frustrating to stronger users who could receive at higher rates but are constrained by the weakest user in the pool.
    \end{enumerate}
\end{itemize}

\subsection{Space Division Multiple Access}

\par 
In response to the limitations inherent in OMA and the growing  demand for higher data rates, improved QoS, and increased network capacity, a new resource dimension - space - has been introduced in modern wireless networks.  This gives rise to the widespread integration of multiple antennas in most wireless access points, signifying the advent of the multiple-input multiple-output (MIMO) paradigm since 4G networks. MIMO has become indispensable in modern and future wireless networks, finding inclusion in nearly all high-rate wireless standards such as WiMAX, 4G LTE, IEEE 802.11n, 5G NR, etc.
By leveraging the spatial dimension, MIMO networks introduce a novel MA known as SDMA \cite{tsefundamentalWC2005, clerckx2013mimo, 1261332}. SDMA empowers multiple users to share the same time-frequency resources, adhering to the interference management principle of precanceling interference at the transmitter and treating interference as noise at the receivers. 

\par 
To mitigate interference at the transmitter, SDMA introduces precoding techniques, which are typically classified into two primary categories:  non-linear and linear precoding. One of the most renowned non-linear techniques  is dirty paper coding (DPC) \cite{DPC1983, clerckx2013mimo}. It  attains the capacity region of MIMO Gaussian broadcast channel (BC) when perfect channel state information is available at the transmitter. However, its application is hindered by the impracticality arising from its high computational demands.
In contrast, multi-user linear precoding (MU-LP) offers a more practical alternative by leveraging linear precoding techniques at the transmitter while regarding multi-user interference as noise at the receivers \cite{clerckx2013mimo,1261332}. Although MU-LP cannot achieve the capacity region  achieved by DPC, it is particularly  useful when users possess semi-orthogonal channels and comparable signal strengths (and also perfect channel state information at the transmitter-CSIT). Therefore, it plays a pivotal role in many transmission techniques underpinning both 4G and 5G networks, including multi-user MIMO (MU-MIMO), massive MIMO, networked MIMO, and other advanced techniques.

\par 
While SDMA remains crucial in modern wireless networks, the ongoing evolution of  wireless technologies compels us to scrutinize SDMA critically. In the following,  we delineate the advantages and disadvantages of SDMA:

\begin{itemize}
    \item  \textit{Advantages:}

    \begin{enumerate}
     \item \textbf{Enhanced spectrum efficiency:} By utilizing the spatial domain, SDMA allows multiple users to efficiently share the same time-frequency resources, thereby enhancing the spectrum efficiency when CSIT is perfect and the network is underloaded.
    \item \textbf{Interference mitigation:} With perfect CSIT and an underloaded network, SDMA effectively eliminates or suppresses multi-user interference, achieving the maximum degrees-of-freedom in underloaded multi-antenna BC \cite{DPCrateRegion03Goldsmith}.
    \item \textbf{Low transceiver complexity:}  Thanks to the utilization of linear precoding at the transmitter and each receiver's ability to directly decode the intended message while treating interference as mere noise, SDMA exhibits a relatively low level of hardware complexity at both the transmitter and receivers. 
    \end{enumerate}
\end{itemize}

\begin{itemize}
    \item  \textit{Disadvantages:}

    \begin{enumerate}

    \item \textbf{High sensitivity to CSIT inaccuracy:}  The effectiveness of SDMA is highly affected by the inaccuracies in CSIT \cite{NJindalMIMO2006, mao2019beyondDPC}.  While SDMA excels with perfect CSIT in underloaded network, its performance significantly decreases with imperfect CSIT due to residual interference from imprecise interference mitigation strategies that are originally designed for perfect CSIT.
     \item \textbf{Limited network load tolerance:} SDMA performs effectively in underloaded networks but experiences degradation in overloaded systems due to the constraints posed by limited spatial resources. To address this issue, user grouping is usually employed in overloaded scenarios, but at the cost of reduced QoS and increased latency.
    \item \textbf{Limited user deployment flexibility:} SDMA is sensitive to user deployment such as the angles and strengths of user channels. It works well for users with orthogonal channels and similar signal strengths. However, its performance significantly degrades when user channels become nearly aligned or exhibit substantial variations in signal strengths. 
    \item \textbf{Complex scheduling:} Due to the limited user deployment flexibility, precise scheduling is imperative for SDMA, leading to extra complexity in user scheduling. Moreover, the  scheduling algorithms to achieve the (near) optimal performance can be challenging to implement and maintain.
    \item \textbf{High signaling overhead:} The requirements for channel estimation, scheduling, and interference management in SDMA leads to substantial signaling overhead, particularly in dynamic and densely populated network environments or when employing a massive number of antennas at the transmitter.
    \end{enumerate}
\end{itemize}

\subsection{Power-domain Non-Orthogonal Multiple Access}

\par 
PD-NOMA leverages the concept of superposition coding (SC) at the transmitter and successive interference cancellation (SIC) at the receivers to facilitate simultaneous sharing of common resources, i.e., time, frequency, code, or space among users \cite{6692652,Mojtaba2019MAbook,7263349}. This is achieved by allocating users with varying  power levels, and enabling signals from different users superposed in the power domain. PD-NOMA ensures the effective decoding of these signals at users by empowering those with weaker power levels to decode the messages of users with stronger power levels. This approach is also referred to as SC-SIC, adhering to the interference management principle of decoding interference. 

\par 
 It is well-established in the literature that PD-NOMA based on SC–SIC achieves the capacity region for single-input single-output (SISO) BC \cite{Tcover1972}, and  for the SISO multiple access channel (MAC) \cite{tsefundamentalWC2005}, it is also capacity-achieving with time-sharing.
 The performance merits of PD-NOMA over OMA demonstrated in SISO BC/MAC has driven research into  MIMO NOMA. In the multi-antenna BC, to exploit  the spatial domain, MIMO NOMA typically  separates users into distinct  groups. Interference from users within the same group is managed by SC-SIC while interference from users in different groups is treated as noise.  When there exists only a single user group, MIMO NOMA reduces to SC-SIC, requiring users to decode and remove all interference \cite{Mojtaba2019MAbook}.

\par 
PD-NOMA  has not yet been incorporated into emerging wireless standards, though it has been investigated as part of a study item in 5G but was not considered any further in 5G because its gains compared to SDMA/MU-MIMO were not found convincing \cite{8972353}. As highlighted in \cite{9451194}, there exists  many confusions and misconceptions about PD-NOMA, which  impels us to look into its advantages and disadvantages in the following: 

\begin{itemize}
    \item  \textit{Advantages:}

    \begin{enumerate}
     \item \textbf{Enhanced spectrum efficiency:} By utilizing the power domain and advanced receiver techniques, PD-NOMA allows multiple users with closely aligned channels and diverse channel strengths in the same time-frequency resources to efficiently share the same time-frequency resources, thereby enhancing the spectrum efficiency when the network is extremely overloaded.
    \item \textbf{Enhanced user fairness:} As PD-NOMA requires power allocation favoring users with weaker channel strength to enable successful interference cancellation, it is capable of enhancing user fairness than SDMA  in extremely overloaded network  with closely aligned  users and large channel strength disparities.
    \end{enumerate}
\end{itemize}

\begin{itemize}
    \item  \textit{Disadvantages:}

    \begin{enumerate}
     \item \textbf{Inefficient use of spatial dimensions:} As shown in \cite{9451194}, the sum multiplexing gain of MIMO NOMA always falls below or equals that of SDMA. This implies that the slope of the sum-rate of MIMO NOMA at high signal-to-noise ratio (SNR) will be lower than that of SDMA. This phenomenon indicates that PD-NOMA makes an inefficient use of the multiple antennas. 
     \item \textbf{Inefficient use of SIC:} In MIMO NOMA, the number of SIC deployed at each user scales proportionally with the number of users in that user group.  Consequently, compared to SDMA, MIMO NOMA introduces two notable issues: a loss in multiplexing gain and an increase in receiver complexity. This observation underscores MIMO NOMA's inefficient utilization of SIC.
     \item \textbf{High sensitivity to CSIT inaccuracy:}  As the inter-group interference management in MIMO NOMA follows the approach employed in SDMA, akin to SDMA,  MIMO NOMA is also sensitive to CSIT inaccuracy.
    \item \textbf{Limited network load tolerance:} PD-NOMA has been demonstrated as a capacity-achieving scheme in SISO BC, rendering it effective in extremely overloaded scenarios. However, as the number of transmit antenna increases, MIMO NOMA suffers from performance degradation primarily due to its inefficient use of spatial dimensions. Therefore, PD-NOMA is sensitive to the network load.
    \item \textbf{Limited user deployment flexibility:} With perfect CSIT and an underloaded network, SDMA effectively eliminates or suppresses multi-user interference, achieving the maximum degrees-of-freedom in underloaded multi-antenna BC.
    \item \textbf{High transceiver complexity:} In addition to the hardware complexity introduced by SIC receivers, PD-NOMA imposes substantial computational complexity at both transmitter and receivers. Specifically, at the transmitter, the joint optimization of user scheduling, user grouping, decoding orders, and precoders is imperative for enhancing performance. However, solving such resource allocation problem is typically challenging with high computational complexity.
    \item \textbf{High signaling overhead:} In PD-NOMA, extra signaling overhead is required to convey information such as the decoding order, user grouping result, and power levels each user should use.

    \end{enumerate}
\end{itemize}

\subsection{Rate-Splitting Multiple Access}

\par RSMA has recently emerged as a promising non-orthogonal transmission strategy for multi-antenna wireless networks,  owning to its capability to enhance the system performance in a wide range of network loads, user deployment, and CSIT qualities \cite{RSintro16bruno,  Mao2022Survey,7555358,mao2017rate}. RSMA is a generalized MA scheme originally proposed in \cite{RSintro16bruno,7555358,mao2017rate} for downlink and in \cite{485709} for uplink. The key concept of RSMA is splitting each user message into sub-messages at the transmitter. For downlink RSMA, sub-messages are categorized into common and private with the common sub-messages designed to be decoded by multiple users, whereas private sub-messages are intended to be decoded exclusively by their respective users. For uplink RSMA, sub-messages of a given transmitter can be decoded at the receiver in a non-consecutive manner. This message split capability endows RSMA with a flexible interference management strategy of partially decoding the interference and partially treating the interference as noise.

\par
Depending on the specific approaches used for message splitting and combining, RSMA has a range of transmission models, including linearly precoded 1-layer RS, 2-layer hierarchical RS (HRS), and generalized RS (GRS), each of which can be further extended to their respective non-linear counterparts \cite{mao2017rate, Mao2022Survey,9374428}, and even to space-time designs \cite{7152864}.

\par
\begin{figure}[t]
\centering
\includegraphics[width=\linewidth]{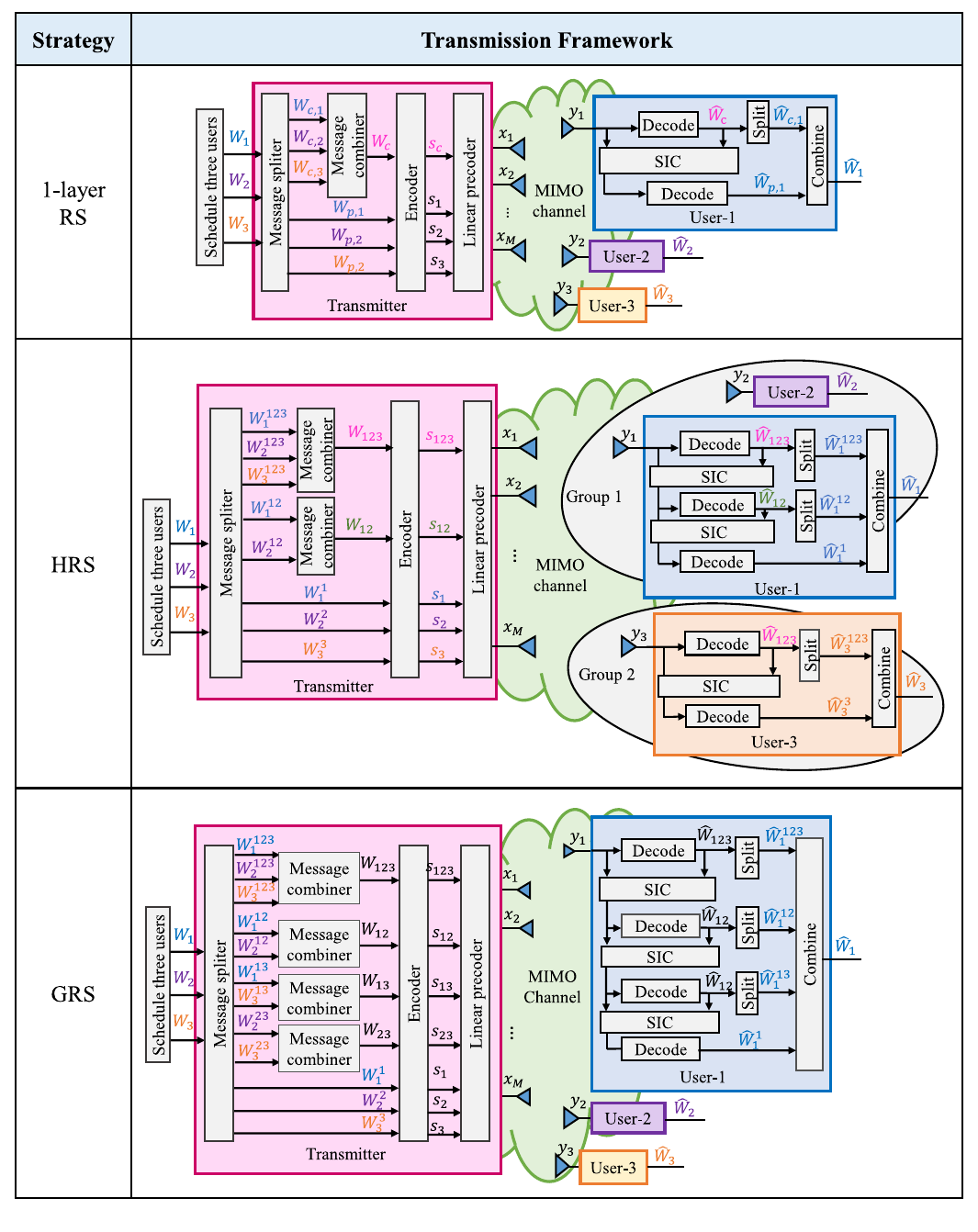}
\caption{Illustration of the downlink transmission frameworks for 1-layer RS, HRS, and GRS with $3$ users.} \label{fig:RScomparison}
\end{figure}

 
The most straightforward and practical downlink RSMA scheme is 1-layer RS \cite{RSintro16bruno,7555358}, which is also the basic building block of almost all existing RSMA schemes. In 1-layer RS, each user message is split into one common sub-message and one private sub-message (Fig. \ref{fig:RScomparison}). All the common sub-messages are combined and jointly encoded as a single common stream to be decoded by all users, whereas the private sub-messages are encoded individually as private streams. 1-layer RS requires only one layer of SIC at each receiver. User grouping and decoding order design is unnecessary since all users decodes the  unique common stream before decoding its private stream. 

\par HRS represents a more encompassing scheme compared to  1-layer RS, as it introduces additional common streams tailored for specific user groups (Fig. \ref{fig:RScomparison}).  In HRS, all users are divided into separate groups, each comprising one or multiple users. At the transmitter,  each user message is split into three sub-messages: an inter-group common sub-message, an inner-group common sub-message, and a private sub-message \cite{7434643}. The inter-group common sub-messages and private sub-messages respectively follow the common and private submessages in 1-layer RS. The primary difference between HRS and  1-layer RS  lies in the inner-group common sub-messages. The inner-group sub-messages for users within the same group are merged into a  group-specific common message, subsequently encoded into a inner-group common stream. The inner-group common stream is decoded exclusively by users within the corresponding user group, following their decoding of the inter-group common stream.  HRS requires two layers of SIC at each receiver. Decoding order design in unnecessary since each user follows the decoding order of inter-group stream, inner-group stream, and private stream. In contrast to 1-layer RS, the user grouping requires to be considered in HRS to achieve a more flexible interference management capability than 1-layer RS.

\par Unlike 1-layer RS and 2-layer HRS, which maintain a constant number of message splits for each user message regardless of the number of users, GRS takes a more comprehensive approach by utilizing all possible message splitting and combining strategies to achieve generalized transmission framework \cite{mao2017rate}. In GRS, during the message splitting and combining process, all potential user grouping are explored and the groups containing different number of users are categorized into different layers (Fig. \ref{fig:RScomparison}). For example, the group in the first layer contains all users while each group in the last layer only contains a single users. Each user message is split into a number of common sub-messages. The common sub-messages in the same group  are combined and encoded following the basic principle of HRS. At the user sides, a number of SIC layers are employed at each user to decode all common streams and the intend private stream. As all possible user grouping has been considered in the message splitting and combining process, only the decoding order among the groups in the same layer need to be carefully designed. In Fig. \ref{fig:RScomparison}, a straightforward comparison of the three-user transmission frameworks among 1-layer RS, HRS, and GRS is illustrated.

\par
\begin{figure*}[t]
\centering
\includegraphics[width=0.9\linewidth]{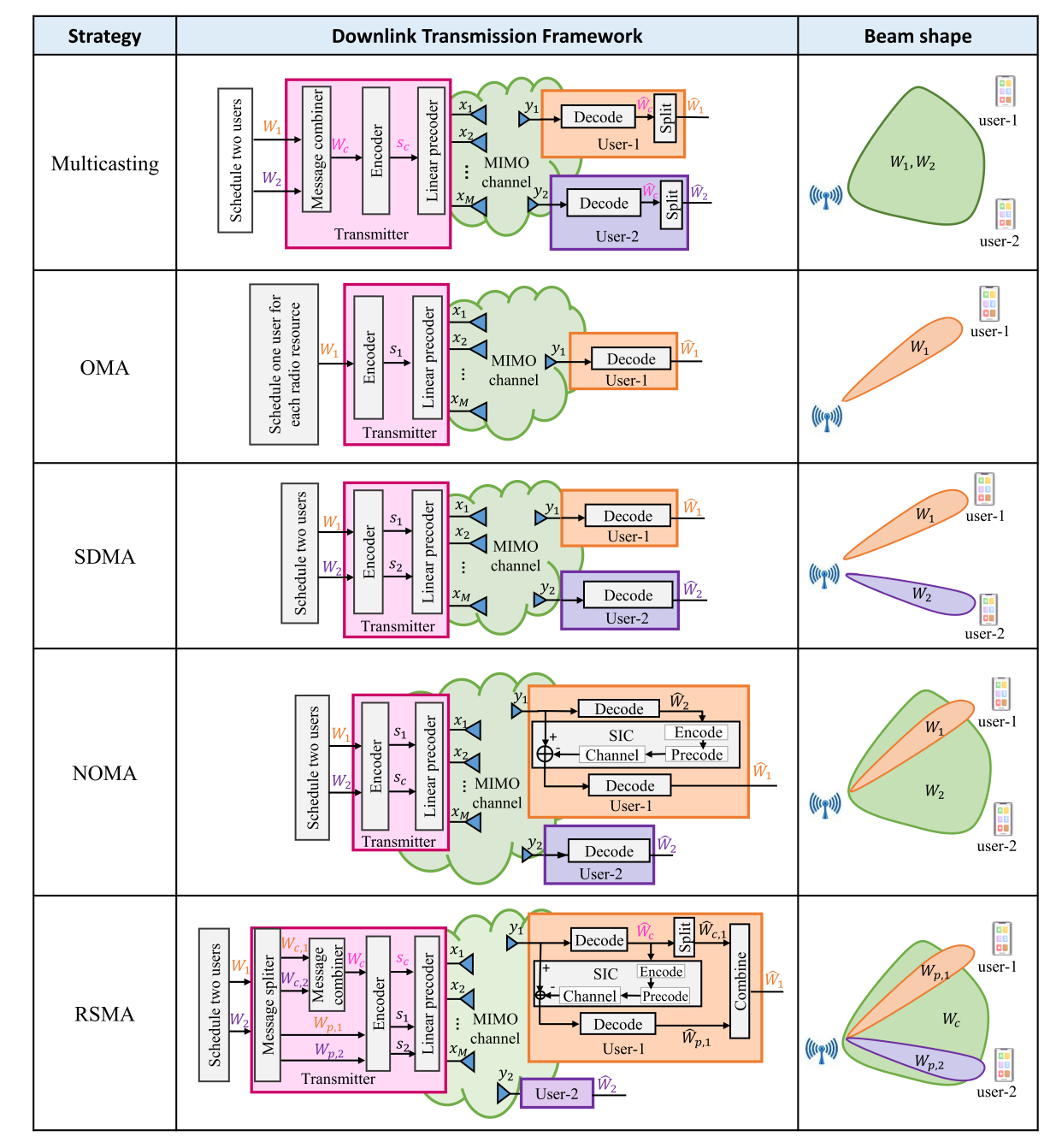}
\caption{Illustration of the two-user downlink transmission frameworks  and beam shapes for multicasting, OMA, SDMA, PD-NOMA, and RSMA.} \label{fig:DL_MA}
\end{figure*}

\begin{figure}[t]
\centering
\includegraphics[width=\linewidth]{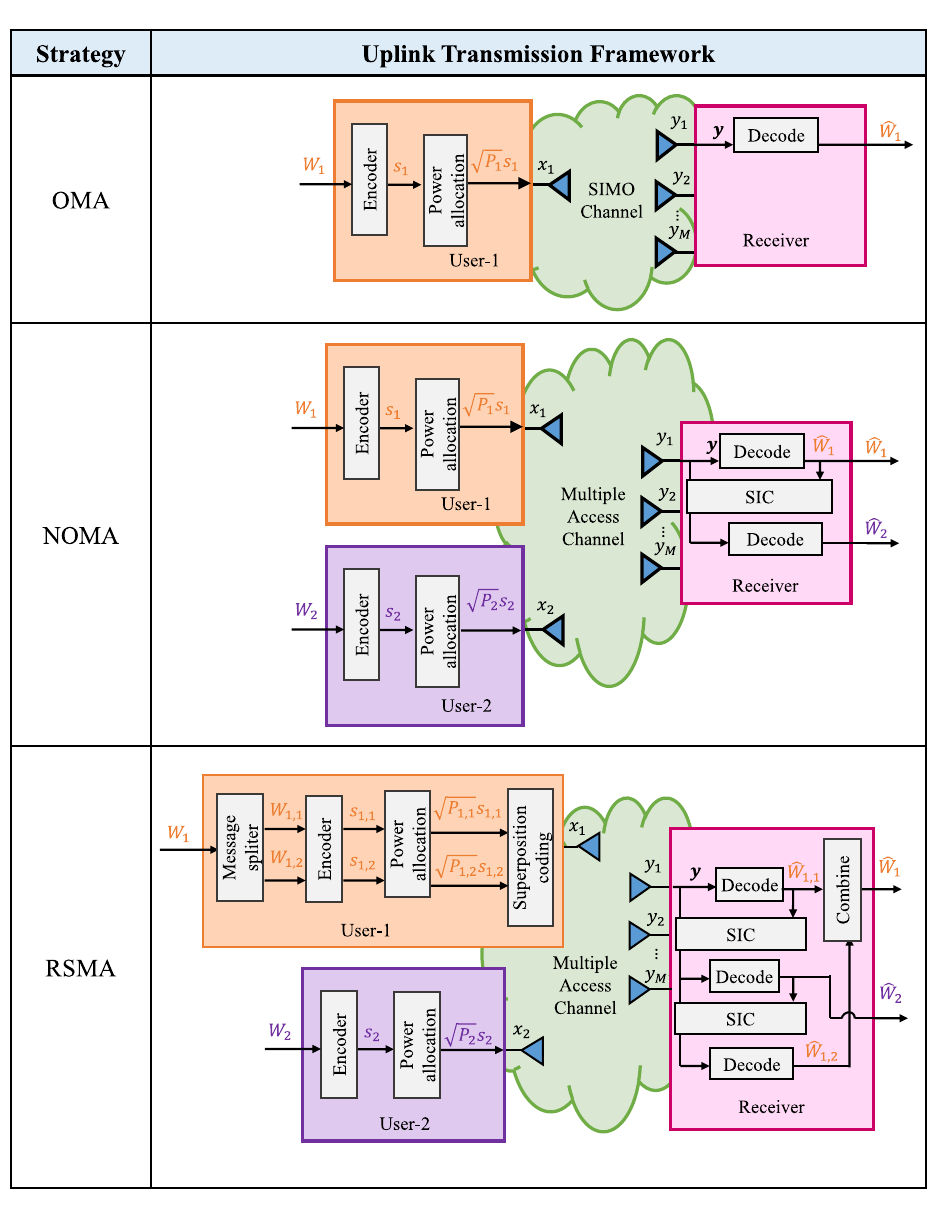}
\caption{Illustration of the two-user uplink transmission frameworks for OMA, NOMA, and RSMA.} \label{fig:UL_MA}
\end{figure}

 RSMA has been shown to generalize multicasting, OMA, SDMA, and NOMA schemes \cite{bruno2019wcl, Mao2022Survey}. Specifically, when all transmit power is allocated exclusively to the common stream or one private stream, RSMA respectively simplifies to the classical multicasting or OMA.  When the transmit power is solely allocated to the private streams, RSMA reduces to SDMA. Furthermore, in the case where each user message is totally encoded into distinct layers of common streams and private stream within the GRS framework, RSMA reduces to PD-NOMA. For this reason, RSMA is a superset and therefore always achieves equal or better performance compared to OMA, SDMA, and NOMA. A simple two-user downlink transmission framework comparison among multicasting, OMA, SDMA, NOMA, and RSMA is delineated in Fig. \ref{fig:DL_MA}, and the corresponding uplink transmission frameworks are illustrated in Fig. \ref{fig:UL_MA}. 

\par 
The main advantages and disadvantages of RSMA are summarized as following:
\begin{itemize}
    \item  \textit{Advantages:}

    \begin{enumerate}
	 	\item \textbf{Universality:} RSMA is a comprehensive MA framework encompassing OMA, SDMA, NOMA, and multicasting as its constituent sub-schemes \cite{bruno2019wcl,mao2017rate}. The universality of RSMA obviates the need for a system to consider switching between OMA, SDMA, NOMA, and multicasting.
	 	\item \textbf{Flexibility:} RSMA is highly adaptable to varying network loads, whether they are underloaded or overloaded, as well as to diverse user deployments characterized by varying channel directions and strengths \cite{mao2017rate,8019852,9382277}. This remarkable advantage of RSMA comes from its capability to flexibly handle multi-user interference through partial interference decoding and treating the remaining interference as noise.
	 	\item \textbf{Robustness:} The rapid growth of RSMA in multi-antenna networks is primarily attributed to its capacity to achieve an optimal spatial multiplexing gain in the multi-antenna BC with imperfect CSIT \cite{7152864,7555358,7513415,7805217,7972900,8000591}. A multitude of extant research endeavors has substantiated RSMA's resilience when confronted with CSIT uncertainties stemming from diverse sources of impairment, such as quantized feedback \cite{7152864,10071962}, pilot contamination \cite{9737523,9942944}, channel estimation inaccuracies, latency in feedback, user mobility \cite{9491092}, and even RF impairments such as phase noise \cite{7892949}.
	 	\item  \textbf{Enhanced spectrum efficiency:} RSMA is guaranteed to achieve the best sum-degree of freedom (DoF) in both perfect and imperfect CSIT, which translates into superior spectral efficiency as well.  In other words, RSMA outperforms other MA schemes in terms of spectral efficiency in both perfect and imperfect CSIT \cite{7555358,mao2017rate}. When CSIT is perfect, the achievable rate region of RSMA is larger than other MA schemes, and it approaches the capacity region achieved by DPC \cite{mao2017rate}. When CSIT is imperfect, RSMA can achieve a larger rate region than DPC \cite{mao2019beyondDPC}. RSMA's enhanced spectrum efficiency is not limited to conventional multi-user multiple-input single-output (MISO) unicast transmissions, but also holds for MIMO unicast \cite{8968439,9663192}, multigroup multicast \cite{8019852,9257433,8926413,9130871}, non-orthogonal unicast and multicast \cite{8846706}, non-terrestrial \cite{9257433,9844445,9684855}, network slicing \cite{10190330}. The benefits have been demonstrated using link-level simulations \cite{Sibo2023,9663192} and real-world experimentation \cite{lyu2023ratesplitting}.
   \item  \textbf{Enhanced energy efficiency:} RSMA not only boosts spectral efficiency but also energy efficiency and their trade-offs across various applications with different network loads and user deployments, as evidenced in many existing works \cite{8846706,9650662}. 
   \item  \textbf{Enhanced QoS and user fairness:} RSMA achieves the optimal max-min fair spatial multiplexing gain in multi-antenna BC with both perfect and imperfect CSIT. This advantage is reflected in the finite SNR regime, where RSMA demonstrates its superior max-min rate compared to the aforementioned MA schemes. Furthermore, the substantial spectral efficiency gain achieved by RSMA becomes even more pronounced when stringent QoS rate constraints are imposed on each individual user.
   \item  \textbf{Low complexity:} 1-layer RSMA stands out for its simplicity at both transmitter and receivers design. Its adaptability to varying network loads and user deployments obviates the need for user ordering, grouping, and scheduling at the transmitter. Each user only necessitates  a single SIC layer to decode and cancel the common stream. This contrasts with MIMO NOMA, which imposes stringent requirements for user grouping and ordering at the transmitter along with multiple layers of SIC at each user. 
   \item  \textbf{Coverage extension:} RSMA with user relaying, also known as cooperative rate-splitting (CRS) \cite{jian2019crs,mao2019maxmin}, has been demonstrated as a promising strategy to amplify the data rates for users located at the cell edge, offering substantial coverage extension benefits \cite{10124174}. CRS involves collaborative user relaying, enabling one user to decode and relay the common stream to other users, thereby enhancing user fairness particularly when jointly serving users with substantial discrepancies in channel strengths. 
   \item  \textbf{Low latency:} RSMA has demonstrated its prowess in improving throughput compared to the aforementioned MA schemes when employing finite-length (e.g. polar) codes, as highlighted in \cite{yunnuo2022FBLRS,yunnuo2023FBLRS,Jiawei2022}. In other words, RSMA achieves the same transmission rate as SDMA and NOMA but with shorter block lengths, resulting in reduced latency. Therefore, RSMA is a promising enabling technology for significantly reducing latency in URLLC services.
   \item \textbf{Security enhancement:} RSMA can adjust the level of confidentiality of its messages and consequently trade off spectral efficiency with secrecy \cite{9967957,Xia:2023}. This is not possible with PD-NOMA since the message of the weaker user is always decoded by a stronger user, which creates a secrecy threat.
    \end{enumerate}

    \item  \textit{Disadvantages:}

    \begin{enumerate}
		\item \textbf{SIC requirement at receiver:} Some RSMA schemes, such as GRS, exhibit a characteristic where the number of SIC layers at each user increases with the number of users. This trend not only leads to considerable burden on receiver complexity but  also introduces the error propagation issues. Promising SIC-free RSMA architectures have recently been developed \cite{Sibo2023}.
		\item \textbf{High encoding complexity:} Compared with the aforementioned MA schemes, RSMA requires more date streams to be encoded due to the additional common streams obtained from message splitting and recombining.
        \item \textbf{High signaling burden:} 
        In RSMA, additional signaling overhead is necessary to facilitate alignment between the transmitter and receivers, ensuring they possess a shared understanding of the methodology for splitting and combining user messages.
		\item \textbf{High optimization burden:} In terms of resource allocation and precoder design, RSMA requires the precoders (accounting for beamformer directions and power allocation) of the common and private streams to be jointly optimized with the common rate allocation. This joint optimization is instrumental in unlocking the complete benefits of RSMA.  However, it's worth noting that this expanded optimization space places a more demanding computational burden on RSMA in comparison to conventional MA schemes. 
    \end{enumerate}
\end{itemize}

\subsection{Code-Domain Multiple Access}

Code-domain NOMA (CD-NOMA) is inspired by the traditional CDMA \cite{Proakis08}. It allows multiple users to share the same time/frequency resources while dedicated interleavers and/or code sequences are employed to multiplex users. However, unlike CDMA, the spreading sequences in CD-NOMA are usually sparse or non-orthogonal low cross-correlation sequences. In general, CD-NOMA can be divided into \emph{sparse NOMA} and \emph{dense NOMA} \cite{9369968,9517908}, depending on the sparsity of the spreading sequences.
In this section, we review some popular CD-NOMA schemes and discuss their main advantages and disadvantages.

\subsubsection{Low-Density Spreading (LDS) MA}
One of the early CD-NOMA techniques is the LDS-CDMA \cite{4471881}, an extension to the classical CDMA. In an LDS-CDMA system, the data symbols of each user are spread by a unique LDS with a small fraction of non-zero entries and then superimposed before transmission. Compared to the classical CDMA, the interference on each chip can be reduced in LDS-CDMA. In particular, the multi-user interference pattern at the receiver entails a low-density graph, where message passing algorithm (MPA) or belief propagation (BP) \cite{910572} can be adopted for efficient symbol detection \cite{4481368}. To design LDS sequences, a structural approach was proposed in \cite{5425243}. Furthermore, the capacity region of LDS-CDMA was investigated in \cite{5910654}, where the impacts of spreading sequence density factor and the maximum number of users associated with each chip on the capacity were revealed. In addition, the idea of LDS-CDMA was extended to OFDM (LDS-OFDM) \cite{5493941} where each user's data symbols are spread across a number of carefully selected subcarriers. Significant improvement of peak-to-average-power ratio (PAPR) the link-level performance was reported for LDS-OFDM over OFDM \cite{6206872,6314235}.

\subsubsection{Sparse Code Multiple Access (SCMA)}
\begin{figure}[t]
				\centering
				\includegraphics[width=\linewidth]{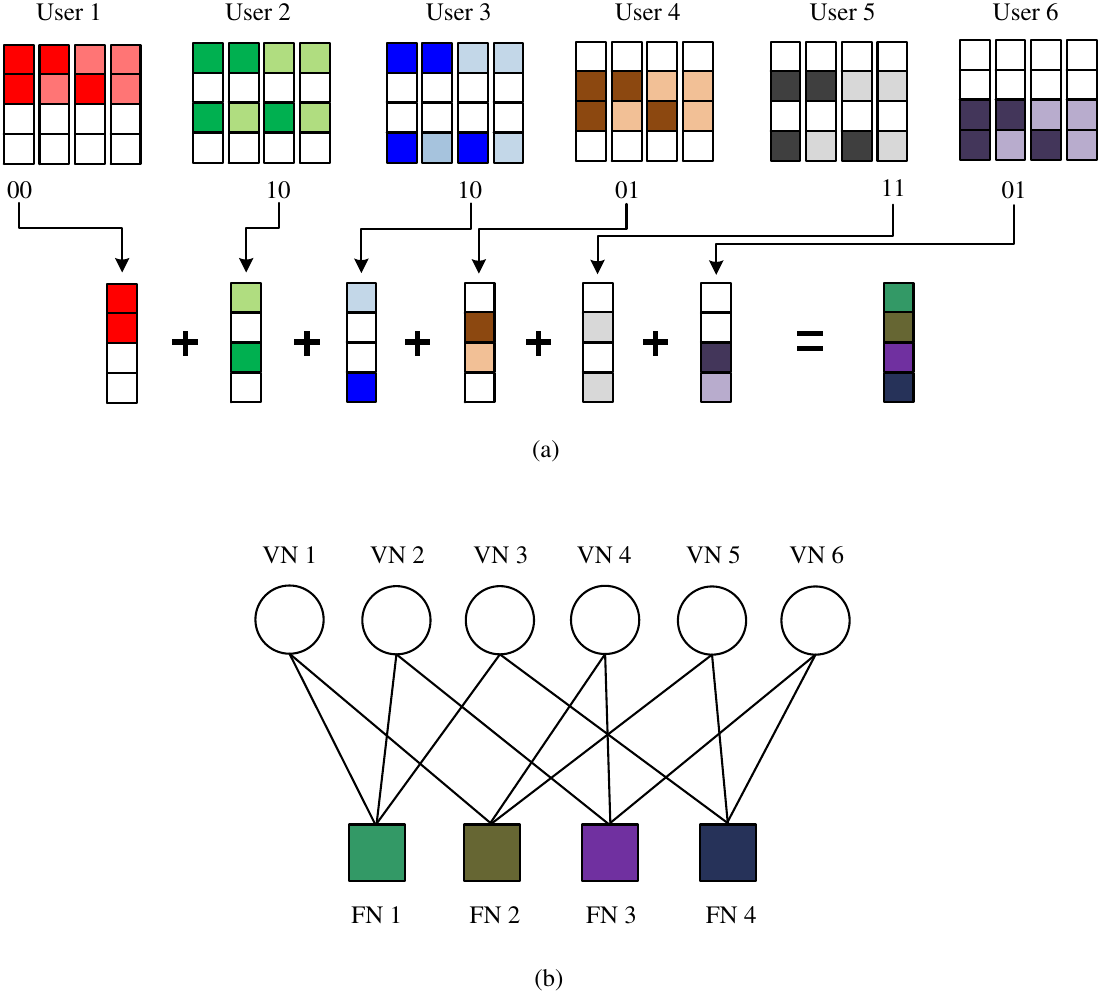}
				\caption{Illustration of a SCMA system with 6 users and 4 resource blocks: (a) mapping between each user's two bits and SCMA codewords; (b) factor graph corresponding to the SCMA system.} \label{fig:SCMA}
			\end{figure}

The concept of code domain multiplexing in LDS-CDMA was extended to SCMA \cite{Nikopour13}. In SCMA, each user is assigned a sparse codeword according to its message. In other words, the modulation symbol mapping and the spreading sequences are merged together such that the bits are directly mapped to a sparse vector of a multi-dimensional constellation. Hence, SCMA improves the spectral efficiency of LDS through shaping gains of multi-dimensional constellations. Meanwhile, it still inherits the benefits of LDS in terms of overloading and moderate complexity of detection. An example of a SCMA system with 6 users and 4 resources is illustrated in Fig. \ref{fig:SCMA}. Each user maps its two bits to its SCMA codeword such that the superimposed transmitted block spread over the 4 resource blocks as shown in Fig. \ref{fig:SCMA}(a). The resultant SCMA system can be represented by a factor graph depicted in Fig. \ref{fig:SCMA}(b), where every circle represents a user or a variable node (VN), every block represents a resource or factor node (FN), and the edge between VN $i$ and a FN $j$ means that user $i$'s data is mapped to the $j$-th resource block $j$ for $i\in\{1,\ldots,6\}$ and $j\in\{1,\ldots,4\}$. The factor graph is regular in the sense that each VN is of degree-2 while each FN is of degree-3. The SCMA codebook design based on lattice constellations \cite{485720} was investigated in \cite{6966170,8688492}. Similar to LDS-CDMA, the SCMA receiver adopts MPA. To reduce the detection complexity of MPA, various detection algorithms such as discretized MPA and sphere decoding can be employed \cite{8688492,9782313}. Potential applications of SCMA in 6G wireless communications systems were discussed in \cite{9409837}. However, the overloading performance is limited by the number of sparse codewords and the sparsity.

\subsubsection{Multi-User Shared Access (MUSA)}
MUSA was introduced in \cite{7504361} to support grant-free Internet of things \cite{9097306}. In MUSA, the data symbols of each user are spread with a short-length spreading sequence. The key principle is that non-orthogonal complex spreading sequences are chosen by multiple users autonomously for enabling grant-free transmissions on the same resources. It is also worth noting that the same user can choose different spreading sequences for different symbols to benefit from interference averaging. The receiver exploits the low cross-correlation properties of spreading sequences and uses SIC decoding. However, unlike LDS multiple access schemes and SCMA, the spreading sequences of MUSA are dense.

\subsubsection{Successive Interference Cancellation Amenable Multiple Access (SAMA)}
SAMA is based on the joint design of the system signature matrix and the SIC-based MPA \cite{7024798}. The symbols of different users are judiciously spread in the frequency, which can be effectively exploited by the SIC-based MPA and to obtain the diversity gain. Different from the aforementioned LDS schemes, the spreading sequences in SAMA have variable sparsity.


We summarize the main advantages and disadvantages of CD-NOMA schemes as follows.
\begin{itemize}
    \item  \textit{Advantages:}

    \begin{enumerate}
     	\item  \textbf{Enhanced spectrum efficiency:} CD-NOMA utilizes the code domain to allow multiple users share the same resources, e.g., time and frequency, efficiently. With advanced multi-user detectors, CD-NOMA improves the spectrum efficiency when CSIT is perfect and the network can be either underloaded or overloaded.
     
    \item \textbf{Enhanced user fairness:} User fairness in CD-NOMA can be achieved by detecting strong users' signals first in the multi-user detection. This is because the early detected users only collect less extrinsic information compared to the late detected users \cite{8713541}. In this way, more extrinsic information together with interference cancellation performed in the early detection steps can improve the detection performance of weak users.

	 	\item \textbf{Flexibility:} CD-NOMA is adaptable to various network loads and channel conditions. This is owing to the effective design of spreading code sequences in terms of the correlation and sparseness for reducing inter-user interference.
 
	 	\item \textbf{Low signaling overhead:} CD-NOMA can allow users to choose their spreading sequences autonomously to reduce the signaling overhead and latency. In this case, the spreading sequences need to be carefully designed for the underlying multi-user detectors to mitigate collisions.

    \end{enumerate}

    \item  \textit{Disadvantages:}

    \begin{enumerate}
		\item \textbf{High receiver complexity:} CD-NOMA requires complex multi-user detection techniques. In particular, to achieve the promised gain of CD-NOMA, iterative receiver architecture is often required. The complexity scales proportionally with the number of users and iteration numbers.

		\item \textbf{High design complexity:} The key ingredient of CD-NOMA is the spreading code sequence. However, finding the optimal design for the spreading sequence is still a difficult problem, particularly when the number of users is large.

		\item \textbf{High channel estimation complexity:} Most existing CD-NOMA schemes assume perfect CSI knowledge. To accurately estimate the channel for each user, channel estimation is often performed jointly with multi-user detection, which incurs a high complexity.
 
        \item \textbf{High sensitivity to synchronization:} Most existing CD-NOMA schemes assume perfect synchronization at the receiver. However, this is not always the case in practice. A delayed signal from a user could disturb the message exchange in the iterative multi-user detection, leading to performance loss.

    \end{enumerate}
\end{itemize}

\subsection{Multiple Access in Other Domains}


In addition to the prominent SDMA, PD-NOMA, RSMA, and CD-NOMA, a range of MA schemes in other domains have been proposed in the literature. In particular, a full list of MA schemes proposed during the study phase of 5G NR were included in \cite{NOMA38812}. In what follows, we will review some typical examples from \cite{NOMA38812} as well as some newly proposed MA schemes.

\subsubsection{Interleave-Division Multiple Access (IDMA)}
IDMA was introduced in \cite{IDMA03} in 2003 and has gained considerable interest in recent years. IDMA can be considered as a spectral case of CDMA with all signature sequences reducing to a single chip \cite{1452830}. In other words, IDMA relies on user-specific interleaving as the only means to distinguish the signals from different users \cite{1618943}. Moreover, the spreading operations in CDMA can be replaced by low-rate forward error correction codes in IDMA to provide increased coding gain. It is also possible to have non-orthogonal interleavers. The IDMA receivers adopt a chip-by-chip elementary signal estimation, such that the total receiver complexity only increases linearly with the number of users \cite{4099341}. Advantages of IDMA over CDMA in terms of performance and complexity under practical considerations were demonstrated in \cite{6092790}. Integration of IDMA with other technologies such as OFDM and massive MIMO were investigated in \cite{4279212} and \cite{7974726}, respectively.

\subsubsection{Pattern Division Multiple Access (PDMA)}
In PDMA \cite{PDMA15}, the transmitted data symbols are mapped to a resource group that can consist of time, frequency, and spatial resources or any combination of these resources according to a pattern. Data of multiple users can be multiplexed onto the same resource group with a different pattern to realize non-orthogonal transmission. In addition, the pattern is designed with disparate diversity order and sparsity such that the BP algorithm can be efficiently employed for detection. To achieve the best possible performance, an iterative turbo receiver architecture \cite{774855} can be adopted, where the outer iteration process between the detector and decoder is performed on top of the inner iteration of the detection and decoder themselves. A gain of 500\% in terms of the number of supported users under the given system packet drop
rate of 1\% was reported for PDMA over OFDMA in the uplink \cite{7526461}. In \cite{8352623}, the patterns of different users are judiciously designed to exhibit appropriate diversity
disparity at the symbol level and power disparity at the physical resource element level. In this way, the appropriate disparity in diversity and power can be
effectively exploited by the low-complexity SIC-based BP detector with reduced error propagation during interference cancellation.

\subsubsection{Compute-Forward Multiple Access (CFMA)}
Compute-and-forward (C\&F) is a relaying strategy introduced in \cite{6034734}, where the relay decodes a noisy linear combination of lattice codewords. In contrast to the classic relaying strategies such as amplify-and-forward and decode-and-forward, C\&F exploits interference to obtain significantly higher rates between users in a network. Motivated by the benefits of C\&F, \cite{7558212} introduced CFMA based on a modified C\&F technique for the Gaussian MAC. The receiver first decodes the sum of the linear combination of lattice codewords. Upon recovering the sum codewords, other users' codewords can be successfully decoded by using the sum as side information. It was proved in \cite{7558212} that whole capacity region of the two-user Gaussian MAC is achievable by CFMA under lattice decoding \cite{1337105}, provided that the SNR of both user is larger than $1+\sqrt{2}$. It is worth mentioning that any rate points on the dominant face of the Gaussian MAC capacity region can be achieved by CFMA, without the need of time-sharing \cite[Ch. 15]{coverbook06} or rate-splitting \cite{485709}. However, random lattice codes were used in \cite{7558212} as the proof techniques, which can be difficult to implement in practice. Practical implementation of CFMA based on off-the-shelf binary low-density parity-check (LDPC) codes and sum-product decoding was conducted in \cite{8485351}.

\subsubsection{Lattice-Partition-Based Multiple Access (LPBMA) without SIC}

	\begin{figure}[t]
				\centering
				\includegraphics[width=\linewidth]{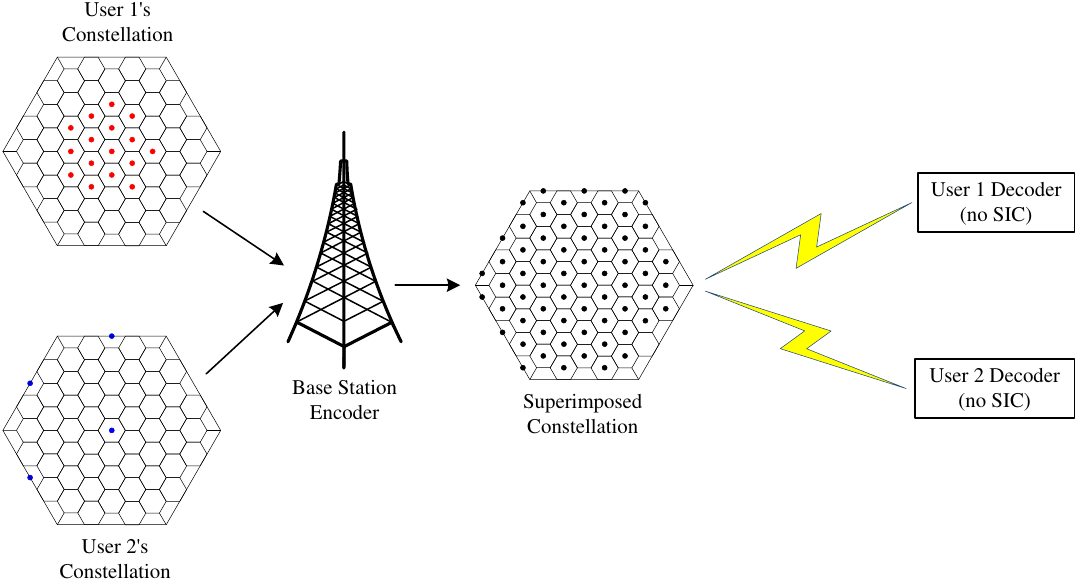}
				\caption{An illustration of the lattice-partition-based multiple access with single-user TIN decoding.} \label{fig:LPBMA}
			\end{figure}

In \cite{8291591}, A lattice-partition framework of MA without SIC, i.e., with single-user treating interference as noise (TIN) decoding, was introduced for the $K$-user Gaussian BC. Each user adopts a linear code \cite{8066336} in conjunction with an appropriately designed constellation carved from a multi-dimensional lattice which includes the commonly adopted one- and two-dimensional constellations, i.e., pulse amplitude modulation (PAM) and quadrature amplitude modulation (QAM), as special cases. Essentially, the design ensures that the superposition of all users' signalings still preserves the lattice structure, which can be exploited to hardness inter-user interference in TIN decoding. A two-user example is illustrated in Fig. \ref{fig:LPBMA}, where both users' signal constellations and the superimposed constellation preserve the same lattice structure. It was proved that the scheme based on discrete signaling and TIN is capable of achieving the whole capacity region within a constant gap independent of the number of users and channel parameters. Several new MA schemes evolving from lattice-partition-based multiple access were proposed for various channels, e.g., the Gaussian interference channel, the Gaussian BC with heterogeneous blocklength and error probability constraints, etc. \cite{8517129,8731926,9535131,10138394}, where carefully designed discrete signalings with TIN decoding are shown to be close to Gaussian signalings with perfect SIC decoding analytically and numerically.

\subsubsection{Spatially Coupled Multiple Access (SC-MA)}

\begin{figure}[t]
				\centering
				\includegraphics[width=\linewidth]{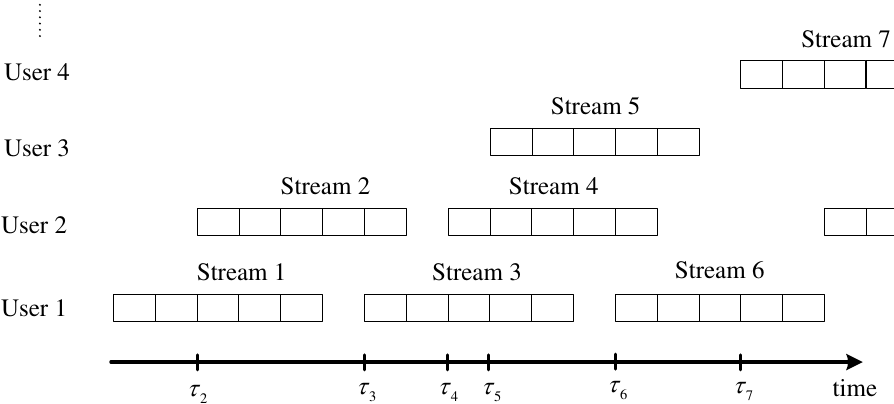}
				\caption{Examples of spatially coupled multiple access where data streams are transmitted with different delays.} \label{fig:Coupled_MA}
			\end{figure}

Spatial coupling \cite{782171} is a code construction technique to boost the performance of uncoupled (weak) codes, e.g., regular LDPC codes and turbo codes, all the way to capacity-achieving \cite{5571910,5695130,8002601,9851473}. Motivated by this, \cite{6620553,8638808} introduced a new signaling formats for the Gaussian MAC, where the modulated data streams are repeated, permuted, and transmitted with regular time offsets (delays). Since the relations between the data bits and modulation symbols transmitted over the channel can be represented by a sparse graph, the receiver observes spatial coupling of the individual graphs which enables efficient demodulation/decoding. It was proved in \cite{6620553,8638808} that coupling data transmission with a two-stage demodulation/decoding in which iterative demodulation based on symbol detection and interference cancellation followed by parallel decoding achieves the Gaussian MAC capacity asymptotically such that the gap to capacity vanishes as the system's SNR increases. Note that the feedback between the decoder and the demodulator is not required. In contrast to most NOMA receivers that have to demodulate/decode the whole symbol block, sliding window demodulation/decoding can be adopted thanks to the spatially coupled graph structure \cite{8525313}. As a result, the demodulation/decoding latency is limited, which can be suitable for streaming services. \cite{9044307} extended the spatially coupled transmission scheme to the case of transmitted blocks with irregular delays. An example for the transmission scheme is illustrated in Fig. \ref{fig:Coupled_MA}, where each user divides its transmission frame into sub-blocks and delays the transmission by $\tau$. However, the spatially coupled transmission scheme can be sensitive to synchronization errors and fractional delay.

\subsubsection{Layered-Division Multiplexing (LDM)}
LDM is a non-orthogonal multiplexing technology adopted in the Advanced Television Systems Committee standard (ATSC 3.0) \cite{7378924}, which has never been implemented in previous broadcast and broadband systems \cite{8626084}. In an LDM system, a layered transmission structure is used to simultaneously transmit multiple signals with different power levels and technologies (channel coding, interleaving, modulation, multiple-antenna, etc.) for different services \cite{8316773}. At the receiver, the decoding of signals from multiple layers requires SIC. The principle of LDM is similar to PD-NOMA. Hence, its gain over the conventional OMA schemes has been well understood. Note that LDM is implemented on top of OFDM such that the superposition of signals take place before OFDM modulations.

\subsubsection{Index Modulation Multiple Access (IMMA)}
Index modulation is a technique of using the indices of resources to carry extra information bits \cite{6587554}. It can achieve higher spectral and energy efficiency than conventional modulation schemes. Motivated by this, IMMA was proposed to further improve the spectral and energy efficiency of the traditional NOMA scheme \cite{8580958}. In IMMA, each user transmits additional information bits (index bits) by partial resources, where the index patterns can be the activation status of time slots, subcarriers, transmit or receive antennas, spreading codes, power levels, etc \cite{10110011}. However, the detectors of these IMMA schemes are joint maximum-likelihood detectors which can be challenging to implement.

\subsubsection{Delay-Doppler Domain Multiple Access (DDMA)}
Orthogonal time-frequency space (OTFS) modulation \cite{7925924,9508932,DD_com_book2002} was introduced for high-mobility wireless applications, which multiplexes data in the delay-Doppler (DD) domain rather than the time-frequency domain as in conventional multi-carrier modulations. OTFS has also stimulated new research on delay-Doppler plane modulations. For instance, the newly proposed orthogonal delay-Doppler division multiplexing (ODDM) \cite{9829188} was proved to achieve orthogonality on the DD plane's fine resolutions with realizable pulses. Owing to the orthogonality, ODDM itself can be potentially employed as an OMA scheme on the DD domain. Another promising research area is that other MA schemes such as SDMA and RSMA can be combined with OTFS/ODDM, which can offer high spectral efficiency with robust performance in high-mobility channels.

\subsubsection{Other Multiple Access}
All the above discussion on MA techniques is not exhaustive with new MA strategies disclosed on a regular basis such as fluid antenna multiple access (FAMA) \cite{9650760}, location division multiple access (LDMA) for near-field communication \cite{10123941}, or orbital angular momentum (OAM) multiplexing and multiple access.

\subsection{Technology Outlook and Future Works}\label{UMA_section}
We draw several observations from the past subsections on common challenges in MA designs and on MA dimensions before giving an outlook of MA technology development toward Universal Multiple Access.

\subsubsection{Common Challenges in MA designs} Modern and emerging MA schemes share a number of common challenges, namely their optimization complexity (of precoders and resource allocation), channel estimation complexity, receiver complexity (involving highly complex operations at the receivers such as SIC), and sensitivity to various impairments (can be synchronization or RF impairments). 

\subsubsection{MA dimensions}
\par Previous discussions highlight that there are numerous dimensions that a MA scheme can exploit, starting from the usual time, frequency, space, power, code, but new dimensions have also appeared in recent years including message, index, pattern, interleaver, etc. We have decided to classify them all into five categories: time, frequency, power, space, signal. Signal is a broad category that encompasses multiple sub-dimensions such as message split and combiner, channel coding, modulation, interleaver, spreading sequence, etc. Table \ref{fig:MA_dimensions_table} displays a non-exhaustive list of dimensions exploited by the aforementioned MA schemes and how aforementioned MA schemes exploit those various dimensions.

\begin{table*}[h]
    \centering
     \caption{MA dimension classification and mapping of MA schemes to MA dimensions. RSMA basic schemes exploit space, power, and message splitting dimensions, though space-time and space-frequency RSMA schemes exist \cite{10038476}, hence exploiting a fourth dimension. IMMA has primarily two different schemes, and the current designs utilize three dimensions at the same time to distinguish users.}
    \label{fig:MA_dimensions_table}
    \begin{tabular}{|c|c|c|c|c|c|c|c|c|c|c|}
        \hline \multirow{3}{*}{\diagbox[width=\dimexpr \textwidth/8+2\tabcolsep\relax, height=0.96cm]{\textbf{ MA} }{\textbf{Dimension}}} & \multirow{3}{*}{\textbf{Time}}& \multirow{3}{*}{\textbf{Frequency}} & \multirow{3}{*}{\textbf{Space}} & \multirow{3}{*}{\textbf{Power}} &\multicolumn{6}{c|}{ \textbf{Signal} } \\
        \cline{6-11}   &  &   &   &    & \emph{Message}  & \emph{Message} & \emph{Channel}  & \multirow{2}{*}{\emph{Modulation}} & \multirow{2}{*}{\emph{Interleaver}} & \emph{Spreading} \\
          &  &   &   &    & \emph{combiner} & \emph{split} & \emph{coding} &  &  & \emph{sequence} \\
        \hline TDMA & $\boldcheckmark$ & & & & & & & & &\\
\hline FDMA & &$\boldcheckmark$ & & & & & & & &\\
\hline OFDMA & $\boldcheckmark$&$\boldcheckmark$ & & & & & & & &\\
\hline SDMA &  &  & $\boldcheckmark$ & & & & & & &\\
\hline PD-NOMA &  &  & &$\boldcheckmark$ & & & & & &\\
\hline PHY Multicasting &  &  & & &$\boldcheckmark$ & & & & &\\
\hline RSMA & \textcolor[rgb]{1.00,0.00,0.00}{$\boldcheckmark$} & \textcolor[rgb]{0.00,0.07,1.00}{$\boldcheckmark$}  & $\boldcheckmark$ \textcolor[rgb]{1.00,0.00,0.00}{$\boldcheckmark$} \textcolor[rgb]{0.00,0.07,1.00}{$\boldcheckmark$}&$\boldcheckmark$ \textcolor[rgb]{1.00,0.00,0.00}{$\boldcheckmark$}\textcolor[rgb]{0.00,0.07,1.00}{$\boldcheckmark$}& $\boldcheckmark$ \textcolor[rgb]{1.00,0.00,0.00}{$\boldcheckmark$}\textcolor[rgb]{0.00,0.07,1.00}{$\boldcheckmark$}& $\boldcheckmark$\textcolor[rgb]{1.00,0.00,0.00}{$\boldcheckmark$}\textcolor[rgb]{0.00,0.07,1.00}{$\boldcheckmark$} & & & &\\
\hline CD-NOMA & $\boldcheckmark$ & $\boldcheckmark$ & & & & & & & & $\boldcheckmark$ \\
\hline IDMA &  &  & & & & & $\boldcheckmark$& &$\boldcheckmark$ &\\
\hline PDMA & $\boldcheckmark$ & $\boldcheckmark$ &$\boldcheckmark$ & & & & & & &\\
\hline CFMA &  &  & &$\boldcheckmark$ & && $\boldcheckmark$& $\boldcheckmark$& &\\
\hline LPBMA &  &  & &$\boldcheckmark$ & & & $\boldcheckmark$ &$\boldcheckmark$ & &\\
\hline SC-MA &  &  & & & & & & &$\boldcheckmark$ &$\boldcheckmark$\\
\hline LDM & $\boldcheckmark$ & $\boldcheckmark$ & &$\boldcheckmark$ & & & & & &\\
\hline IMMA & \textcolor[rgb]{0.00,0.07,1.00}{$\boldcheckmark$}\textcolor[rgb]{1.00,0.00,0.00}{$\boldcheckmark$} &\textcolor[rgb]{0.00,0.07,1.00}{$\boldcheckmark$}  & \textcolor[rgb]{1.00,0.00,0.00}{$\boldcheckmark$} & & && &\textcolor[rgb]{0.00,0.07,1.00}{$\boldcheckmark$}\textcolor[rgb]{1.00,0.00,0.00}{$\boldcheckmark$} & &\\
\hline DDMA & $\boldcheckmark$ & $\boldcheckmark$ & & & & & &$\boldcheckmark$ & &\\
        \hline
    \end{tabular}
\end{table*}


\par Table \ref{fig:MA_dimensions_table} also highlights that MA techniques exploit one or multiple of those dimensions. For instance, SDMA and PD-NOMA exploit the space domain and the power domain, respectively, while RSMA by opening the door to the message dimension exploits the message combiner, message split, power, and space domains to generalize and unify SDMA, NOMA, and physical layer multicasting. PDMA exploits space, time and frequency domains and CD-NOMA schemes exploit the spreading code domain together with time and frequency. All five domains would be exploited by OFDM(A)-RSMA \cite{10236464,sahin2023ofdmrsma}. 

\subsubsection{RSMA - A First Step Toward Unification in 6G}



The beauty of RSMA is that it \textit{unifies} into a single MA scheme the seemingly unrelated strategies of PD-NOMA, SDMA, and physical layer multicasting. This unification capability translates in RSMA to be a superset of those three strategies and can boil down to any of them by turning off some of the streams. This was illustrated in Fig. \ref{fig:DL_MA} for a two user scenario and has been discussed extensively in the NOMA and RSMA literature \cite{Mao2022Survey,10038476,9451194} and demonstrated by the well known message-to-stream mapping of \cite{bruno2019wcl,10038476}. Consequently, RSMA can softly bridge those three strategies, explore operating points that are not achievable by any of them, and outperforms them all. This gives RSMA an edge over other MA schemes as demonstrated in over 40 applications in 6G \cite{10038476,Mao2022Survey}, including in multi-functional networks such as ISAC \cite{9531484}.

\par
\begin{figure*}[t]
\centering
\includegraphics[width=0.8\linewidth]{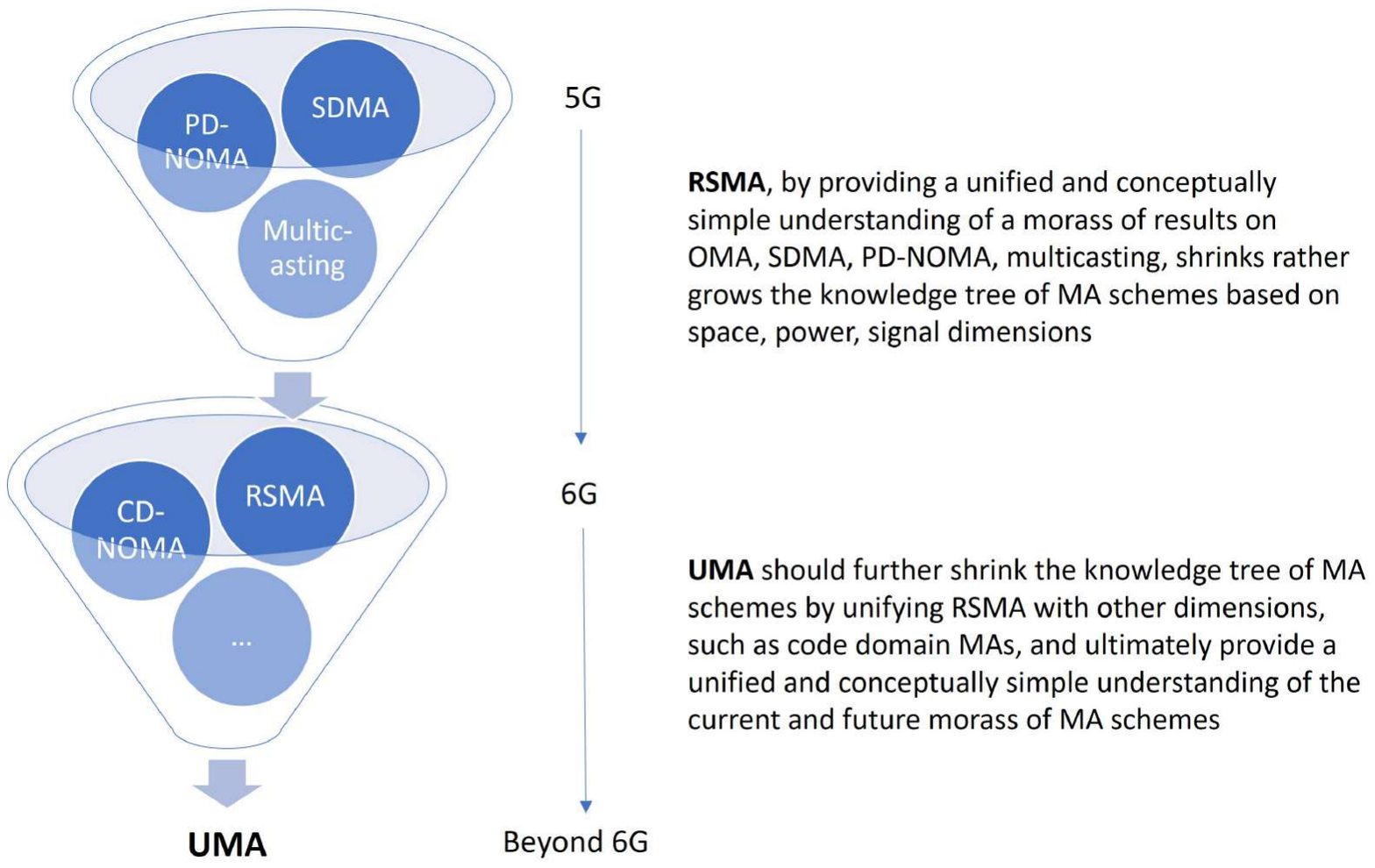}
\caption{From RSMA to UMA: How to further shrink the knowledge tree.} \label{fig:RSMA_UMA_filter}
\end{figure*}

 The capability of RSMA to \textit{unify} MA schemes is crucial for the long term research in the theory and practice of communications and wireless systems. Recall indeed the wise words in the acknowledgments section of the book of Prof. David Tse and Prof. Pramod Viswanath ``Bob Gallager’s research and teaching style have greatly inspired our writing of this book. He has taught us that good theory, by providing a unified and conceptually simple understanding of a morass of results, should shrink rather than grow the knowledge tree." \cite{tsefundamentalWC2005}. In our context, as illustrated in Fig. \ref{fig:RSMA_UMA_filter} and by the RSMA branch in Fig. \ref{fig:RSMA_tree}, \textit{RSMA, by providing a unified and conceptually simple understanding of a morass of results on SDMA, PD-NOMA, physical layer multicasting, shrinks rather grows the knowledge tree of MA schemes based on space, power, signal dimensions.} The capability of RSMA to unify and therefore be more universal than other MA schemes makes practical implementation and operation easier. Indeed a single unified and general MA scheme would be easier to implement and optimize than a combination of multiple MA schemes, each optimized for specific conditions. This is increasingly important in multi-functional 6G and beyond networks where the range and diversity of services, use cases, and deployments explode. 

\subsubsection{Toward Universal Multiple Access in Beyond 6G}
\par Though RSMA provides a good example of a unified theory of MA schemes, RSMA does not unify all MA schemes and therefore does not exploit all dimensions of time, frequency, power, space (e.g. antennas, beams), signal (e.g. messages, codes, etc). 

\par This would then bring the central question for future research on MA in beyond 6G: ``\textit{What is Universal Multiple Access (UMA)?}". Following the same wise philosophy of Prof. Gallager, Prof. Tse, and Prof. Viswanath, \textit{UMA should further shrink the knowledge tree of MA schemes by unifying RSMA with all other dimensions, such as code domain MAs, and ultimately provide a unified and conceptually simple understanding of the current and future morass of MA schemes.} This question, vision, and research philosophy is illustrated in Fig. \ref{fig:RSMA_UMA_filter} and Fig. \ref{fig:RSMA_tree}.
Following the above question, come the natural and important questions ``\textit{How to design UMA?}" and ``\textit{Why and when do we need UMA?}". 
\par
\begin{figure}[t]
\centering
\includegraphics[width=\linewidth]{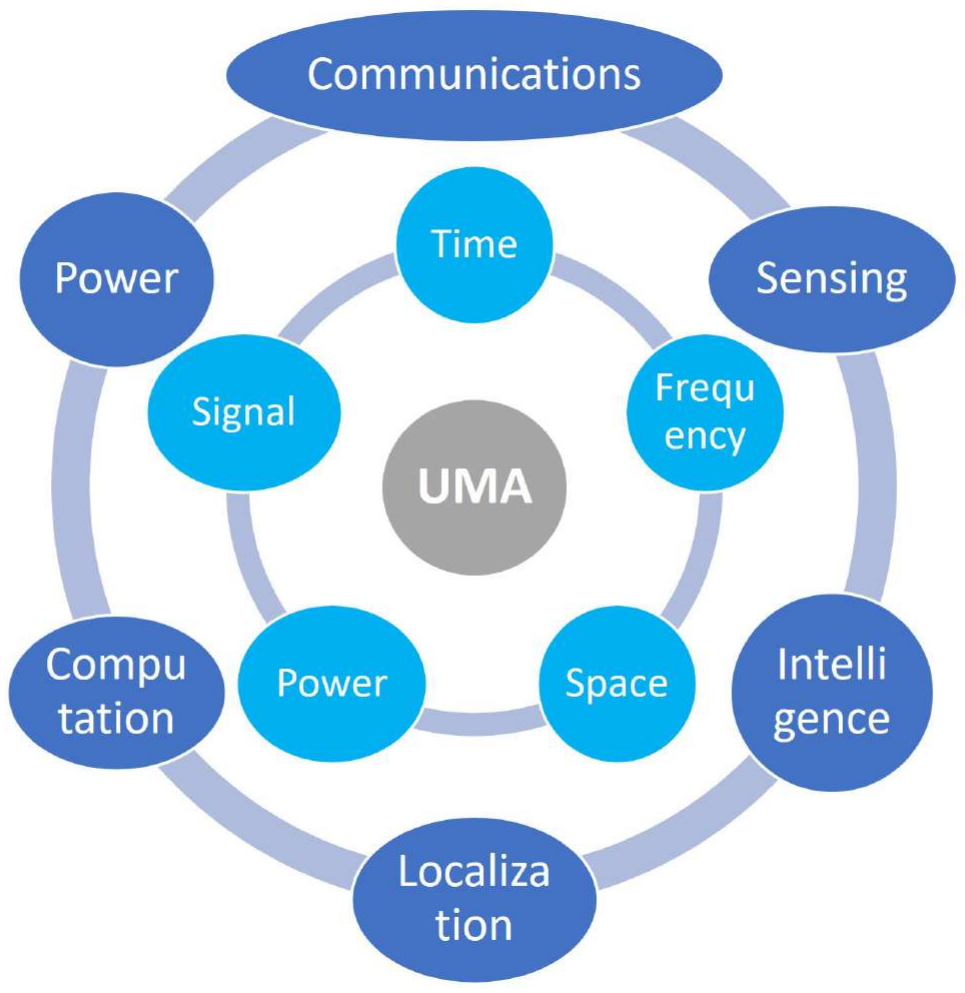}
\caption{UMA exploiting all five dimensions to enable multi-functional 6G.} \label{fig:UMA_dimensions_multifunctional}
\end{figure}

 There are no answers to those quesitons yet since UMA does not exist. Those  questions nevertheless aim to trigger discussions, give research
directions, and motivate further research. Addressing them is very timely for 6G and beyond for two main reasons. First, the literature and the number of MA schemes have exploded in the past decade and grown non-organically, fueled by the sheer interest in making new or better use of MA dimensions and the limited resources but also by the large number of new wireless services offered by multi-functional and intelligent 6G. Many of those schemes claim to exploit new dimensions and presenting new ideas, though it remains to be seen whether this is true or whether those are a rebranding or twist of known concepts. Second, the multi-functionality of wireless networks calls for a more efficient, flexible and robust use of resources, better management of interference, better handling of heterogeneity of wireless services, unified and simplified hardware and software architectures. As illustrated in Fig. \ref{fig:UMA_dimensions_multifunctional}, UMA should shrink the knowledge tree by truly understanding the essence of a unified MA design at exploiting in the most simple way all five dimensions (time, frequency, space, power, signal) to efficiently provide intelligence and multi-functionality (communications, sensing, localization, computation, energy transfer and harvesting) in 6G and beyond network.

\par Many MA schemes exploiting the code domain have been classified as CD-NOMA with LDS and SCMA being instance of sparse CD-NOMA and MUSA and SAMA of dense CD-NOMA. However, CD-NOMA here only stands for a collection MA schemes exploiting various properties of codes but CD-NOMA is not a MA scheme itself that unifies LDS, SCMA, MUSA, SAMA, i.e., there is no MA scheme that unifies LDS, SCMA, MUSA, SAMA and enables to softly bridge them all. Hence CD-NOMA does not have the same unification capability as RSMA. Nevertheless, it is important to observe that RSMA and CD-NOMA span different MA dimensions, with e.g., RSMA not exploiting the spreading code dimension. Shrinking the knowledge tree of MA schemes to identify UMA would require a better understanding of how all CD-NOMA schemes are linked to each other and whether they form particular instances of a more general class of MA schemes, and a better understanding of the interplay between RSMA and CD-NOMA. Unfortunately such research avenues remain largely unexplored. Indeed, a number of works have attempted to combine CD-NOMA with other MA schemes. The combination of PD-NOMA and SCMA was investigated in \cite{8115155,8647770,8689104}. The motivation is to allow the same SCMA codeword to be used by multiple users simultaneously, thereby increasing the number of supported users. At the receiver, MPA combined with SIC decoder is employed. However, the complexity would be higher than that of a single SCMA or a PD-NOMA scheme. In \cite{9400763}, CD-NOMA is combined with SDMA in the massive MIMO setting, where the cases of CD-NOMA offering spectral efficiency improvement were identified. The integration of CD-NOMA into RSMA was suggested in \cite{10038476}. Since the split common and private messages in RSMA can be seen as virtual users, each virtual user can be assigned a dedicated spreading sequence for enabling code-domain multiplexing. However, such an idea has not been explored further.

\section{Artificial Intelligence for Multiple Access} \label{Section_AI_for_MA}
In this section, we demonstrate how AI techniques can be used to address some of the challenges highlighted in the previous section. We start by providing an overview of AI methods and delve into how they can be applied for resource allocation and optimization for different MAs. We then discuss AI-empowered channel estimation, receiver design, and user behavior predictions, and finish the section by providing some outlook of AI for UMA.

\subsection{AI-empowered MA Resource Allocation and Optimization}


\par 
In every wireless communication system, there is a constant challenge: harnessing limited resources to achieve improved performance while satisfying various QoS requirements such as rate, latency, and reliability.  Therefore, the allocation of wireless resources is of critical importance in the pursuit of optimized network performance and the fulfillment of the multifaceted needs inherent to wireless communication systems.
Different MA schemes introduce distinct resources for allocation. For instance, OFDMA introduces the challenge of subcarrier allocation among users, TDMA involves time slot allocation, NOMA introduces user pairing, grouping, decoding order, and power allocation, SDMA incorporates beamforming, power allocation, and user scheduling, RSMA centers on beamforming, power allocation, common rate allocation design, and SCMA deals with sparse code allocation among users. Essentially, the choice of the optimal resource allocation approach relies on various factors, including the specific MA scheme, available resources, system requirements, and the intricate interplay of these key factors.

\par  
Conventional resource allocation algorithms typically rely on optimization theory and game theory. While these methods can yield mathematically optimal, sub-optimal, or Nash equilibrium solutions, they often come with a high computational cost.
In the future 6G wireless networks, the substantial proliferation of communication devices and antennas, along with the exponential growth in data traffic, brings formidable optimization challenges at a large scale. As a result, these conventional resource allocation algorithms will face an increased computational burden.

\par
\begin{figure*}[t]
\centering
\includegraphics[width=0.9\linewidth]{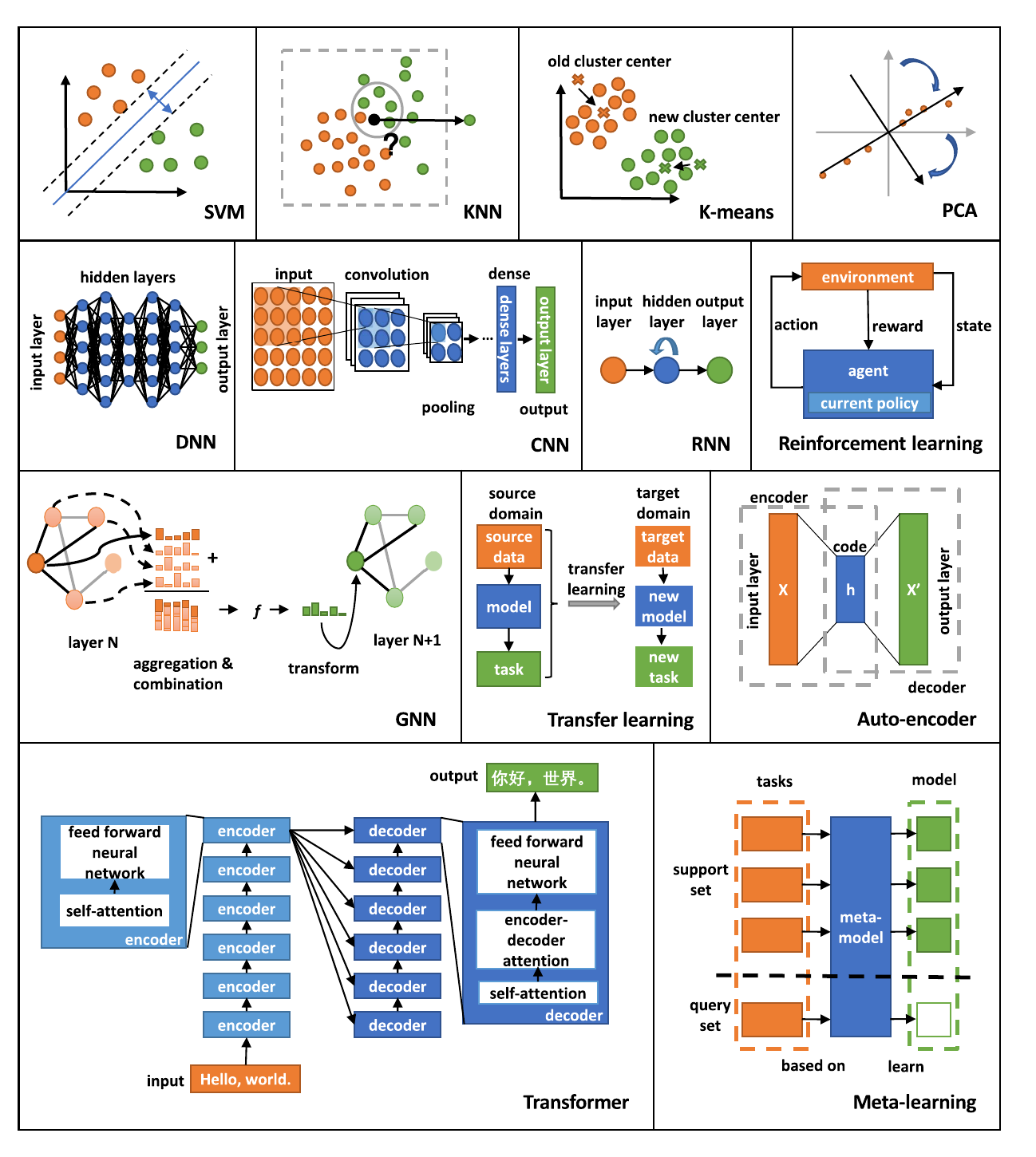}
\caption{A non-exhaustive search of AI methods that can be applied for resource allocation.} \label{fig:AImethods}
\end{figure*}

 
The rapid advancement of AI in the past decades  has opened up a new approach to tackling the complex, non-convex problems often encountered in resource allocation. This approach allows the resource allocation scheme to be learned directly from data samples or environment, eliminating the need for complex mathematical models.
Fig. \ref{fig:AImethods} shows a non-exhaustive search of the potential AI methods that can be applied for resource allocation.  Broadly, AI-based resource allocation methods fall into three categories: traditional ML, reinforcement learning (RL), and deep learning (DL).  In the following, we provide a concise overview of these three methodologies and how those AI-based approaches can be applied to diverse resource allocation scenarios, followed by a summary of the advantages and disadvantages for using AI-based approaches.

\subsubsection{Machine learning} The term ``machine learning (ML)" originally referred to the development of algorithms that enabled machines to learn and solve specific problems. With the growing development of neural network-related methods, ML now primarily refers to traditional learning approaches that do not rely on  neural network. Common ML methods include support vector machine (SVM), K nearest neighbors (KNN), and K-means clustering and principal component analysis (PCA). 
\par     
SVM is often used for resource allocation scenarios that involve binary data classification. For instance, it can  be applied to spectrum sensing in dynamic spectrum access (DSA) systems. In this application, SVM is used to classify segments of the spectrum as either occupied or available for secondary users based on received signal characteristics, enabling efficient spectrum utilization \cite{SS2020TCOM}. 
KNN works by finding the K closest data points in the training data to a test data point, then predicting or classifying based on the majority label or aggregated values of those nearest neighbors. It's non-parametric, so it doesn't make any assumptions about the data distribution. In applications like user grouping, KNN helps clustering or categorizing users based on their similarity  in terms of channel conditions or interference patterns \cite{KNN2019TCOM}.
SVM and KNN are both supervised learning algorithms, and they share common challenges when dealing with large datasets due to their high computational complexity. For KNN, it suffers the ``curse of dimensionality" wherein data points tend to become equidistant from each other in high-dimensional spaces, posing difficulties in identifying meaningful nearest neighbors.
In contrast, both K-means clustering and PCA are unsupervised learning algorithms. K-means clustering  can be applied to wireless resource allocation problems such as user grouping, spectrum allocation. For example, a K-means based  user clustering algorithm is developed in \cite{Kmeans2018TWC} for NOMA by exploiting the channel correlation among users.  PCA is a dimensionality reduction technique and itself is not a direct resource allocation algorithm for wireless networks. However, it plays a crucial role as a preprocessing and analysis tool that enhances the quality of data used by resource allocation algorithms \cite{PCA2022WCL}.
\par 
All the aforementioned non-neural network-based ML algorithms offer straightforward interpretations of results. However, their effectiveness is  pronounced when dealing with data characterized by straightforward relationships. They may encounter difficulties in capturing intricate hierarchical representations within complex data relationships. In such case, DL or neural works demonstrate superior performance.

\subsubsection{Deep Learning} 
The DL approach, powered by  neural networks, is one of the most renowned techniques in the field of AI  for its success in addressing various challenges, such as image recognition and natural language processing. Within the realm of wireless resource allocation, DL has shown to be valuable thanks to its capacity to efficiently process vast and intricate datasets. It facilitates intelligent and adaptive resource allocation, continually optimizing network performance in real-time. Popular neural networks mainly consist of dense neural networks (DNN), convolution neural networks (CNN), recurrent neural networks (RNN), and graph neural networks (GNN). Each of these architectures has demonstrated its prowess in addressing specific resource allocation challenges for different MA schemes.

\par 
\textit{DNN}: DNN is the most fundamental neural network architecture, often referred to as multi-layer perception (MLP). Within this architecture,  neurons are associated with input weights and incorporate an activation function to produce outputs. The computation unfolds layer by layer, with parameter updates facilitated through the process of backpropagation. It's worth noting that this structural framework finds application in various other types of neural networks as well.
DNNs have been employed to approximate the weighted minimum mean square error (WMMSE) algorithm, a versatile non-convex optimization technique for resource allocation, in an end-to-end manner \cite{DNN2018TSP}. This pioneering approach signifies a paradigm shift in optimization known as ``learn to optimize". It has demonstrated that a three-layer DNN can closely approximate WMMSE while achieving significant computational time reductions, often spanning orders of magnitude. Considering the widespread adoption of WMMSE in wireless resource allocation tasks such as beamforming and power allocation for different MA schemes, this achievement implies that DNNs can effectively address a broad range of resource allocation challenges \cite{DNN2020WCL}.

\par \textit{CNN}: 
In contrast to DNNs, CNNs leverage convolutional and pooling layers for specialized operations. By stacking these convolutional and pooling layers alongside dense layers, CNNs significantly reduce the number of parameters required to achieve equivalent performance to that of DNNs. CNNs excel in feature extraction and are notably easier to train compared to DNNs. This advantageous has led to the widespread adoption of CNNs in addressing resource allocation challenges.  CNN has been trained to approximate WMMSE for power allocation in \cite{CNN2018CL}, and for SDMA beamforming design in \cite{CNN2020TVT}. In \cite{CNNNOMA2022TVT}, the power control problem of NOMA is also addressed by CNN.
 
\par \textit{RNN}: 
RNNs take into account the impact of  inputs from the previous time steps. This distinctive feature allows RNNs to maintain memory of prior inputs through their unique network architecture. Therefore, RNNs are well-suited for handling time-dependent data. 
This characteristic of RNNs make them highly valuable to capture the temporal correlations of data.
In \cite{RNN2018TCOM}, RNN is used by each base station to predict the spectrum allocation. Additionally, in \cite{RNN2022TMC}, a graph attention RNN is proposed for traffic prediction. 
The RNN family includes various common variants, including long short-term memory (LSTM) and gated recurrent unit (GRU), among others. These variants have found highly effective in tackling various resource allocation challenges.  LSTM is used to forecast the user mobility, and enhance the performance of resource allocation \cite{LSTM2020CL}. 
In terms of beamforming design for SDMA, LSTM  emerges as a powerful tool for integrating information from multiple preceding steps in order to refine the current step, as evidenced in \cite{meta2023TNNL}. This contrasts to conventional alternative optimization algorithms, which typically rely solely on information from the present and the immediately preceding step to influence their output.
For GRU, it can serve as a  robust predictive model capable of consistently delivering high levels of accuracy in forecasting resource requests \cite{GRU2020TCCN}.

\par \textit{GNN}: GNNs are specifically built upon graph structures, offering the capability to handle non-Euclidean data. GNNs exhibit remarkable flexibility in accommodating input data of varying dimensionality, making them a natural fit for the dynamic nature of the wireless communication domain, where the number of users and antennas can vary significantly. 
By constructing directed/undirected graphs based on the structure of the considered wireless networks with devices as nodes and channels as edges, GNNs are capable to learn the wireless network and enhance resource allocation \cite{GNN2021TWC, GNN2021JSAC}. 
As an alternative, GNNs can construct graphical models based on the mathematical formulation of specific resource optimization problems by defining node sets and incorporating parameters as features, as evidenced in \cite{GNN2022TCC, GNN2022WC}.
By such means, GNNs are promising for addressing large-scale resource allocation problems for various MA schemes while enjoying a high computational efficiency. 

\par \textit{Other Methods}:
In addition to the previously mentioned standard neural network architecture, numerous advanced DL approaches have gained widespread adoption in addressing resource allocation problems. These encompass a range of techniques, such as the \textit{transformer} \cite{transformer2023JIoT}, \textit{auto-encoder} \cite{Autoencoder2022TWC},  \textit{transfer learning} \cite{transfer2021TWC}, \textit{meta-learning} \cite{meta2023TNNL}. 
The transformer takes the advantage of the attention mechanism, which sets it apart from RNNs. It is suitable for parallel computing, leading to substantial performance enhancements when compared to RNNs.
An autoencoder, a neural network for unsupervised learning, contains two parts: an encoder that compresses input data into a simpler form, and a decoder that recreates the original input from this simpler representation. In wireless resource allocation, autoencoders are used to learn efficient data representations that capture important patterns in communication data. These learned representations improve the efficiency of resource allocation methods like power control, channel assignment, or user scheduling by giving decision-making algorithms a more concise and informative input.
The fundamental concept behind transfer learning involves extracting essential features from the source domain and fine-tuning the pre-trained model for application in the target domain. By leveraging its capability to transfer valuable prior knowledge to new scenarios, transfer learning offers a promising solution to address the challenge of task mismatch encountered in practical wireless communication systems.
The key to meta-learning is the ability to ``learn-to-learn" so that the trained model can  improve its learning ability and adapt to new tasks or domains with minimal or no human intervention. This approach facilitates rapid adaptation to new environments, thus enabling more effective resource allocation decisions. In this context, meta-learning was found to be a powerful computationally efficient alternative to WMMSE-based optimization techniques to optimize precoders and resources allocation in RSMA \cite{Loli:2023}. The performance obtained with meta-learning is comparable to convex optimization-based approaches but achieves a significantly lower running time. This is particularly helpful for large scale RSMA network settings with many antennas and users, where WMMSE approaches are not feasible due to the astronomic computational time, and low complexity precoding algorithms achieve poor and clearly sub-optimal performance.

\par 
Additionally, algorithm-driven methods like deep unfolding/unrolling and learning to branch and bound (BB) have also played significant roles in wireless resource allocation for different MA schemes. 
As the resource allocation problems are typically non-convex, one line of algorithm design is to approximate the original non-convex problem to a sequence of convex problems, and use an iterative algorithm to solve the approximated problems. 
Deep unfolding/unrolling extends this idea by unfolding the iterations of an iterative resource allocation algorithm into a neural network architecture, allowing for end-to-end training and optimization. Such approach has been widely studied in beamforming design, user scheduling, power allocation, etc \cite{unfolding2021TWC, unfolding2023TCOM}.  
Another standard method to deal with the non-convex optimization problem is to find the  global optimal solution based on BB, which is of high computational complexity. To address this issue, a learning-based BB is proposed that use neural network to learning the BB algorithm \cite{BB2020TVT}.

\subsubsection{Reinforcement Learning} 
RL involves  the interaction between agents and the environment, with the goal of optimizing a long-term objective. 
Among the most widely embraced model-free RL algorithms, Q-learning is the most well-known algorithm for computing an optimal policy that maximizes the long-term reward. 
However, as system complexity deepens, particularly when there exists hidden system states, calculating and maintaining all Q-values becomes exceedingly impractical. This challenge, often referred to as the ``curse of dimensionality", makes Q-learning less suitable for the intricate demands of ultra-dense and complex wireless networks  \cite{DRL2019VTM}.
To address this issue, deep RL (DRL) emerges as a powerful solution that integrates traditional  RL with deep learning techniques.
\par 
Recent advances in DRL has made it a  promising frontier for resource optimization in wireless network for different MA schemes \cite{DRLSDMA2020TCOM, DRLNOMA2019JSAC, DRLRSMA2021WCL, DRLRSMA2023TMC}. DRL has found applications in a wide range of resource allocation problems, such as user scheduling, power control, beamforming design, and bandwidth allocation. DRL demonstrates remarkable capability in tackling the intricate challenges of sequential decision-making within dynamic and large-scale networks, all without relying on explicit models of the transmission environment. This unique feature empowers agents to continually refine their decision policies through interactions with the unknown environment.

\begin{table*}[htbp]
\caption{AI-empowered resource allocation for different multiple access schemes.}
\label{TABLE_AI}
\centering
\begin{tabular}{|c|c|c|c|}
\hline
MAs                   & AI Method & Reference & Resource Allocation Problem\\ \hline
\multirow{3}{*}{OFDMA} &Deep Transfer Reinforcement Learning&\cite{DRL2022OFDMA}&beamforming and subcarrier assignment\\ \cline{2-4}
&\multirow{3}{*}{DNN}&\cite{DNN2023OFDMsubcarrier}&subcarrier assignment \\ \cline{3-4}
&&\cite{DNN2023OFDMpower}&power allocation\\\cline{3-4}
&&\cite{DNN2023OFDMscheduling}&user scheduling\\ \cline{2-4}
&DRL&\cite{DRL2020OFDM}&subcarrier assignment and power allocation\\
\hline


\multirow{2}{*}{TDMA}                 &  RNN (LSTM)                        &\cite{TDMA2018RNN}                      & radio resource assignment             \\ \cline{2-4}
&DRL&\cite{TDMA2022DRL}&time-frequency resource block allocation \\\cline{2-4}
  &  DRL (Q-learning)                        &\cite{TDMA2017DRL}                      & power allocation            \\ \hline
\multirow{6}{*}{SDMA}  & K Nearest Neighbors& \cite{SDMA2018KNN} & beamforming  \\ \cline{2-4} 
& K-means& \cite{SDMA2018Kmeans} & user grouping   \\ \cline{2-4}

& \multirow{3}{*}{CNN}        &  \cite{SDMA2020CNNHybridBF} &   hybrid beamforming  \\ \cline{3-4} 
                      &  &  \cite{SDMA2020CNNBF1}                    & \multirow{5}{*}{beamforming}  \\ \cline{3-3} 
                      &&\cite{CNN2020TVT}&\\
                      \cline{2-3} 
                      & \multirow{3}{*}{GNN}        &                 \cite{SDMA2023GNN1}     &               \\ \cline{3-3} 
                      &                          &    \cite{SDMA2023GNN2}                  &               \\ \cline{3-3} 
                      &                          &   \cite{SDMA2023GNN3}                   &               \\ \hline
\multirow{3}{*}{NOMA} &{GNN}        & \cite{NOMA2023GNN}      &   user scheduling\\ \cline{2-4} 
&{DQN}&\cite{NOMA2023DQN}&user pairing and power allocation\\\cline{2-4}
                      & {DNN}                 & \cite{NOMA2023DNN}                     &      joint beamforming and SIC decoding                
                      \\ \hline
\multirow{3}{*}{RSMA}                   &  Transformer                        &    \cite{RSMA2022transformer}               & \multirow{3}{*}{beamforming}              \\ \cline{2-3} 
            &  DNN and autoencoder                        &    \cite{RSMA2022DNN}               &               \\ \cline{2-3} 
                &   Meta-Learning                    &    \cite{Loli:2023}               &               \\ \cline{2-4} 
          &DRL  &    \cite{RSMA2021DRL}               & power allocation              \\\hline

\end{tabular}
\end{table*}

\subsubsection{Advantages and Disadvantages}
\par A non-exhaustive search of existing works that use AI methods to allocate resources for difference MA schemes is summarized in TABLE.\ref{TABLE_AI}. The advantages and disadvantages of these AI-empowered algorithms are summarized as follows.

\par AI-empowered resource allocation algorithms offer the following advantages: 
\par \textit{Superior Performance}: Thanks to their data-driven features and the well developed neural networks, AI-empowered resource allocation can discover communication patterns and relationships that are too complex to identify  precise mathematical model. Therefore, AI-empowered algorithms often achieve better performance compared to traditional model-based algorithms.

\par \textit{Generalization}: AI-empowered resource allocation algorithms can generalize from historical data to make predictions or decisions on uncharted  data. This capability empowers these algorithms to flexibly adapt to rapidly changing wireless environment and efficiently allocate resources.

\par \textit{Scalability}: 
AI-empowered resource allocation algorithms can  efficiently handle and process larger volumes of data. They possess the ability to accommodate growing demands without compromising performance or efficiency. Therefore, these algorithms are generally proficient in tackling resource allocation challenges of diverse sizes and complexities.

\par \textit{Low computational complexity}: 
AI inference demands only a limited set of basic  operations and can be executed in real-time. This appealing characteristic allows these algorithms to reduce processing time and require fewer computational resources, making them practical and cost-effective solutions for both online and offline resource allocation problems. 

\par \textit{Robustness}: 
As AI-empowered algorithms are data-driven, they are typically robust to handle uncertainty in data or environment, allowing them to make reliable decisions despite incomplete or noisy information. This robustness is particularly valuable in resource allocation problems because it ensures that the algorithm can adapt and make sound  decisions even when operating in unpredictable or dynamic environments.

\par  Although AI-empowered resource allocation solutions offer numerous advantages, they also come with certain disadvantages and challenges:

\par \textit{Lack of Interpretability}: 
Directly replacing the conventional resource allocation algorithm with a generic neural network in an end-to-end fashion is often referred to as a ``black-box" approach, which lacks interpretability regarding the decisions it makes. To address the issue, one can combine data-driven and model driven algorithms, i.e., by unfolding an iterative algorithm, it becomes possible to enhance interpretability. 

\par \textit{Data Dependency}: AI-empowered algorithms have a strong dependence on data for both training and inference. When the training data is biased, incomplete, or not representative, it can result in suboptimal resource allocation decisions. As the wireless environment is changing rapidly,  user channels, typically used as input data, may shift to different distributions over time. This can significantly impact the ability of the trained neural network to generalize effectively to new conditions and scenarios.

\par \textit{Training Overhead}: Developing and training AI models for wireless resource allocation can be time-consuming and resource-intensive. This is especially evident when utilizing traditional resource optimization algorithms, known for their high computational complexity, to generate the training data, leading to substantial time requirements. Furthermore, to maintain optimal performance, periodic model updates may also be a necessity.
 
\subsection{AI-empowered MA Channel Estimation}




CSI plays a pivotal role in wireless communications. On one hand, precise CSIT facilitates the implementation of adaptive resource allocation and user scheduling, leading to substantial improvements in performance and efficiency. On the other hand, accurate CSI at the receiver (CSIR) empowers coherent detection and decoding processes. 
For acquiring CSIR, downlink channel estimation is a necessity, where the transmitter sents pilot signals to the receivers. This task becomes particularly vital and challenging in OFDM systems due to the temporal variability and frequency-selective characteristics of wireless channels.
In terms of CSIT acquisition, in time division duplex (TDD) systems, CSI is obtained at the transmitter through the exploitation of reciprocity. In this case, uplink channel estimation is required, where users send pilot signals, enabling the base station to estimate the uplink CSI. However, tackling this task in  massive access is challenging due to the limited number of orthogonal pilot resources \cite{CommMag2014TDDCS}. 
In frequency division duplex (FDD) systems, acquiring CSIT involves sending the CSI estimated at the receiver back to the transmitter through a feedback link.  However, this becomes challenging in massive MIMO setups due to the large dimensions of the CSI, resulting in a substantial increase in feedback overhead. To mitigate the overhead, it becomes imperative to efficiently compress the CSI estimate at the receiver or employ a codebook-based approach for quantizing the CSI \cite{JSAC2008LimitedFeedback,Access2016FDDCE}.
\par 
AI can play a pivotal role in tackling the aforementioned challenges and improving the quality of CSI obtained at either the transmitter or receiver. This can be achieved through two primary directions \cite{JSAC2019ML4CE}:
\begin{enumerate}
    \item Channel estimation: By modeling the channel estimation problem as a regression task, we can create a neural network-based channel estimator.
    \item CSI compression: By modeling CSI compression as a dimension reduction problem, we can develop AI architectures specifically tailored to address this issue.
\end{enumerate}

These two directions in general apply to different MA techniques. Note that for different MA enabled channel estimation, some additional problems might introduce. For example, in TDD uplink NOMA, the same pilot might be shared by multiple users. Additional power control issue is introduced, which can also be addressed by AI methods \cite{TVT2018NOMACE}. In the following, we will respectively discuss the existing works on these two directions for different systems with different MA schemes. 

\par 
\textit{AI-enabled channel estimation:} Deep learning enabled channel estimation is first studied in \cite{WCL2018DL4CE} for a point-to-point OFDM system, where a DNN is employed to learn the characteristics of frequency selective wireless channels. After that, many existing works focus on using neural networks, such as CNN, generative adversarial network (GAN) to tackle channel estimation of MIMO networks \cite{CL2019CECNN,OJVT2020CE,TWC2020ChannelEstimation}. 
As summarized in the survey paper \cite{OJCOM2022CEsurvey}, most of existing works on channel estimation consider point-to-point communication networks, only limited studies have delved into the application of deep learning to uplink channel estimation in TDD multi-user MIMO systems for achieving less pilot-aided channel estimation. In \cite{CL2019CEmuMIMO}, a deep learning based joint pilot design and channel estimation scheme is proposed for uplink multi-user MIMO. To mitigate inter-user interference received at the base station, uplink NOMA is considered at the base station, which employs SIC for channel estimation. To jointly design pilot and channel estimation schemes,  a two-layer neural network is proposed for non-orthogonal pilot design and another deep neural network is constructed for non-linear channel estimation. The proposed approach is shown to achieve better performance than the traditional MMSE estimator and the deep learning approach without SIC.
\cite{TVT2020CEmuMIMO} further simplified the joint pilot signals and channel estimator design by removing the use of SIC at the base station. 
Furthermore, considering a millimeter wave massive multi-user MIMO systems with limited number of RF chains, a  prior-aided Gaussian mixture learned approximate message passing (GM-LAMP) approach is proposed in \cite{TCOM2021CEmmWave} to  exploit the sparsity
of beamspace channels and deal with the beamspace channel estimation problem.  Simulation results show that the proposed GM-LAMP network considering the prior distribution can enhance estimation accuracy with a low pilot overhead.
Considering a grant-free NOMA transmission network, the authors in \cite{TWC2023CENOMA} propose to deploy a DNN at the base station for identifying the active devices and estimating  corresponding channels. Numerical results show that such approach achieves a higher user activity detection success probability and more accurate channel estimation compared to the conventional approaches, especially in the small packet transmission scenarios for the grant-free NOMA system.
\par 
\textit{AI-enabled CSI compression:} 
In FDD massive MIMO systems, to reduce the CSI feedback overhead, one line of works focuses on  CSI compression methods and codebook designs.
While compression theories such as compressive sensing have been applied to simply the CSI feedback or codebook design \cite{TSP2014CS,JCN2016CS},  the  feedback signaling overhead in these approaches remains heavy since the overhead typically increases linearly with the number of antennas, imposing practical limitations.
Deep learning has shown  to be highly effective in addressing data compression challenges across various domains, including images, audio, and video \cite{balle2017end,lombardo2019deep}. Recognizing CSI as an additional data source for compression, deep learning techniques have been successfully applied to the compression and feedback of CSI for massive MIMO systems \cite{WCL2018CSICompression,WCL2019CSICompression,TWC2021CSIcompression,JSAC2021CEmmWave,TWC2022CSIcompression,TWC2022CSIcompression2,TCOM2023CSIcompress}. 
In \cite{WCL2018CSICompression}, an autoencoder-based deep learning network called CsiNet is proposed to handle CSI compression and feedback. By deploying an encoder at each user  to compress the CSI into a unique codeword for efficient feedback and a decoder at the base station to reconstruct the CSI from the received codeword, CsiNet shows significant compression efficiency enhancement compared with conventional compressive sensing-based approaches. 
Built upon CsiNet, various deep learning based approaches have been investigated for massive MIMO CSI compression and feedback to enhance CSI reconstruction accuracy while aligning with practical requirements.
Taking temporal dynamics into account, \cite{WCL2019CSICompression} further extends CsiNet to handle time-varing channels by proposing a LSTM-based deep learning network called CsiNet-LSTM. 
To address the difficulties of transmitting continuous codeword values,  recent works incorporate feedback quantization into network training so as to improve the network performance with low-resolution feedbacks \cite{TWC2021CSIcompression,TWC2022CSIcompression2,TCOM2023CSIcompress}.

\subsection{AI-empowered MA Receiver Design}



\par Aforementioned MA schemes often require advanced receivers. For instance, PD-NOMA and RSMA commonly rely on SIC, though other types of receivers can be used, as demonstrated recently in \cite{Sibo2023} for RSMA where the tradeoff between performance and receiver complexity of various interference cancellation and joint demodulation and decoding architectures have been compared.

\par One commonality in the design of advanced receivers for RSMA or other types of MAs is the reliance on models and assumption for the noise, interference, signal propagation, etc \cite{9523403}. It is nevertheless challenging to mathematically relax those assumptions while maintaining a tractable model. Indeed interference is often non-Gaussian with finite constellations, SIC is imperfect leading to error propagation, CSIR is imperfect, decoding delay and latency increases chance of errors, etc. Those model-based receiver algorithms have so far performed quite well \cite{Sibo2023} but take the risk to rely on accurate prior model knowledge and potentially perform poorly if it is not accurately acquired \cite{9523403}.

\par AI/BL-based receivers are able to directly extract meaningful information
from the unknown channel solely on observations, which is a major advantage. Therefore,
AI/DL is well suited for scenarios in which the underlying
mathematical channel model is unknown, its parameters cannot be acquired with precision, or when it is too complex
to be studied by model-based algorithms with low
computational resources \cite{8786074}. Nevertheless, employing conventional AI/DL techniques for wireless receivers presents substantial challenges. Firstly, the appeal of model agnosticism in AI/DL approaches necessitates a complex network with numerous nodes and layers, along with a sizable training set to acquire a specific mapping. Consequently, this results in a substantial computational burden at the receiver during the training phase. Secondly, sharing a training set between the transmitter and receiver introduces significant training overhead, potentially causing delays in transmitting actual useful data to communication users. Lastly, due to the time-varying and dynamic nature of wireless channels, periodic training of the DNNs at the receivers becomes essential to accommodate channel variations. However, this requirement renders the transmission of large training sets highly impractical.

\par
\begin{figure*}[t]
				\centering
				\includegraphics[width=\linewidth]{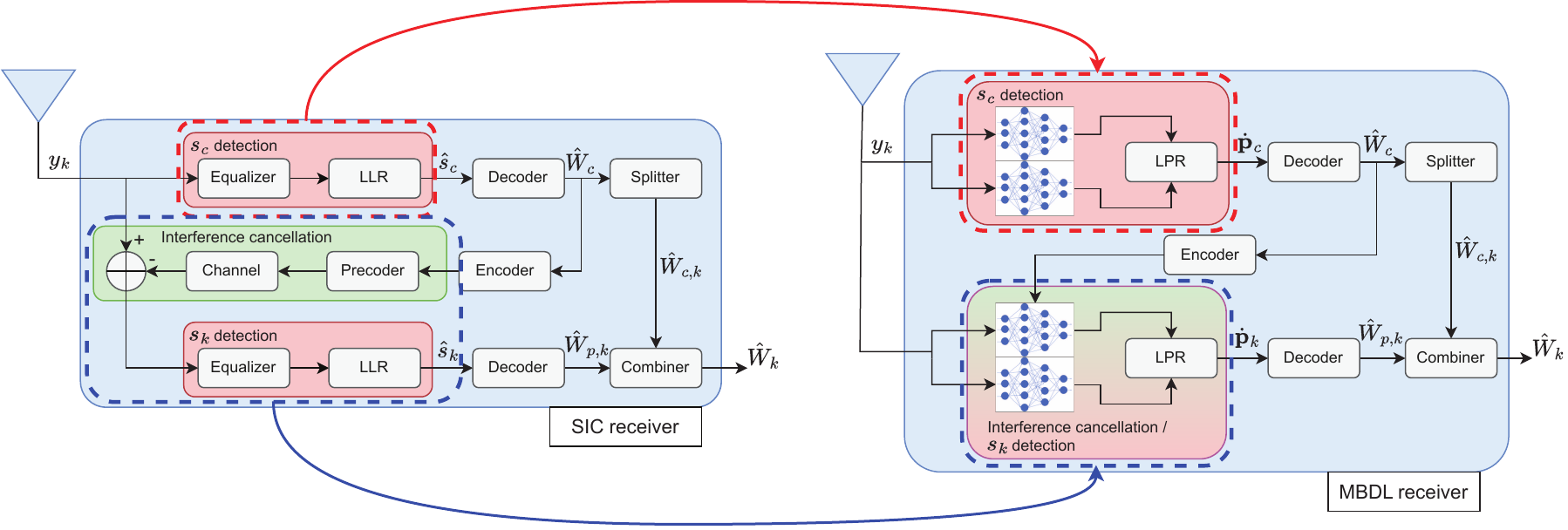}
				\caption{1-Layer Model-Based Deep Learning RSMA receiver and relationship with 1-Layer SIC RSMA receiver \cite{Loli:2023_TWC}.} \label{fig:AI_for_RSMA}
\end{figure*}

 In an attempt to tackle the aforementioned challenges and leverage the advantages of both conventional model-based algorithms and model-agnostic DL approaches, model-based deep learning (MBDL) has been introduced. This innovative approach aims to combine the simplicity inherent in traditional model-based algorithms with the model-agnostic nature of DL techniques, as highlighted in \cite{9523403}.
To realize this integration, MBDL systems are implemented by substituting specific steps and computations in model-based algorithms, which rely on precise channel model knowledge, with compact neural networks. These neural networks demand smaller training data sets compared to conventional DL systems. In the realm of wireless receivers, prior adaptations of the SIC receiver using MBDL have been proposed for uplink and downlink NOMA in \cite{9348107,8827912}, as well as for downlink RSMA in \cite{Loli:2023_TWC}. In these studies, specialized DNNs are employed for tasks such as interference cancellation and symbol classification. The findings demonstrate that the MBDL approach surpasses the performance of the conventional model-based SIC receiver specifically in the context of NOMA and RSMA.
The illustration in Fig. \ref{fig:AI_for_RSMA} showcases an MBDL receiver designed for RSMA, offering a visual comparison with the conventional SIC approach.

\subsection{AI for MA User Behavior Predictions}

Due to the multi-functionality of wireless networks and the heterogeneity of users, various user and communication service requirements exist. 
Different services usually lead to different QoS requirements and different resource management schemes.
However, conventional resource management strategy adopts static average resource allocation methods and ignores the disparity of service from different users, which makes it hard to meet the QoS requirements of all services with limited wireless resource. 
Thus, it is important to investigate the user service behavior prediction to match the wireless communication system configuration with the diverse service QoS demands, thus improving system efficiency, decreasing network delay, and guaranteeing the QoS demands of each user. 

According to  \cite{balachandran2002characterizing}, a base station captures user behavior information such as user identity number, user Internet access traffic costs, terminal brands, mobile product brands, user call fees, network request base station latitude, network request base station longitude, value-added service fee, mobile phone operating system, call start time, call end time, call base station latitude, call base station longitude, and network request time. Through analyzing the obtained history user mobile broadband information with the help of ML technology, the future user behavior can be predicted and then future communication QoS requirements can be further obtained \cite{yang2022federated}.
Based on the predicted communication QoS requirements, resource management such as bandwidth allocation, base station on-off control, and transmit beamforming, can be configured in time, thus providing seamless coverage. 
Moreover, to support a wide diversity and number of users, different MA schemes (including OMA, PD-NOMA, SDMA RSMA, CD-NOMA) could be applied, which also pose design challenges in ML for user behavior predictions due to different signal transmission schemes in different MA schemes.

Based on the types of predicted user behavior information, the main research areas in ML for user behaviour predictions in MA mainly include four aspects:
user activity prediction,
wireless traffic prediction,
content popularity prediction, 
and  user mobility prediction. 
In the following, we introduce these four aspects with pointing the design difference in the aspect of MA.

\subsubsection{User Activity Prediction}
User activity indicates that the user is active or not. 
In the communication system, the user activity information can reflect the total number of served users. User activity prediction has first been studied in cognitive radio networks, where, to efficiently utilize dynamic spectrum resource in such networks, ML techniques can be used to  predict the primary users’ activity and the secondary user can choose the idle channel \cite{agarwal2016learning,8935716,7962476}. ML techniques for user activity prediction in OMA cognitive radio networks such as TDMA are mainly supervised ML, including artificial neural networks (ANN) and SVM \cite{bkassiny2012survey}.
Furthermore, for the multi-channel multi-hop mobile cognitive radio networks, the Probabilistic Suffix Tree based ML algorithm is used to learn the primal user's activity pattern 
\cite{7564910}.
Using the history space-time access records in mobile broadband data, the scalable online expectation-maximization based algorithm to fast learn latent Dirichlet allocation has been studied for user activity level prediction \cite{luo2016telco}. 
With the location-based social network data, \cite{app13063517} shows the feasibility of using ML algorithms for user activity prediction, which include gradient-boosted trees, deep learning, logistic regression, and generalized linear model.

\par Different from OMA, there exists interference among users with NOMA. 
In particular, users need to be grouped for forming NOMA transmission, which require the user activity information.
However, due to the dynamic channel gain, the user grouping and user activity are always dynamic and coupled, thus making it hard to predict the user activity in NOMA.
Through taking user activity as an unknown mapping, the work in 
\cite{8387468} uses the LSTM to approximate this mapping.
The framework that integrates LSTM into NOMA is shown to improve the user activity performance in terms of reliability and complexity.

\subsubsection{Wireless Traffic Prediction}
The wireless traffic includes the traffic duration time and traffic data requirements, and can be explored via reinforcement learning and other AI techniques. Using reinforcement learning, in \cite{song2020semi}, the Dueling deep-Q network is applied to obtain the wireless traffic prediction and server load balance in OMA system.  
The wireless traffic prediction models with AI are categorized into statistical, ML, and DL models
\cite{jiang2022cellular}. The ML models include random forest, LightGBM, Gaussian progress regression, multiple linear regression (MLR), and Prophet, while the deep learning models \cite{9076110,9615104} include feed-forward neural networks (FFNNs), CNNs, and RNNs.

For random access (see Section \ref{IoT_section}), supervised learning-based method is proposed to improve the access performance, which uses the history data to obtain the time-varying trend of wireless traffic \cite{9569275,9422330}. 
Besides, the LSTM-based prediction
algorithm can be used to characterize the time related property of traffic, which is helpful in predicting the locations of aerial base stations, such as unmanned aerial vehicles (UAVs) \cite{8647209,9367523,10092916}.
With the predicted UAV locations, various uplink MA schemes are considered in \cite{10092916}, such as TDMA, FDMA, NOMA, and RSMA. To minimize the total transmit power, it is found that RSMA can achieve the best performance among all schemes since the RSMA can achieve all points in the capacity region. 

\subsubsection{Content Popularity Prediction}
The basic principle of content-based content prediction is to obtain the user's interest preferences based on the user's historical behavior, and recommend objects similar to his interest preferences to the user. Content prediction generally requires three steps, namely constructing user feature representation based on user information and user operation behavior, constructing subject feature representation based on subject information, and representing user recommended subject based on user and subject characteristics.
Having predicted the content popularity, the most preferred content can be cached near the intended users in advance, thus improving the system performance including delay and energy \cite{feng2019content}. 

To design a comprehensive popularity prediction scheme for OMA systems, a multi-head attention based popularity prediction model is introduced in 
\cite{9220908} to extract the multi-dimension features.
Since multiple users can be served simultaneously in NOMA and the content storage capacity is always limited, these exists a tradeoff between the resource for NOMA transmission and content caching \cite{yan2019joint}.
To solve this tradeoff, an autoencoder-based algorithm is proposed in NOMA enabled fog radio access networks for content popularity prediction in NOMA system 
\cite{yan2019joint}.

\subsubsection{User Mobility Prediction}
%
The user mobility information including speed, direction, and angle, plays a crucial role in the prediction of the wireless environment \cite{8570749}.
Due to the mobility of users and limited coverage area of one base station, handover is needed to maintain the QoS requirement during the whole moving process of the user. 
In \cite{8570749}, state-of-art ML approaches for user mobility prediction are summarized including the movement predictability, prediction outputs, and performance metrics.  
Considering the wireless communication protocol and standard, a ML based mobility prediction frameworks is proposed in \cite{9392779} in 5G core network.
Further for supporting the emerging applications such as automotive, the framework of RNN together with LSTM is proposed in
\cite{fattore2020automec}.
The time-varying property of user movements can be viewed as a time-varying graph and a graph based ML network can be used in \cite{9428552} for user mobility prediction.

Since the user movement is always related to the physical environment, high-accuracy prediction of user movement requires AI technology to obtain the relationship between the user movement and environment information \cite{cheng2020joint,chaalal2022new,8896109}. In order to solve the high complexity and spatio-temporal correlation of user mobile behavior, multi-dimensional and multi-scale user behavior prediction method based on ML aims to fully capture the dynamic and complex spatio-temporal relationship between individual mobile behavior and the physical environment.

\begin{figure*}[t]
\centering
\includegraphics[width=0.9\linewidth]{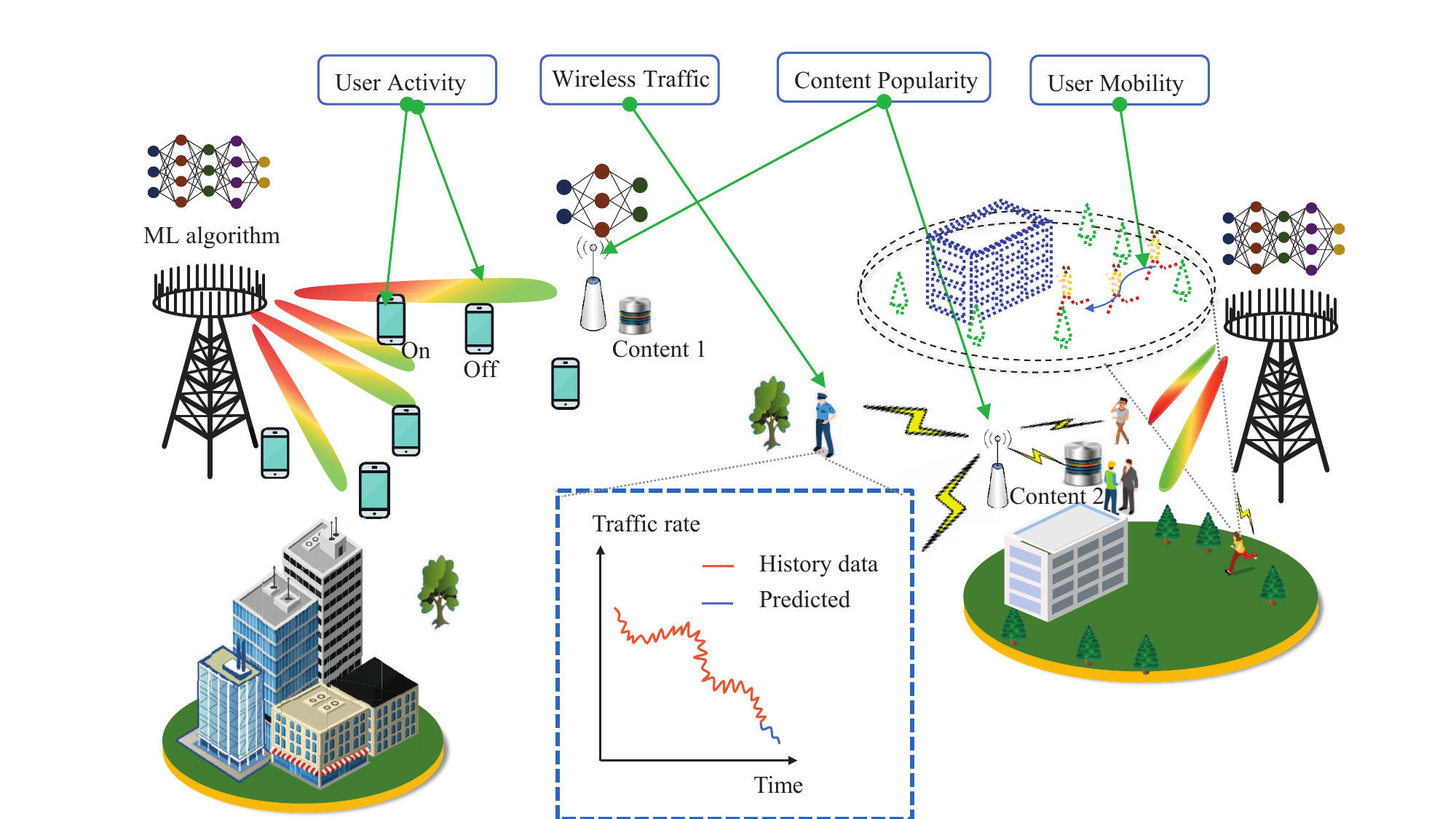}
\caption{An illustration of ML for user behavior predictions.} \label{fig:userbehavior}
\end{figure*}

In summary, by analyzing the history user behavior  data in different MA schemes, the ML approaches can predict various kinds of behavior information, such as user activity, wireless traffic, content popularity, and user mobility, as shown in Fig.~\ref{fig:userbehavior}. 

\subsection{Technology Outlook and Future Works}

AI-empowered resource allocation, optimization, channel estimation, user behaviour predictions have been investigated partially for some relatively simple MA schemes such as OMA, but more needs to be done for more advanced MA schemes, such as RSMA and CD-NOMA, for which the literature still remains quite limited with a lot of new research challenges and opportunities on the horizon.

AI-empowered resource allocation, optimization, and channel estimation is likely to play an even bigger role for UMA for which the optimization space is enlarged due to the exploitation of all five dimensions (and sub-dimensions). Moreover, AI could also be helpful to complement communication and information theory to discover and identify key ingredients of UMA designs and architectures. 

Building upon RSMA and CD-NOMA, UMA is likely to benefit even more from AI-empowered receiver design for the same reasons as RSMA but also because of the additional complexity increase incurred by the the exploitation of the code dimension.

AI for user behaviour predictions will also be helpful to UMA to predict in advance the user activity, wireless traffic, content popularity and user mobility and accordingly feed those information to the AI-empowered resource allocation and receiver architecture for most efficient use of all MA dimensions of UMA.

\section{Multiple Access for Artificial Intelligence}\label{Section_MA_for_AI}

We now switch the role and assess how MA schemes can be designed and tailored to serve AI applications. We focus on MA for federated learning (FL), edge intelligence and over the air computation.


\subsection{MA for Federated Learning and Edge Intelligence}

FL represents a decentralized method for training ML models. It eliminates the need for transferring data from client devices to global servers. Instead, the model is trained locally using raw data on edge devices, thereby enhancing data privacy. One challenge of deploying FL \cite{PNASchen} over wireless networks is the communication bottleneck, which arises from several edge devices transmitting large sized model parameters to a parameter server or a base station. Moreover the heterogeneity of edge devices computing capabilities, communication rates, and amount and quality of data can affect the training performance in terms of accuracy, fairness and convergence time. Researchers have attempted to reduce the resultant communication latency using different approaches such as excluding slow devices (``stragglers") \cite{AirComp_a1} 
or compressing FL model parameters by exploiting their sparsity. In this section, we provide a detailed literature review on the use of MA schemes for supporting FL deployment over realistic wireless networks, with a specific emphasis on OMA/SDMA/PD-NOMA/RSMA-based networks for wireless FL performance optimization. 

\subsubsection{OMA for FL}

Using OMA, devices must use different spectrum resource for FL parameter transmission. Due to limited resources in wireless networks, the wireless resource (i.e., power, computational power, spectrum) allocated to each user is limited, which may significantly limit the number of devices that can participate in FL and increase FL convergence time \cite{chen2019joint,9264742,9562559}. Consequently, it is necessary to optimize resource allocation such that devices can efficiently complete the FL training process. 
In \cite{Abad:ICASSP20}, the authors implemented FL over a hierarchical network architecture and they showed that local training with the help of small base stations (SBSs) can not only speed up the learning process but also increase communication efficiency by frequency reuse across multiple SBSs. The works in \cite{8737464,8664630} investigated the trade-off between local and global ML model updates to minimize the total energy consumption used for local ML model transmission and update. In \cite{9154285}, the authors used gradient statistics to select the devices to join FL training per communication round. The authors in \cite{8851249} analyzed how user scheduling affects the FL convergence and then optimized spectrum resource for the selected users. The work in \cite{9261995} designed a FL algorithm for devices with non independent and identically distributed data and then optimized resource allocation to improve FL convergence and training loss. In \cite{9207871}, the authors jointly optimized resource allocation and device scheduling to maximize the model accuracy within a training time constraint. The authors in \cite{9142401} designed a multi-armed bandit based method to select FL participating devices without the knowledge of wireless channel state information and data information of devices. To further improve FL performance in OMA based networks, current researchers also study the joint optimization of wireless resource allocation and other advanced technologies such as RIS \cite{10032291,mao2023roar}, MIMO \cite{9625822}, compression \cite{10181115}. The main drawback of OMA is that it does not scale well with the number of devices. Specifically, the required radio resources increase linearly with the number of transmitters, or else the latency will grow linearly. This calls for the integration of FL with other MA schemes.

\subsubsection{SDMA for FL} To solve OMA drawback and enhance the applications of FL, one can use MIMO and spatial domain techniques, which enable edge devices to use the same time-frequency resources for FL parameter transmission thus improving wireless resource utilization and FL convergence speed. 
The authors in \cite{9124715}  studied the deployment of FL over a cell-free MIMO based network and jointly optimized the local FL training accuracy, transmit power, data rate, and users' processing frequency to minimize FL convergence time. The authors in \cite{9856665} studied the joint use of MIMO and compression techniques for FL parameter transmission and analyzed the reconstruction error of the designed method and its impact on the FL convergence rate. The work in \cite{10215316} considered two unique characteristics of massive MIMO -- channel hardening and favorable propagation, and designed a novel communication design for FL in a MIMO wireless system. In \cite{9155479}, the authors considered the application of FL in wireless networks featuring uplink multi-user MIMO, and aimed at optimizing the communication efficiency during the aggregation of client-side updates by exploiting the inherent superposition of RF signals.

\subsubsection{PD-NOMA for FL}
To further improve the FL parameter transmission efficiency, one can use NOMA to separate edge devices into distinct groups and managed the interference in each group by SC-SIC. Currently, the authors in \cite{9207963} exploited PD-NOMA together with adaptive gradient quantization and sparsification to facilitate efficient FL parameter transmission from edge devices to the parameter server. In \cite{9252950}, the authors studied power allocation of PD-NOMA for distributed gradient descent in wireless FL with the aim of minimizing the learning optimality gap under privacy and power constraints. The work in \cite{9844152} jointly optimized IoT device scheduling, transmit power allocation, and computation frequency allocation for FL in a PD-NOMA and Relay enabled wireless network. In \cite{9718086}, the authors studied the PD-NOMA assisted FL in which a group of edge devices form a PD-NOMA cluster to send their locally trained models to the cellular base station for model aggregation and the base station adopts wireless power transfer to power edge devices for their data transmission and local training. They jointly optimized the wireless power transfer of the base station, PD-NOMA based FL parameter transmission schemes, the processing rates of the base station and edge devices, as well as the training accuracy of the FL, with the objective of minimizing the system-wise cost accounting for the total energy consumption as well as the FL convergence latency.

\subsubsection{RSMA for FL}
\begin{figure}[t]
				\centering
				\includegraphics[width=\linewidth]{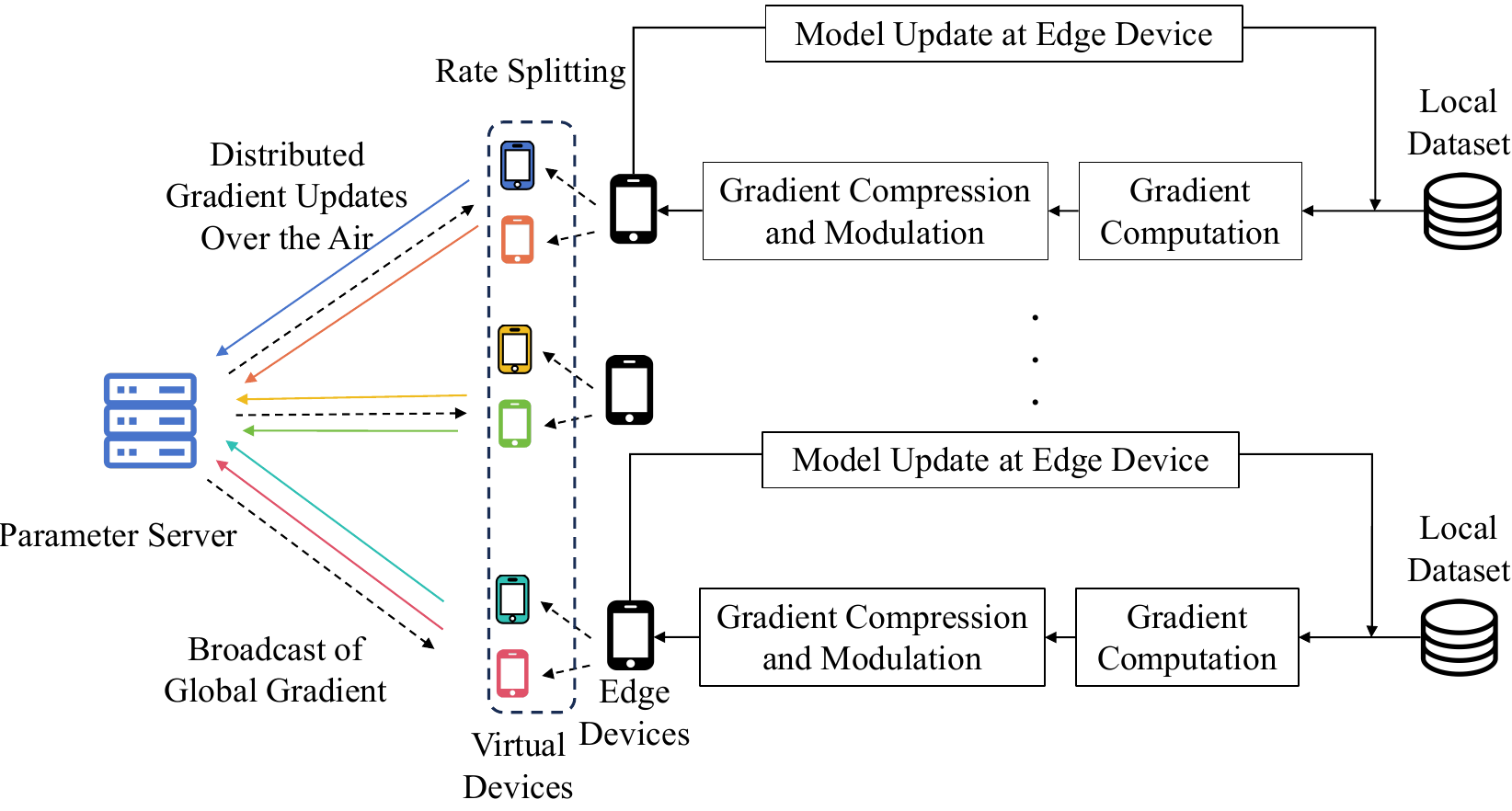}
				\caption{An illustration of an RSMA-based FL.} \label{fig:RSMAFL}
			\end{figure}

Since PD-NOMA decodes the interference of all other users, it lacks in flexibility and the designed FL algorithms in NOMA based networks may not be used for scenarios where each user needs to transmit with a strict rate demand due to the high dimension of model gradient. 
Moreover, since the FL process requires multiple transmission rounds and accurate knowledge of the CSI in all those time slots is unlikely, the use of MA schemes robust to imperfect CSI is important in FL applications. However, NOMA cannot be well adapted to the imperfect CSI scenarios \cite{9451194}. To address these issues and benefit from the space and power domains, one can use RSMA to efficiently balance the signal and interference and cope with imperfect CSI. In particular, due to the flexibility of signal design and decoding, RSMA can achieve robust performance with imperfect CSI, which is suited to the long-term transmission in FL. Fig. \ref{fig:RSMAFL} shows an illustration of an RSMA-based FL. In this figure, each device is considered as two virtual devices that jointly transmit local FL model with different data rates.

\par Currently, several works have focused on the use of RSMA for FL performance optimization. In particular, the authors in \cite{10201874} designed an RSMA based uplink radio access scheme for wireless clustering FL where the edge devices performing local training are closely located in a cluster. By exploiting the message split capability, RSMA has been shown not only to significantly reduce the latency of clustering FL in comparison to systems employing TDMA and NOMA, but also to be more energy efficient, reaching a lower latency with less power than the latter alternatives. 
Another promising scenario of RSMA for FL is in a fog radio access network (F-RANs). F-RAN leverages proximity edge nodes (ENs) to collect local model parameters of wireless devices \cite{9790067}. By splitting the message in two parts in uplink RSMA as in Fig. \ref{fig:RSMAFL}, one part is intended to be decoded by the ENs and the other part by the cloud. By adjusting the content and the transmit power of each part, one can softly combine the edge-decoding and cloud-decoding approaches and consequently achieve a better tradeoff between fronthaul overhead (communication cost) and learning accuracy. The work in \cite{10032163} designed an RSMA-based Internet of Vehicles solution that jointly optimized platoon control and FL training.



\subsection{Over The Air Computation}



In this section, we introduce an alternative MA approach~-- over-the-air computation (AirComp) for wireless FL performance optimization. In particular, AirComp enables edge devices to achieve over-the-air FL parameter aggregation via the waveform superposition property of a multi-access channel \cite{8870236}. Hence, when edge devices implement AirComp, the FL transmission delay per iteration will not depend on the number of edge devices thus improving FL convergence speed. 
Note that AirComp uses ``interference'' among device transmission for functional computation via uplink devices’ cooperation. This contrasts with the usual uplink MA schemes that aim at suppressing interference among transmitted messages as devices transmit independent data and do not collaborate. In this section, we first introduce the basic concept AirComp. Then, we introduce the state-of-art of using AirComp for FL.

\begin{figure}[t]
				\centering
				\includegraphics[width=\linewidth]{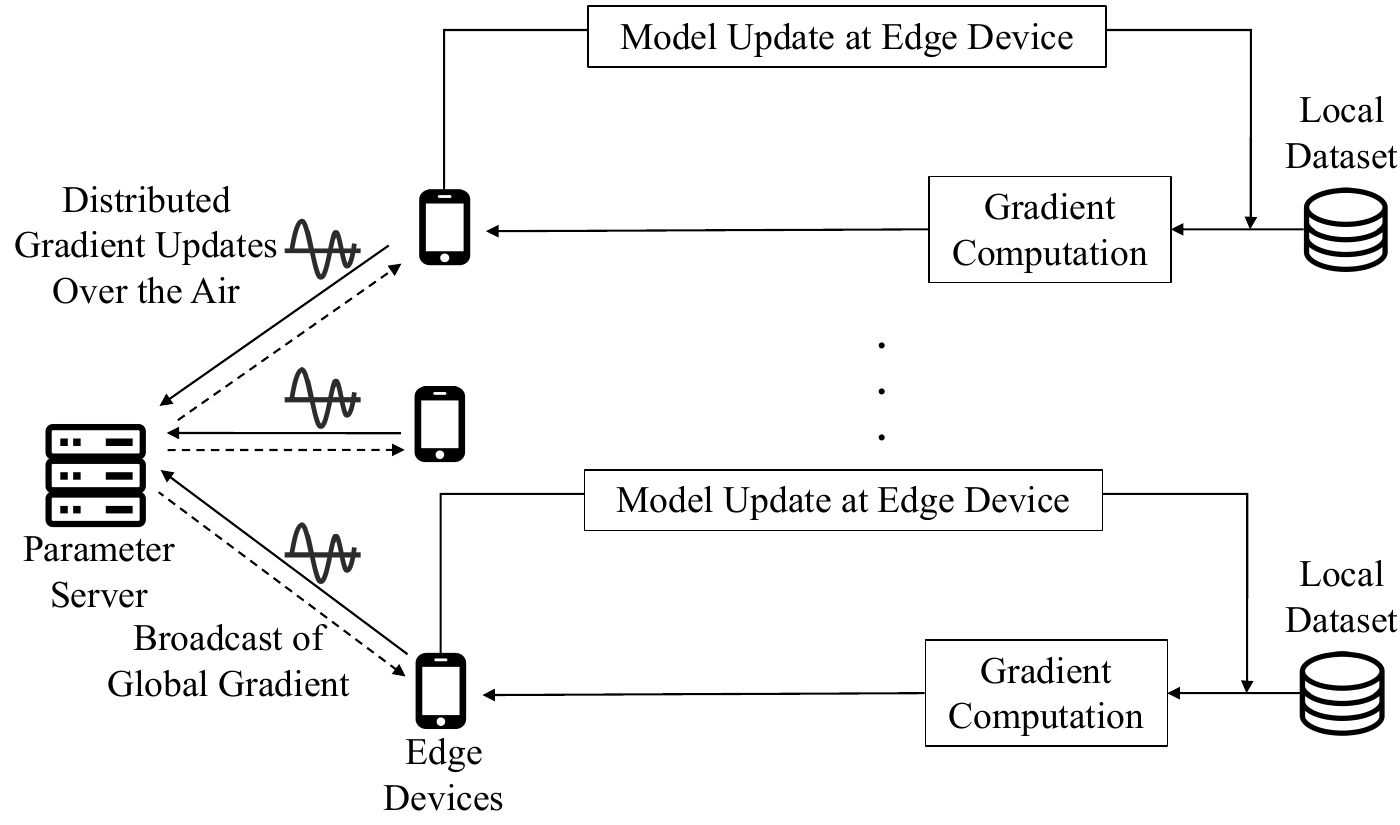}
				\caption{An illustration of an analog AirComp FL.} \label{fig:AnalogFL}
			\end{figure}

\subsubsection{Analog AirComp} The main idea of AirComp is summarized as follows. First, AirComp requires all edge devices to simultaneously transmit FL parameters to a parameter server (PS) such that their transmitted FL parameters can be superimposed over-the-air and their weighted sum (global FL parameters) is received by the PS and the weights of each edge device's FL parameters is determined by channel coefficients, as shown in Fig. \ref{fig:AnalogFL}. To make the weights of edge devices uniform and eliminate the bias of model aggregation, AirComp based FL requires each device to modulate its signal using linear analog modulation and to invert its fading channel by transmit power control. AirComp based FL can compute a broad class of nomographic functions \cite{Goldenbaum:TC:13} such as polynomial, Euclidean norm, arithmetic mean, weighted sum, and geometric mean. It was shown in \cite{AirComp_b2} that Analog AirComp can be optimal in terms of minimizing the mean squared error (MSE) distortion if all the multi-access channels and sources are Gaussian and independent.

\par A key requirement for implementing AirComp based FL is time synchronization of devices' transmissions. To address this challenge, one can use timing advance methods which have been used in several existing practical communication systems (e.g., LTE). The timing advance technique requires each edge device to estimate the FL parameter propagation delay and then transmitting in advance to ``cancel'' the delay. Thereby, different signals can overlap with sufficiently small misalignment. The signal distortion of AirComp stems from channel noise and interference of analog modulated signals. Therefore, channel noise and interference will affect the model updates of AirComp based FL. The impacts of channel noise and interference on FL model updates can be evaluated by the corresponding FL performance metrics. 
\par The authors in \cite{9563232} analyzed the convergence and optimized the performance of analog AirComp based FL over device-to-device communication networks. In \cite{9793704,9382094}, the authors optimized device selection and power allocation for analog AirComp based FL. The work in \cite{9502547} exploited the use of intelligent reflecting surface to assist FL model parameter transmission for analog AirComp based FL. In \cite{10109680}, the authors designed the FL model parameter retransmission schemes for analog AirComp FL. Finally, the authors in \cite{10215316} introduced a novel random orthogonalization scheme for FL parameter transmission while considering two unique characteristics of massive MIMO - channel hardening and favorable propagation.

\begin{figure}[t]
				\centering
				\includegraphics[width=\linewidth]{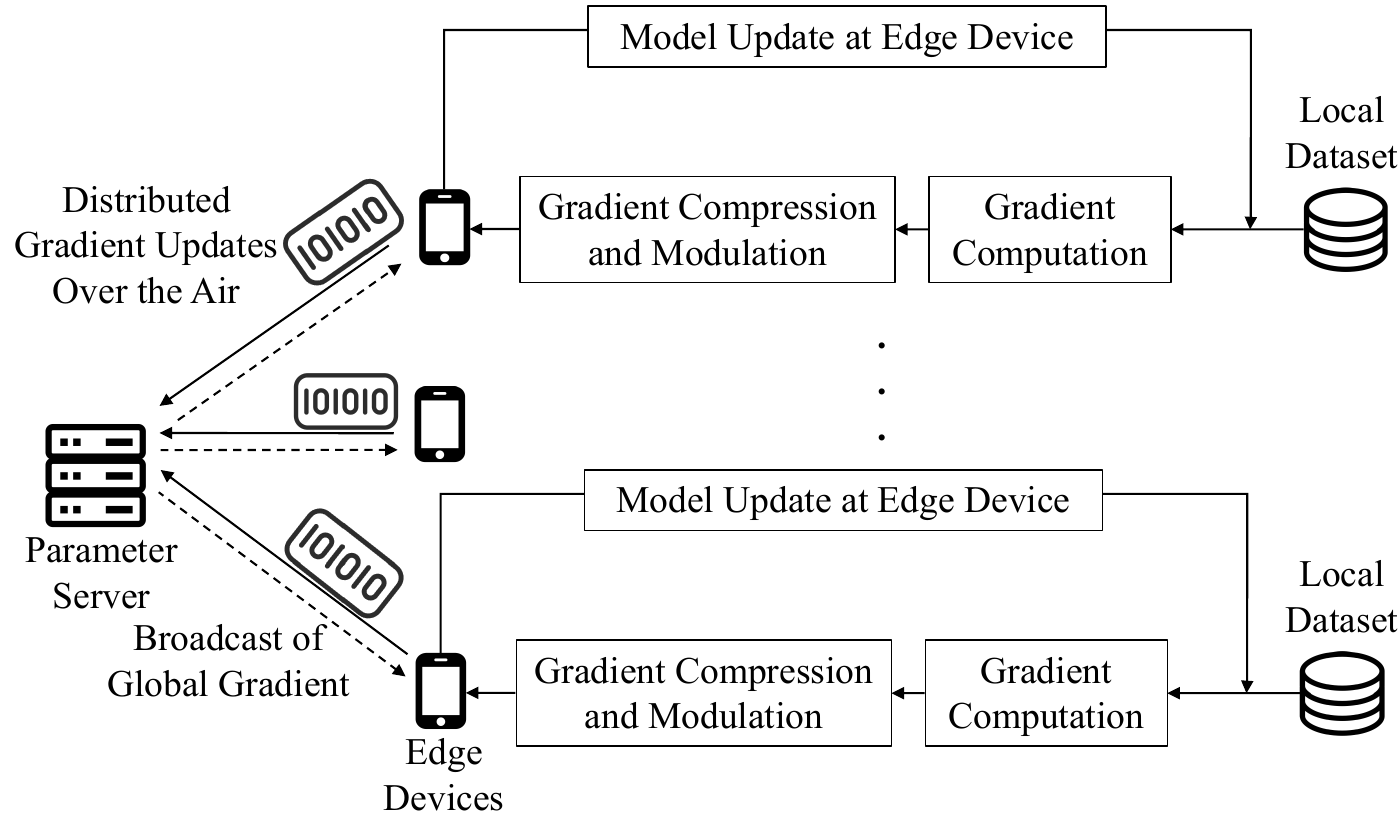}
				\caption{An illustration of a digital AirComp FL.} \label{fig:DigitalFL}
\end{figure}

 \subsubsection{Digital AirComp}
 Analog AirComp uses discrete-time analog transmission for FL parameter transmission and aggregation. However, the performance of analog AirComp is sensitive to the phase offsets among the super imposed signal. In particular, if the signals cannot be synchronized, the errors can also be accumulated thus leading to significant aggregation errors. To address this problem, digital AirComp is proposed which enables multiple devices to use non-orthogonal subcarriers for data transmission. Different from analog AirComp that relies on accurate channel precoding, digital AirComp leverages digital modulation and channel codes to combat channel impairments and phase misalignments, thereby achieving accurate model aggregation even when the signal phases of multiple edge devices are misaligned at the parameter server. In particular, for digital AirComp based FL, local FL model parameters must be preprocessed by quantization, source coding, channel coding, and modulation. Then, the pre-processed FL parameters will be transmitted over wireless channels, and superimposed in the air. The receiver will receive the aggregated FL parameters and decode them instead of decoding the FL parameters transmitted from several edge devices. Here, the receiver will first demodulate phase-asynchronous symbols, and then removes the protection of channel codes to get aggregation results. An illustration of a digital AirComp FL is shown in Fig. \ref{fig:DigitalFL}.

\par Several works have studied the design of novel digital AirComp FL. Specifically, the authors in \cite{9520774} analyzed the convergence rate of digital AirComp based FL. In \cite{wang2023cross}, the authors studied the use of MIMO and beamforming techniques to aggregate FL model parameters over the air. Different from the works on analog AirComp based FL that do not process the FL model signals that will be transmitted to the server, the work in \cite{wang2023cross} considered parameter quantization and modulation before FL signal transmission and hence it can be considered as digital AirComp based FL. The authors in \cite{10215506} presented an orthogonal frequency-division multiplexing (OFDM)-based digital AirComp system for wireless FL, where multiple edge devices transmit model parameters simultaneously using non-orthogonal OFDM subcarriers, and the server aggregates data directly from the superimposed signal. The work in \cite{9272666} designed a digital broadband over-the-air aggregation, which features one-bit gradient quantization followed by digital quadrature amplitude modulation (QAM) at edge devices and over-the-air majority-voting based decoding at edge server. 

\subsection{Technology Outlook and Future Works} Different MA techniques have their own advantages and disadvantages as discussed in Section \ref{section_II_MA_overview}. For FL, the selection of MA techniques depend on the number of participating edge devices, wireless resource (i.e., spectrum, transmit power), computational power, and hardware (i.e., CPU or GPU), FL model architecture of each device. To find the optimal MA schemes, one must first analyze the convergence of FL and figure out how different MA schemes affect FL convergence. However, in practice, the selection of MA schemes must be performed before FL implementation. Hence, the central controller (i.e., base station or parameter sever) may not be able to know channel state information and FL model setting information (i.e., gradient information), which may increase the difficulty of MA scheme selection. Therefore, MA scheme selection also depends on the available wireless and FL setting information. The above discussion further motivates the design of unified and universal schemes such as RSMA and ultimately UMA, which would completely eliminate the need for selection of MA schemes and whose design and optimization could be fully integrated with FL. Also, the use of over the air computation opens the door to other types of uplink MA schemes that brings back analog computation into MA design picture. 

\section{Multiple Access for Integrated Sensing and Communications}\label{Section_MA_for_ISAC}


In this section, we discuss the use and design of MA schemes for ISAC, including joint sensing and communications, multimodal sensing-aided communications, digital twin-assisted communications, communication-aided sensing/localization systems

\subsection{Joint Sensing and Communications}

ISAC has gained increasing interest in recent years. This is primarily motivated by the potential of sharing the same spectrum and, hence, providing access to more bands for both the communication and sensing systems. Operating at the same band, however, requires careful coordination of the various wireless resources and necessitates the development of new communication and sensing systems as well as joint processing frameworks. For multiple access, and when a base station or access point is jointly serving communication users and performing sensing functions, it is critical to efficiently allocate the available resources in time, frequency, space, code, power, etc., and optimize the waveform for both the communication and sensing objectives. Next, we summarize a few key directions for multiple access in ISAC systems and highlight important challenges in the ISAC multiple access design.

\subsubsection{SDMA-Assisted ISAC}
Based on the waveform diversity of MIMO radar and the spatial degrees of freedom of SDMA, \cite{7347464} embeds information stream into radar pulses for target detection at the mainlobe and performing communication tasks at the sidelobe. The communication symbols at the sidelobe level can be modulated via amplitude shift keying (ASK) \cite{7347464} or phase shift keying (PSK) \cite{10.1049/iet-rsn.2015.0484}. That said, the communication rate is limited by the Pulse Repetition Frequency (PRF) in those works. To overcome this limitation, \cite{8288677} proposed a joint MIMO radar-communication system, where a joint waveform shared by both radar and communication users is transmitted to serve both sensing and communication purposes. To design the probing beampattern, the precoders are designed using zero-forcing (ZF) beamforming under the communication SINR constraints. \cite{9124713} extended the framework of \cite{8288677} for probing multiple radar targets and communicating to multiple users simultaneously, where individual communication and radar waveforms are jointly precoded to achieve the maximum DoF of the MIMO radar waveform. The advantages of integrated dual-functional radar-communication (DFRC) MIMO transmission over separated antenna deployment for communication and sensing in terms of better trade-off between communication and sensing performance have been demonstrated in \cite{9737357} and the references therein. However, the ISAC systems based on SDMA may not perform well in overloaded scenarios and the impacts of imperfect channel estimation on sensing and communication performance trade-off have not been fully investigated.

\subsubsection{PD-NOMA Assisted ISAC}
The concept of PD-NOMA has been applied to ISAC for the uplink multiple access channel and the downlink broadcast channel \cite{10024901}. ISAC schemes based on PD-NOMA can be characterized into two types. The first type exploits the transmitted superimposed signals intended for communications for sensing. As a design example, \cite{9668964} considered maximizing the weighted sum of the communication throughput and the effective sensing power by designing the beamformers. The second type further exploits sensing signals for communications, where extra information is included in the sensing signals whose interference to the communication users is removed by SIC. ISAC schemes based on PD-NOMA can provide a better trade-off between communication and sensing performance compared to the conventional ISAC schemes based on OMA. However, the number of SIC scales proportionally with the number of users, which can increase the decoding latency and complexity and may not be suitable for delay-sensitive sensing and communication applications. In addition, SIC by fully decoding interference as in PD-NOMA may not achieve the optimal multiplexing gain when combined with multiple antennas \cite{9451194}, which leads to a performance loss of PD-NOMA Assisted ISAC compared to other multi-antenna MA schemes \cite{9832622}.

\subsubsection{RSMA-Assisted ISAC}
\begin{figure}[t]
\centering
\includegraphics[width=\linewidth]{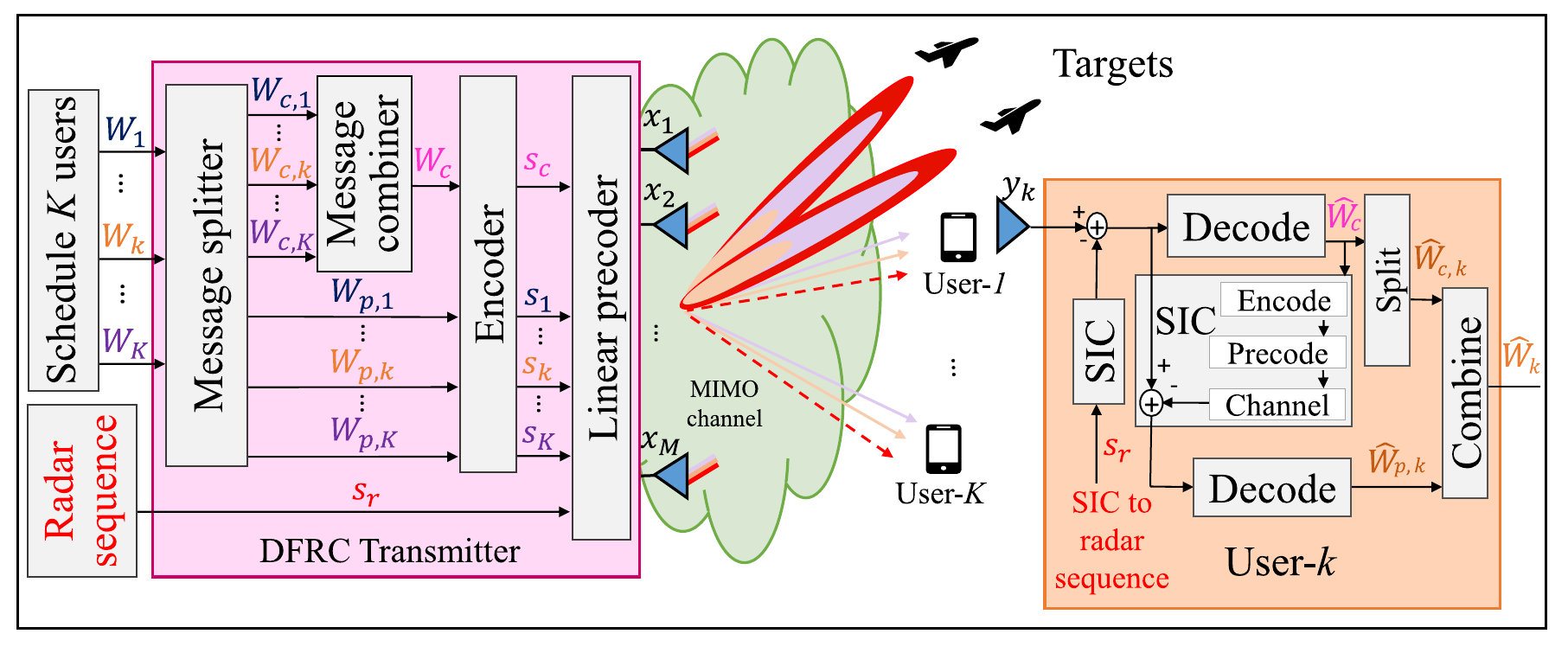}
\caption{RSMA-assisted dual-functional radar-communication systems with one base station serving $K$ users while tracking a target's direction \cite{9145168}.} \label{fig:ISAC_RSMA}
\end{figure}

Inspired by the advantages of RSMA in robust interference management, spectral efficiency, and user fairness, the interplay between RSMA and DFRC systems was proposed in \cite{9531484}. As illustrated in Fig. \ref{fig:ISAC_RSMA}, a radar sequence and RSMA-processed common and private communication streams are transmitted to achieve the dual function of probing targets and communications. The precoder and common stream of RSMA were carefully designed to achieve optimal sensing and communication performance. In \cite{9832622}, it was shown that RSMA-assisted ISAC has been extended to the scenario with partial CSIT and users with mobility, satellite systems with integration of communications and moving target sensing in space networks \cite{9771644}, and the DFRC systems with low-resolution digital-to-analog converter (DAC) on each RF chain \cite{9797869}. These works demonstrate that RSMA-assisted ISAC is very effective in managing multi-user/inter-beam interference while simultaneously performing radar functionality with the required mean square error constraints. Importantly, the RSMA-assisted DFRC framework unifies and embraces existing SDMA and NOMA-based DFRC frameworks as subschemes. Consequently, RSMA-assisted DFRC outperforms the SDMA and NOMA counterparts \cite{9832622}. However, the joint optimization of precoding, common, and private stream rate allocation can be a demanding computational burden at the transmitter side.

\subsubsection{Technology Outlook and Future Works} The design of efficient MA techniques for ISAC systems is a challenging task. First, the joint optimization of the resources and waveforms is typically associated with high computational complexity.  This complexity mainly stems from the differences in the communication and sensing objectives and from the various hardware and operation constraints on these systems.  This becomes more complicated in large-scale MIMO systems which are main components in 5G, 6G, and beyond. Going from nodes to networks, it becomes critical to design multiple access approaches for networks of integrated sensing and communication nodes. This requires accounting for the possible interference between different nodes in both the sensing and communication signals. With that, it becomes interesting to explore the trade-off between coordination performance gains and complexity in multiple access ISAC networks. This results in a set of approaches that range from no coordination to fully coherent ISAC operation where the full network acts as one cell, leading to cell-free ISAC systems \cite{demirhan2023cellfree}. How to design multiple access in these coordinating ISAC networks is an interesting research direction. Finally, the fundamental trade-off between sensing and communication performance of general ISAC systems with multiple access has not been fully understood. Only a few prior works have characterized the trade-off for very basic channel models with small numbers of users, e.g., \cite{9785593}. Hence, to gain more insights into coding, signaling, and waveform designs for more general channels, it is also important to study the fundamental trade-off in terms of achievability and converse for these channels. Last but not least, though RSMA-assisted ISAC has been shown to outperform SDMA/NOMA-assisted ISAC thanks to RSMA universality and flexibility advantages, it would be important to explore how other MA domains could be exploited for ISAC and get a complete picture of what UMA-assisted ISAC would look like.

\subsection{Multimodal Sensing-aided Communications}\label{subsec:multi-modal}

\begin{figure}[t]
\centering
\includegraphics[width=\linewidth]{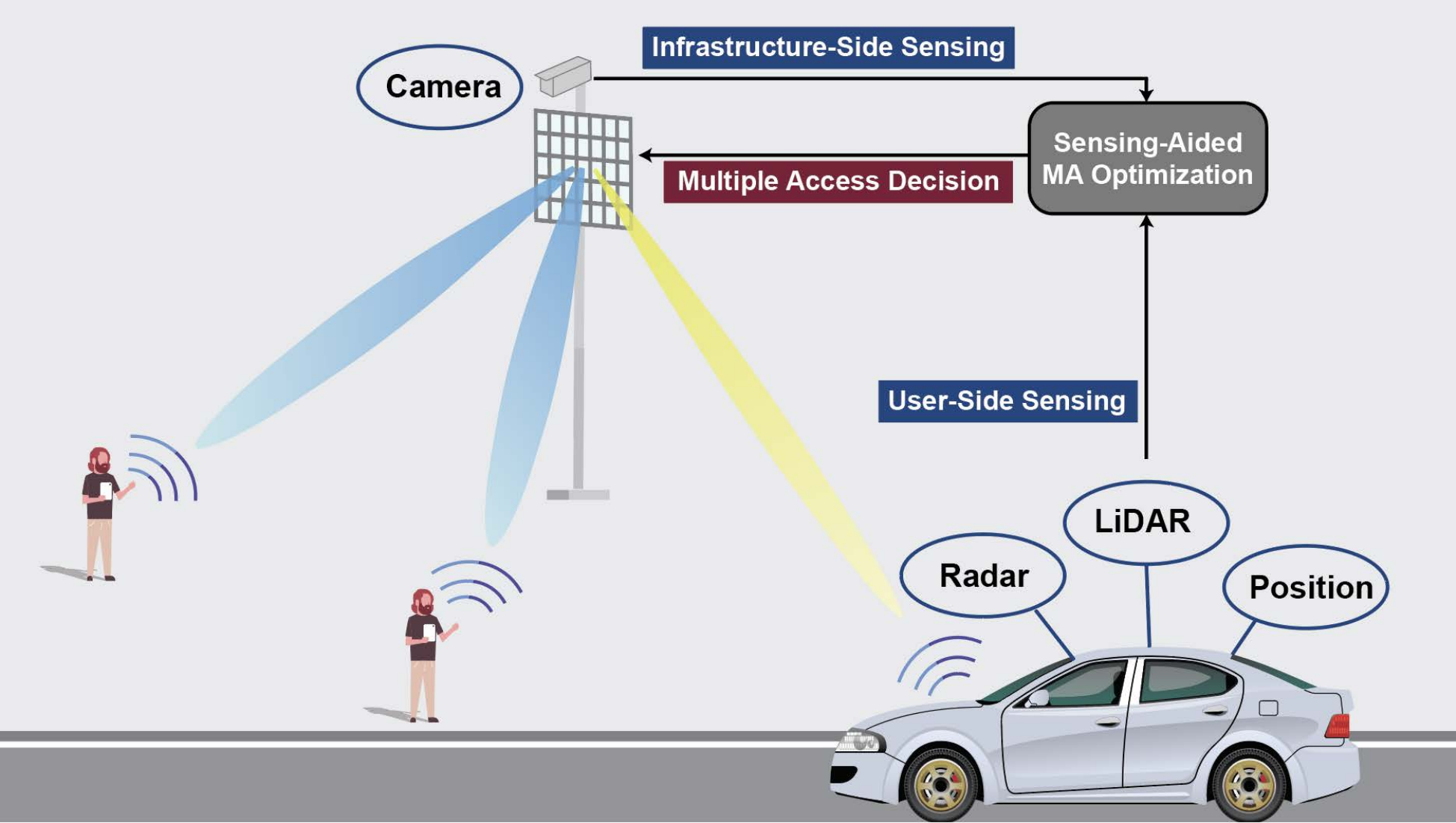}
\caption{The figure illustrates the key idea of multi-modal sensing-aided multiple access, where infrastructure-side and user-side multi-modal sensing information from cameras, LiDARs, radars, position sensors, etc., is leveraged to optimize the multiple access decisions.} \label{fig:MA_sensing}
\end{figure}

Two key trends in current and future communication systems are (i) the use of large numbers of antennas and (ii) the increasing dependence on high frequency bands. The motivation for using large antenna arrays ranges from increasing the multiplexing gain, which is a main goal for sub-6GHz massive MIMO, to ensuring high the array gain, which is a key objective for millimeter wave (mmWave) and sub-terahertz (THz) MIMO. Beyond the base stations, the use of large antenna arrays is also a feature of the new network nodes such as reconfigurable intelligent surfaces and holographic MIMO repeaters. The other key trend is captured by the increasing adoption of the mmWave band as a key pillar for 5G systems, and with the potential use of sub-THz bands in 6G and beyond. The key trends, while promising high data rates, pose multiple challenges for the design and operation of future communication systems. For example, the use of large numbers of antennas increases the overhead of estimating the channels or training the beams, making it hard for these systems to support highly mobile applications. Further, the use of high frequency bands makes the signal propagation very sensitive to blockages, which challenges the reliability and latency requirements for these networks.

While the use of large antenna arrays and high frequency bands impose new challenges on these communication systems, they also make the communication links very dependent on the user location and on the  geometry of the surrounding environment. Therefore, the knowledge of the user position and surrounding environment could potentially aid the communication system decisions.  This motivated the use of various multi-modal sensory data, collected for example  from visual cameras, LiDAR, radar, position/IMU sensors, etc., to assist future communication systems \cite{DeepSense}. Over the last few years, researchers investigated the potential gain of leveraging multi-modal sensing data in multiple problems. For example, \cite{demirhan2022radar_beam,charan2022vision,jiang2022lidar,charan2022towards,morais2022position, Zecchin,Rezaie_2020,Va_2019} showed that position, visual, radar, and LiDAR data  can guide the beam selection/alignment at the base station and user, leading to significant reduction in the beam training overhead. This was then extended to beam tracking where the sensing data is utilized for  predicting future beams, providing a practical solution for relaxing the data capture, processing, and prediction latency requirements \cite{jiang2022computer,luo2023millimeter}.  Multi-modal sensing has also been leveraged to proactively predict future blockages \cite{wu2022blockage,wu2023proactively,charan2021vision,charan2022computer,demirhan2022radar,demirhan2022radar_ICC,10011630}, predict channel statistics \cite{xu2021deep}, design codebooks \cite{chen2022computer}, design MIMO precoding without traditional CSI acquisition and feedback \cite{10011630}  among other interesting applications.

Multi-modal sensing provides some perception about the location and geometry of the user devices, infrastructure nodes, and various elements of the environment. This perception can enable more efficient multiple access operation in several ways. First, multi-modal sensing may enable the base stations and access points to intelligently select and schedule users to maximize the multiple access system performance. For example, leveraging the absolute or relative user position information, obtained by GPS, radar, LiDAR, or camera, the base station may schedule the users that are well-separated on the same time-frequency resource to have less interference. In dense deployment, the sensing information may also be utilized to assign users to base stations/access points or to proactively predict handover, which yields higher reliability, less latency, and better session continuity. Beyond user selection and scheduling, the underlying information captured by multi-modal sensing can guide the  multiple access resource allocation decisions. For example,  in PD-NOMA,  multi-modal sensing data can potentially aid the optimization of the transmit power of the different devices.  Further, in OMA, this sensing data may enable  adaptive allocation of the time, frequency, and spatial resources based on the users relative positions, mobility patterns, and dynamic scatterers in the environment. All that motivates investigating the various promising applications of multi-modal sensing in next-generation multiple access systems.

Realizing the potential gains of multi-modal sensing aided multiple access in practice, however, requires overcoming a number of challenges. First, a fundamental challenge in using sensors such as cameras, LiDARs, and radars is identifying and differentiating the communication users in the sensing scene. For example, for a given visual scene (photo) taken by a camera installed at a base station, how could the base station identify the mobile user from the other objects in the scene? This task, which is defined as user identification \cite{charan2022user}, is a key for enabling multi-modal sensing-aided communication in realistic environments. Further, with multiple access, the scene could typically have multiple users, which makes it important for the communication system to differentiate between the different users. Initial approaches for user identification rely on augmenting other user-specific measurements such as sparse/limited channel or beam power measurements \cite{charan2023camera,charan2022vision}.  In addition to user identification and differentiation, the coverage mismatch between the multi-modal sensing and communication systems raises critical challenges. Overcoming that motivates the extension to distributed multi-modal sensing which involves the design and coordination of distributed sensors to optimize the multiple access objectives. Addressing these important challenges defines interesting directions for future research in the interplay between multiple access techniques and multi-modal sensing systems.

\subsection{From Multimodal Sensing to Digital Twin-Assisted Communications}

Multi-modal sensing information provides the mobile devices and communication network with some awareness about what is happening in the environment. This awareness could be leveraged, as discussed in Section \ref{subsec:multi-modal}, to aid the various communication decisions. Building upon this and targeting a more comprehensive perception of the communication environment, the concept of (near) real-time digital twins of the physical environment has been proposed \cite{Alkhateeb2023}. In these digital twins, the multi-modal sensing information is fused with the 3D maps to construct real-time or near real-time maps of the surrounding environment. Running real-time ray-tracing, which could potentially be machine learning enhanced, on these 3D maps leads to (near) real-time digital twins of the physical communication system. Using these digital twins, the communication systems can pre-train ML models for the different communication tasks, optimize the various network procedures, or directly make predictions for the real system. This is particularly valuable when these digital twins are sufficiently calibrated with the site-specific measurements. 

The relationship between wireless communication and digital twins is mutually beneficial. With more efficient communication of the sensory data, more accurate and updated digital twins can be constructed. Further, as presented in \cite{Alkhateeb2023}, the digital twins with real-time 3D maps of the environment can be jointly leveraged to optimize the communication networks and to assist other applications such as autonomous driving, proactive collision prediction, traffic management, surveillance, and many other applications. Towards this objective, multiple access techniques play a central role in communicating the data from the distributed sensors in autonomous vehicles, IoT devices, distributed infrastructure side units, etc., needed for constructing and maintaining the digital twins. For example, it is important to develop digital twin-specific multiple access approaches that optimize the resource allocations among different sensors and modalities with the required digital twin fidelity and freshness as objectives \cite{ruah2023digital}. This is an interesting and challenging research problem since these different communication channels for the different sensors/modalities may have different QoS requirements, especially when conditioned on the final digital twin construction objectives. This optimization needs also to be done in a way that ensures the scalability of the network and the adaptability to the varying digital twin requirements for the different tasks.

On the other side, these environment digital twins could also be very beneficial for the optimization of the communication networks and the design of multiple access solutions. In particular, the digital twins provide the communication network with comprehensive perception, that is potentially near real-time, about its surrounding environment including the locations of their users, the nearest stationary/dynamic scatterers, sources of interference/blockages, etc. This perception can assist the various resource allocation decisions. For example, the digital twins could be leveraged to decide on what users to schedule in a specific MA transmission based on their relative locations, mobility patterns, and interaction with the nearest blockages. They can also be leveraged to develop adaptive and site-specific MA techniques that dynamically change their parameters based on the specific site characteristics. The perception provided by these digital twins can also enable the communication networks to make proactive multiple access and resource allocation decisions as they can extrapolate to how the environment will change in the next few tens or hundreds of milliseconds. This highlights the promising gains when leveraging environment digital twins to assist multiple access in future communication systems.

\subsection{Communication-aided sensing/localization systems}




Localization technologies have been embedded in mobile communication systems for decades \cite{9665436}. It is well known that satellite-based positioning technologies, e.g., Global Navigation Satellite Systems (GNSS), have been popular. However, they suffer from severe degradation in indoors and urban areas. To address these challenges, cellular-based localization methods have been considered due to their natural advantages of wide coverage, easy deployment, and low cost \cite{8226757}. As we have seen in the previous sections, multiple access schemes are capable of providing larger spectral and energy efficiencies in multi-user communications. Hence, under limited resources, multiple access techniques can contribute to the potential improvement of positioning accuracy and coverage for multiple users via efficient resource and interference management. In what follows, we review the latest multiple access schemes for communication-aided localization and discuss the research challenges.

\subsubsection{SDMA-Assisted Localization}
MIMO positioning \cite{7849233} becomes popular since it provides high-angle resolution, which results in a more accurate estimation of angle-of-arrival and angle-of-departure. In \cite{8415758}, 3D MIMO was employed to simultaneously support high-rate data communication and mobile 3D positioning for multiple users in the downlink \cite{8415758}. The transmit beamforming was optimized by minimizing the sum-localization error under data rate, total transmit power constraints, and imperfect CSI assumption. However, individual localization accuracy may not be guaranteed in this case. Moreover, \cite{8415758} assumes that localization is performed during data transmission after pilot transmission. In \cite{9540385}, mmWave MIMO system with cooperative base stations is considered for joint communication and localization under imperfect CSI and hybrid beamforming assumptions, where localization is performed together with channel estimation during the pilot transmission period. A joint beamforming and power allocation optimization problem for maximizing the weighted sum rate under individual localization accuracy constraints was formulated and solved. A larger rate-accuracy region was demonstrated in \cite{9540385} compared to the conventional schemes.

\subsubsection{PD-NOMA-Assisted Localization}
PD-NOMA was considered for a communication-localization integration system in \cite{9681857}. Specifically, a multi-scale PD-NOMA system was introduced in the multi-cell scenario for supporting continuous positioning waveform and flexible configurations for users with different locations to obtain higher positioning accuracy and less resource consumption. A convex optimization problem of power allocation was formulated to minimize the error rate for communication and the ranging accuracy for positioning. Performance gains over the benchmark TDMA-based localization were demonstrated. In \cite{9097407}, PD-NOMA was combined with mmWave MIMO for localization to further improve the positioning accuracy, where the users within one beam form a local PD-NOMA group. However, these works assume perfect SIC, which may not always be the case in practice. SIC also introduces latency in localization. In addition, PD-NOMA by fully decoding interference does not achieve the optimal multiplexing gain in multi-antenna settings \cite{9451194}. Hence, further improvements on the schemes, e.g., by using RSMA, are possible.

\begin{figure}[t]
\centering
\includegraphics[width=\linewidth]{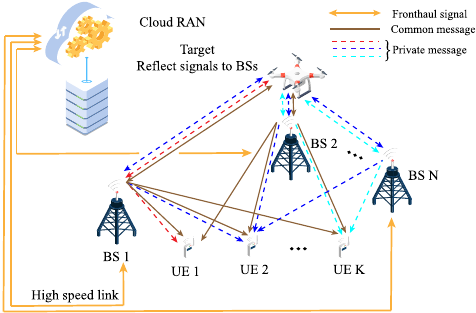}
\caption{RSMA assisted direct localization with $N$ cooperative base stations serving $K$ users while estimating a target's location  \cite{10032141}.} \label{fig:Localization_RSMA}
\end{figure}

\subsubsection{RSMA-Assisted Localization}
The use of RSMA to assist direct localization was investigated in \cite{10032141}, where multiple base stations cooperatively communicate with multiple users and directly estimate a target's 3D locations simultaneously through the RSMA framework via efficient interference management.  An example with $N$ cooperative base stations and $K$ users is shown in Fig. \ref{fig:Localization_RSMA}, where the splitting of common and private messages are illustrated, and all base stations are connected to a central processor via high-speed links. A Pareto optimization framework is formulated to characterize the sum rate of multiple users and the positioning error bound of the radar target. In particular, communication-centric and localization-centric optimization problems in their proposed cooperative ISAC scheme are investigated to find the optimal performance trade-offs. It was shown that the RSMA-assisted cooperative ISAC system can achieve the best trade-off performance among various baselines, e.g., the SDMA-assisted cooperative ISAC system.



\subsubsection{Technology Outlook and Future Works}
Using multiple access for assisting localization is a new research area. In light of the above works, a number of challenges are identified. First, the above works assume static users. However, the impacts of user movements and synchronization errors in MA-assisted localization have not been fully investigated. Second, the performance metric for measuring positioning accuracy is generally based on the Cramér–Rao bound. Explicit designs of practical localization estimators to achieve this bound are still lacking. Third, the MA-assisted localization schemes based on SIC could cause some decoding latency, which may not be desirable for delay-sensitive applications. To fully exploit the benefits of MA-assisted localization in wider communication scenarios, these challenges need to be addressed in future works. Additionally, it is crucial to understand how one can exploit all MA dimensions in the design of MA-assisted localization beyond the SDMA/PD-NOMA/RSMA approaches investigated so far and ultimately aim for a UMA-assisted localization.

\section{Multiple Access for Emerging Intelligent Applications}\label{Section_MA_for_Intelligent_Applications}

\subsection{Semantic Communications}
 




\begin{figure*}[t]
\centering
\includegraphics[width=1\linewidth]{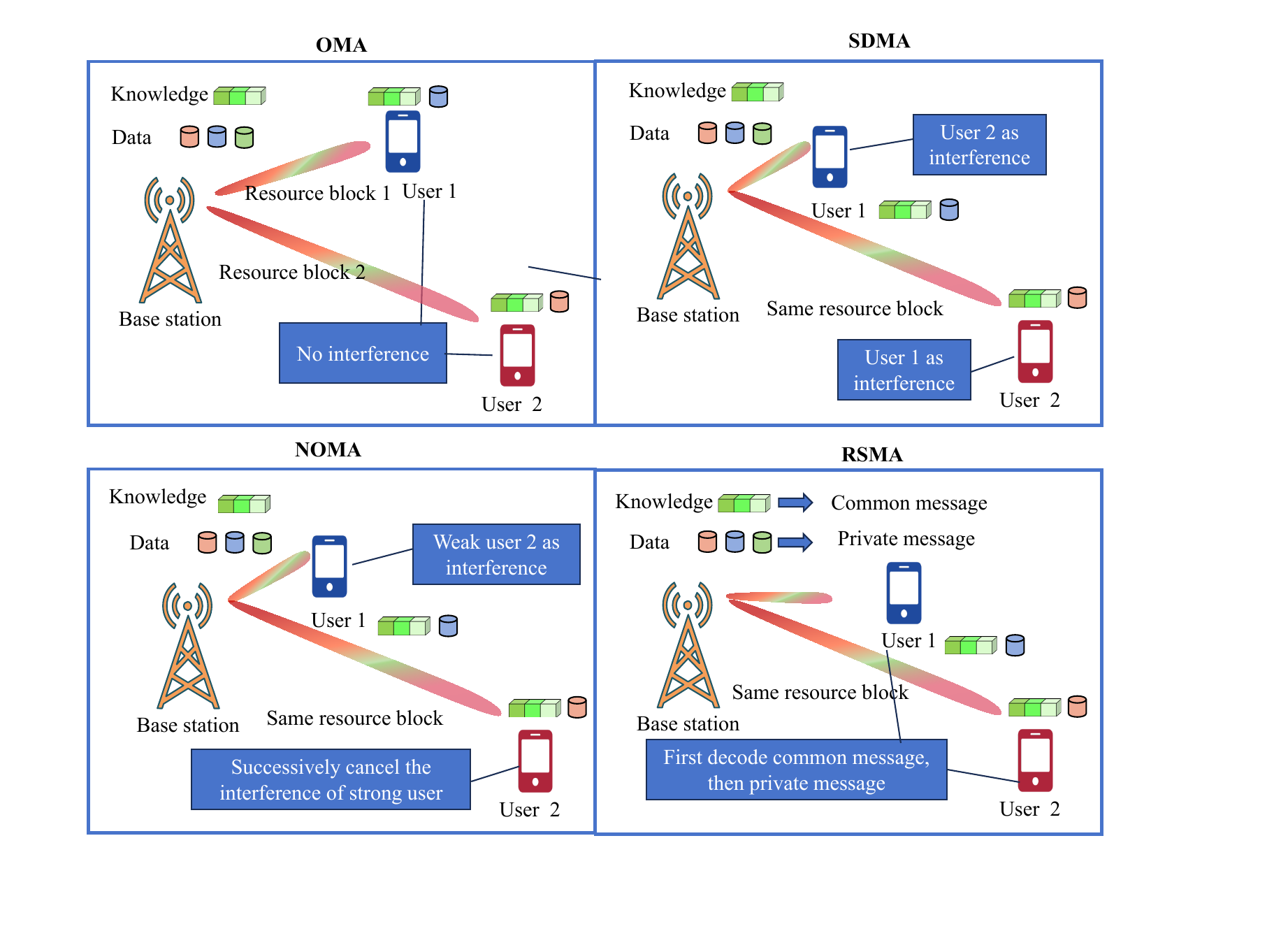}
\vspace{-5em}
\caption{An illustration of SeCom with multiple MA schemes.} \label{fig:semantic}
\end{figure*}

Semantic communication (SeCom), regarded as an advancement beyond the Shannon paradigm, strives to successfully transmit semantic information from the source.  Unlike the conventional Shannon paradigm that focuses on precision in receiving individual symbols or bits regardless of the meaning, SeCom greatly enhances the communication efficiency and reliability by prioritizing the transmission of meaningful information 
\cite{shi2021semantic,bao2011towards,qin2021semantic,gunduz2022beyond,yang2022semantic,tong2021federated,niu2022towards,lin2023blockchain,luo2022semantic,lu2022rethinking,ding2023distributed}.
In SeCom, it is typical for both the transmitter and receiver to possess a shared knowledge base. This shared knowledge enables the extraction of compact semantic information at the transmitter and the reconstruction of  the original information from compressed semantic signal at the receiver \cite{xu2023edge}. 
To handle additional knowledge information, it is important to design the communication system to smartly synergize MA schemes and SeCom. 
Besides, considering the process of SeCom, proper MA design can greatly enhance the performance of SeCom. 

SeCom typically involves AI-related design. Unlike conventional communication systems, it usually lacks a well-defined performance metric, such as the Shannon capacity. To determine the appropriate MA for SeCom, it is important to identify the key performance indicator (KPI) for SeCom. Broadly, there are two main kinds of KPI, i.e., simulation-based KPI and analysis-based KPI.
In simulation-based KPI, the semantic rate expression is regraded as a function of SINR. This function is obtained through data fitting 
in the simulation results about semantic rate versus SINR \cite{yan2022resource,mu2022heterogeneous,hu2023multiuser}.
In analysis-based KPI for SeCom, 
one involves utilizing the probability graph shared by both the transmitter and the receiver,  referred as the knowledge base \cite{zhao2023semantic,yang2023secure}. The transmitter extracts the small-size semantic information based on the probability graph, while the receiver recovers the original information based on the shared probability graph. By employing the probability graph \cite{zhao2023semantic}, the semantic rate is derived by modifying the Shannon capacity expression in the following two aspects. In the first aspect, the reciprocal of the semantic compression ratio is multiplied in front of the Shannon capacity expression. Here, semantic compression ratio is defined as the ratio of  the bits of the extracted semantic information to the bits of the original information. 
The reason of multiplying the reciprocal of the semantic compression ratio  is that one bit of obtained semantic information can convey more than one bit in the original information.
In the second aspect, the SINR expression is modified by reducing the communication power.
This adjustment is necessary as some power needs to be allocated for the computation involved in semantic information extraction.
For a fair comparison between the conventional communication and SeCom, it is essential to maintain a fixed total power including both communication and computation power. The precise SINR expression in in SeCom is therefore obtained by theoretically demonstrating that the computation power for semantic information extraction follows a linear segment function concerning the semantic compression ratio. Furthermore, the communication power in semantic communication can be derived by reducing the computation power within the total power budget.

\par Based on the aforementioned KPIs, various existing works delve into the MA design for SeCom. In general, the currect MA schemes for SeCom encompass OMA, PD-NOMA, SDMA, RSMA, and some customized MA schemes, which are summarized as follows: 


\subsubsection{OMA-aided SeCom}
In OMA-aided SeCom, the base station uses   orthogonal resources to transmit both common knowledge and data information to different users as shown in Fig.~\ref{fig:semantic}. 
For OMA-aided SeCom with simulation-based KPI, 
the  spectral efficiency is maximized through optimizing the channel assignment and number of semantic symbols \cite{yan2022resource,10001594}.
Additionally, to address the energy efficiency issues, \cite{nguyen2023joint,yang2022performance} explores the optimization of total energy consumption under latency constraints. 
Combining with mobile edge computing (MEC), the energy consumption is minimized through jointly optimizing the semantic-aware division factor, as well as communication and computation resource management \cite{cang2023online,cang2023resource}. 
For OMA-aided SeCom with analysis-based KPI \cite{zhao2023semantic2,zhao2023semantic}, the total energy of the whole system can be optimized through optimizing the semantic compression ratio. 
The advantage of OMA-aided SeCom is easy to deploy, while the disadvantage is the number of served users is limited by the wirelress resource. 

\subsubsection{PD-NOMA-aided SeCom}
In PD-NOMA-aided SeCom, the base station can simultaneously transmit the semantic information to different users  as shown in Fig.~\ref{fig:semantic}. 
Moreover, as illustrated in \cite{mu2022heterogeneous,mu2023exploiting}, PD-NOMA can support the coexistence of
 conventional bit-interested user and semantic information-interested user.
In particular, the far-user exploits the SeCom, while the near-user employs the conventional PD-NOMA-aided communication \cite{mu2023exploiting}. 
Furthermore, a unified framework for NOMA-aided  SeCom  with general  datasets and  data modalities is presented in \cite{10225385}.
The PD-NOMA-aided SeCom system achieves the system performance gain through jointly considering both conventional and SeCom transmissions.

\subsubsection{SDMA-aided SeCom}
In SDMA-aided SeCom, the semantic information can be simultaneously transmitted to multiple users through the strategic design of precoding in the multiple-antenna assisted system as shown in Fig.~\ref{fig:semantic}. 
To solve the semantic performance optimization problem for SDMA-aided SeCom, the generalized power iteration precoding algorithm is applied in \cite{kim2023semantic}.
Considering the signal transmission characteristic of wireless communication, the multimodal SDMA-aided SeCom system is investigated in 
\cite{luo2022multi}.
In SDMA-aided SeCom, the precoding scheme enables efficient semantic information transmission for multiple users through exploiting the space domain characteristics.



\subsubsection{RSMA-aided SeCom}
In RSMA-aided SeCom, the knowledge intended to multiple users can be encoded into the common stream, while the individual data intended to a specific user is encoded into the private stream \cite{yang2023energy}, as shown in Fig.~\ref{fig:semantic}.
In NOMA-aided SeCom and SDMA-aided SeCom, the common knowledge for each user is transmitted and received in a unicast way, even though the common knowledge is the same for all users.
In RSMA-aided SeCom, the common knowledge is transmitted in a multicast way. 
Using multicast, the resource allocated for common knowledge can be greatly reduced in RSMA-aided SeCom than NOMA-aided SeCom and SDMA-aided SeCom.
As a result, the overall performance of RSMA-aided SeCom outperforms NOMA-aided SeCom and SDMA-aided SeCom \cite{yang2023energy}.
Taking into account the URLLC, the weighted sum semantic information transmission rate maximization problem is formulated in \cite{zeng2023task} through jointly optimizing the semantic information extraction, delivery duration, rate splitting, and transmit beamforming.
Besides, the authors in \cite{cheng2023interest,cheng2023resource} focus on optimizing the  quality of experience (QoE) for image transmission in the RSMA-aided SeCom system. 
In RSMA-aided SeCom, the common knowledge is transmitted in multicast, while the individual data are transmitted in unicast.
Due to the flexible combination of multicast and unicast, the system performance in RSMA-aided SeCom is superior over NOMA-aided SeCom and SDMA-aided SeCom.

\subsubsection{MDMA-aided SeCom}
 A novel MA called model division multiple access (MDMA) is proposed in \cite{zhang2023model}. This innovative approach involves simultaneously serving multiple users with distinct model information space resources.
In MDMA-aided SeCom, through analysing the resource of the semantic domain in high-dimensional semantic space, the information of different users can be recovered in the model domain \cite{zhang2023model}.
It is shown that MDMA-aided SeCom can improve the performance over traditional FDMA and NOMA. 
Due to the superior performance of MDMA, MDMA-aided SeCom is applied  in a point cloud based SeCom system to solve the three-dimensional representation problem
\cite{liu2023semantic}.
The realization of MDMA requires learning-based design 
and the authors in 
\cite{10288558} utilize deep learning-based multiple access (DeepMA) method by training SeCom models to construct MDMA-aided SeCom.

\subsubsection{Technology Outlook and Future Works}
In general, OMA can be easily applied to SeCom and satisfy the requirements of knowledge information transmission and semantic computation. 
By leveraging multiple antennas, SDMA-aided SeCom can be used to further enhance the system performance.
For the weak user with low transmission rate, PD-NOMA-aided SeCom can be flexibly applied. 
Recognizing the inefficiency of knowledge transmission in PD-NOMA and the superior performance gain of RSMA over OMA, SDMA, and PD-NOMA, RSMA-aided SeCom is shown to be more effective in balancing the interference during knowledge information and data information transmissions.
Moreover, leveraging the unique properties of semantic space, new MA schemes such as MDMA can be tailored for SeCom. 
Future directions of MA schemes for SeCom should explore the integration of MA design in semantic information extraction and resource allocation for emerging MA schemes including NOMA-aided SeCom, RSMA-aided SeCom, MDMA-aided SeCom, and ultimately UMA-aided SeCom to exploit all MA dimensions.

\subsection{Metaverse}



The Metaverse is considered to be the new generation of the Internet, aiming to build a digital world where people can meet and interact in real-time by integrating various emerging technologies, such as  digital twin (DT), extended reality (XR), and holographic \cite{wang2022survey,kang2023adversarial,xu2022full,du2023rethinking}.
In particular, the physical devices in the physical world transmit the sensed information to the mobile edge servers, which partial/completely compute the received information \cite{luo2023privacy,jagatheesaperumal2023semantic}. Then, mobile edge servers transmit the computed information to the cloud server for further inference. In Metaverse, the information of the digital world is updated in the cloud server to  meet the real condition of the physical world. Besides, the simulated information of the digital world in Metaverse including the communication system schedules and physical devices configurations can be transmitted to the physical devices through communication system. 
To support the immerse and intelligent-related requirements of emerging applications in Metaverse including education, military, healthcare, real estate, and manufacturing,
MA schemes must be considered 
\cite{brik2023guest}.
Existing MA schemes for the Metaverse applications include OMA, NOMA, SDMA, and RSMA, which are respectively specified as follows.

\subsubsection{OMA-aided Metaverse}
In OMA-aided Metaverse, users are supported with orthogonal resource. 
To support the immerse applications in Metaverse, a blockchain based OMA scheme is proposed for ubiquitous access controls
\cite{gai2022blockchain}.
For VR applications in the Metaverse,  a learning-based system can be applied \cite{9838736}.
Further considering the security in the Metaverse \cite{10001331}, the covert communication technique is used for OMA-aided Metaverse.

\subsubsection{NOMA-aided Metaverse}
Considering multiple physical devices and limited wireless resource, NOMA-aided Metaverse can serve more physical devices than OMA-aided Metaverse.
The QoE model for Metaverse application is explored in \cite{jiang2023qoe} by incorporating factors such as the virtual distance and network effect.    
To maximize the QoE utility within the Metaverse system, a two-step joint resource allocation for NOMA user channel allocation 
and Metaverse service selection scheme is proposed.
In the first step, the resource allocation for NOMA is determined by the one-to-many matching game, while a hedonic coalition formation game in the second step is applied to solve the Metaverse service selection.
The two-step scheme is verified to improve the average QoE utility of the system. 
In NOMA-aided Metaverse, the massive access problem in Metaverse can be solved through serving multiple users with the same time and frequency resource.

\subsubsection{SDMA-aided Metaverse}
In SDMA-aided Metaverse, the beamforming technique is utilized to support the communication between the physical devices in the physical world and the digitial devices in the Metaverse. 
Considering the advantages of RIS, the 
RIS-assisted Metaverse system is explored in \cite{rahman2023spectral}. 
Given the substantial communication rate demands in the Metaverse, mmWave and THz communication systems are required.
To solve the beam prediction problem introduced by the high-frequency system, the authors in \cite{nie2023vision} propose a multimodal deep learning framework based on three-dimensional convolutional transformers using both optical and rad data. 
In SDMA-aided Metaverse, the huge communication rate demand can be solved through proper beamforming design in the mmWave and THz communication systems.

\subsubsection{RSMA-aided Metaverse}
Through dynamically treating the power of power of interference and noise, RSMA-aided Metaverse is a promising way to flexibly handle the dynamic channel conditions between the physical world and digitial world. 
In uplink RSMA-aided Metaverse, the multi-user capacity region can be approximated with lower complexity than the conventional scheme such as dirty paper coding, which is promising in the systems with  a large number of users.
In downlink RSMA-aided Metaverse, thanks to robustness of RSMA towards CSIT inaccuracy and user mobility, RSMA achieved by RSMA can well support Metaverse applications in particular for high-mobility scenarios. 

\subsubsection{Technology Outlook and Future Works}
The Metaverse applications require high-data transmission, immerse experiences, and intelligent services.  
To meet the unique characteristics of Metaverse, MA schemes can be applied to support massive connectivity, enhance low-latency immerse experience, high spectral efficiency, and robustness.
Comparing different MA schemes for Metaverse,  NOMA can support more physical devices than OMA at the cost of additional decoding computation complexity. 
RSMA-aided Metaverse is more suited to the Metaverse applications than other MA schemes especially  for high-mobility scenarios.
Future directions of MA-aided Metaverse include the cross-layer design of considering both the physical-layer OMA/NOMA/SDMA/RSMA resource allocation and the application-layer scheduling, the design of MA schemes that account for  considering multiple time slots in the Metaverse, 
as well as MA schemes that consider both low-latency and prediction accuracy in the digital world. Ultimately, the goal is to design UMA-assisted metaverse.

\begin{figure*}[t]
\centering
\includegraphics[width=1\linewidth]{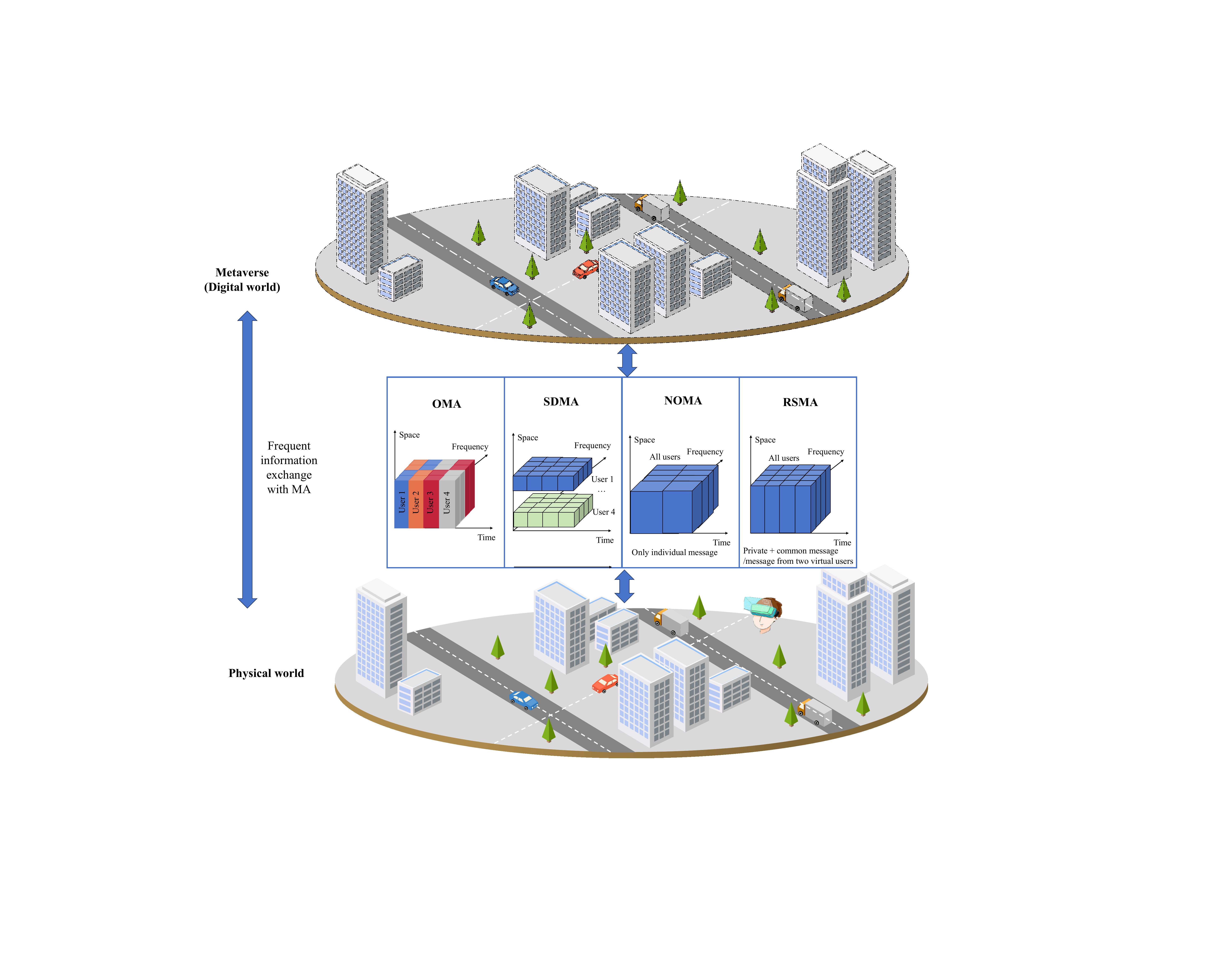}
\vspace{-9em}
\caption{An illustration of Metaverse over communication systems with multiple MA schemes.} \label{fig:metaverse}
\end{figure*}

\subsection{Virtual Reality}

In VR video streaming, an immersive viewing experience is provided for users by rendering 360-degree videos through their head-mounted displays (HMDs). To accommodate the growing demand for VR streaming and the increasing number of VR users, different challenges should be addressed in wireless systems. The bit rate of the 360-degree videos transmitted in VR streaming is much higher than that of conventional two-dimensional (2D) videos. In VR streaming, users may experience motion sickness due to a motion-to-photon delay (from the user's head movement to the time at which the image gets updated on the VR screen) larger than $20$ milliseconds (ms). This delay occurs when a user turns its head to watch another part of the 360-degree video frame, but that part has not been rendered for the user yet~\cite{elbamby2018toward}. To meet the required low latency and high bit rate for VR video streaming, different components of wireless systems, such as radio resource allocation algorithms, MA schemes, and network infrastructure, must be optimized and integrated. Three MA schemes are widely used in the literature for VR video streaming, namely OMA, NOMA, and RSMA.

\subsubsection{OMA-aided VR} Both TDMA and OFDMA have been used as OMA in VR video streaming. OMA scheme provides VR users with a simple receiver, which can subsequently lead to lower weight HMDs and extended battery life. A number of works have focused on TDMA and OFDMA VR systems discussed in the sequel.
\par \textit{TDMA} was investigated in~\cite{long2020optimal} for a 360-degree VR video streaming with multicast opportunities. The authors aimed at minimizing energy consumption by optimizing transmission resource allocation and quality level selection.  In~\cite{guo2018optimal} a downlink TDMA system for multicast transmission of VR video tiles from a server to multiple users is considered. A globally optimal closed-form solutions for two formulated non-convex optimization problems is provided: one aimed at minimizing average transmission energy while maintaining video quality, and the other focused on maximizing received video quality within a given transmission energy budget.
\par \textit{OFDMA} was studied in~\cite{guo2021power} for 360-degree VR video streaming across wireless networks. A multi-group multicast optimization problem was formulated aiming at finding the optimal beamforming, subcarrier allocation, transmission power, and rate allocation. In ~\cite{zhang2021buffer}, the resource allocation problem in a buffer-aware OFDMA VR streaming network was formulated as a stochastic game, where users bid for their desired data rates to enhance their VR viewing experience. The authors proposed a DRL algorithm for each user to solve this problem in a distributed manner. In~\cite{feng2023qoe} OFDMA was used in a DT-enabled wireless VR system. The authors formulated an optimization problem to maximize the max-min QoE and ensure fairness of the resource allocation among the users. In~\cite{yang2019cmu}, a VR video streaming using cooperative multicast and unicast in an OFDMA-based heterogeneous cloud-radio access network~(H-CRAN) was studied, consisting of one macro base station  and several remote radio heads~(RRHs). The authors formulated a mixed-integer nonlinear problem to maximize users' QoE when the macro base station transmits a specific version of the VR video to all users via multicast, and the RRHs transmit enhanced versions of tiles based on the predicted viewport to their corresponding users. The authors in~\cite{comcsa2020innovative} proposed a ML-based scheduling and resource allocation algorithm when an unmanned aerial vehicle~(UAV) records live events with its 360-degree spherical camera, sends the recorded video to a mobile edge computing~(MEC) server, and the live video is streamed from the MEC to users using OFDMA. The authors aimed to maximize the QoS of the live video for the users while maintaining an acceptable QoS for other existing traffic types in the network. In~\cite{zhang2022uav}, OFDMA for VR video streaming was used in a UAV-aided MEC wireless network, which includes one macro base station and multiple UAVs. The authors considered that each VR video comprises a base layer and several enhancement layers, and aimed to maximize the users' QoE by optimizing the UAV placement, VR video layer assignment, and MEC and radio resource allocation. In~\cite{eltobgy2020mobile}, the authors considered live multicast streaming of 360-degree videos to mobile users by allocating the available resource blocks in an OFDMA-based wireless network. The authors first partition the users into multicast groups and allocate the available resource blocks among them. Once the resource blocks are allocated to each group, the users' QoE is maximized by solving a tile quality selection problem. In the aforementioned works, each orthogonal radio resource (i.e., time slot or time-frequency resource block) is allocated to only one VR user for unicast transmission. Thus, the number of VR users that can be supported simultaneously is restricted by the number of available orthogonal radio resources.

\subsubsection{PD-NOMA-aided VR} In NOMA, VR video tiles can be transmitted to multiple users using the same radio resource. Thus, compared to the OMA scheme, NOMA can support a greater number of VR users and has been the focus of several works discussed next.
\par In~\cite{yu2023user}, a downlink PD-NOMA system is considered for transmitting VR frames from a Metaverse server to multiple VR users. The authors proposed a multi-agent DRL algorithm called user-centric critic with heterogeneous actors~(UCHA) for optimizing channel access and downlink power allocation. In~\cite{xiao2022transcoding}, PD-NOMA is used to deliver VR videos in an edge-enhanced wireless network, employing a multi-agent DRL algorithm as a cooperative caching scheme in edge base stations. To enhance content delivery efficiency, the authors considered a combination of unicast and multicast, and proposed a base station-multicast group matching algorithm. In~\cite{li2023noma}, a wireless network consisting of a base station, UAVs, and VR users is considered. In this network, PD-NOMA is employed to transmit video tiles from the base station to UAVs, and also from the UAVs to VR users. The authors aimed to optimize the selection of tiling patterns at the base station, the grouping of VR users and UAVs, as well as the allocation of computational resources and the duration of NOMA transmissions. In~\cite{zhang2021joint}, authors consider a UAV-assisted cellular network in which PD-NOMA is employed for content delivery from the UAVs to users. In particular, two users that request the same content for their AR and normal multimedia applications are grouped for PD-NOMA transmission. In this work, the authors aimed to minimize the content delivery delay by optimizing the user association, power allocation of PD-NOMA, and UAVs deployment. Since the PD-NOMA scheme manages the multi-user interference with SIC, the receiver complexity is increased as the number of VR users increases. Additionally, the grouping of VR users and the decoding order play a crucial role in the performance of the PD-NOMA scheme. Furthermore, due to the use of SIC, the VR users' physical layer privacy may be at risk in a PD-NOMA system, necessitating additional privacy protection measures. Therefore, most of the existing works in the literature have focused on utilizing PD-NOMA for VR video multicasting. However, finding the optimal user grouping is a challenging problem because it requires optimizing the number of groups and the number of users within each group.

\subsubsection{RSMA-aided VR} 
\begin{figure}[t]
\centering
\includegraphics[width=\linewidth]{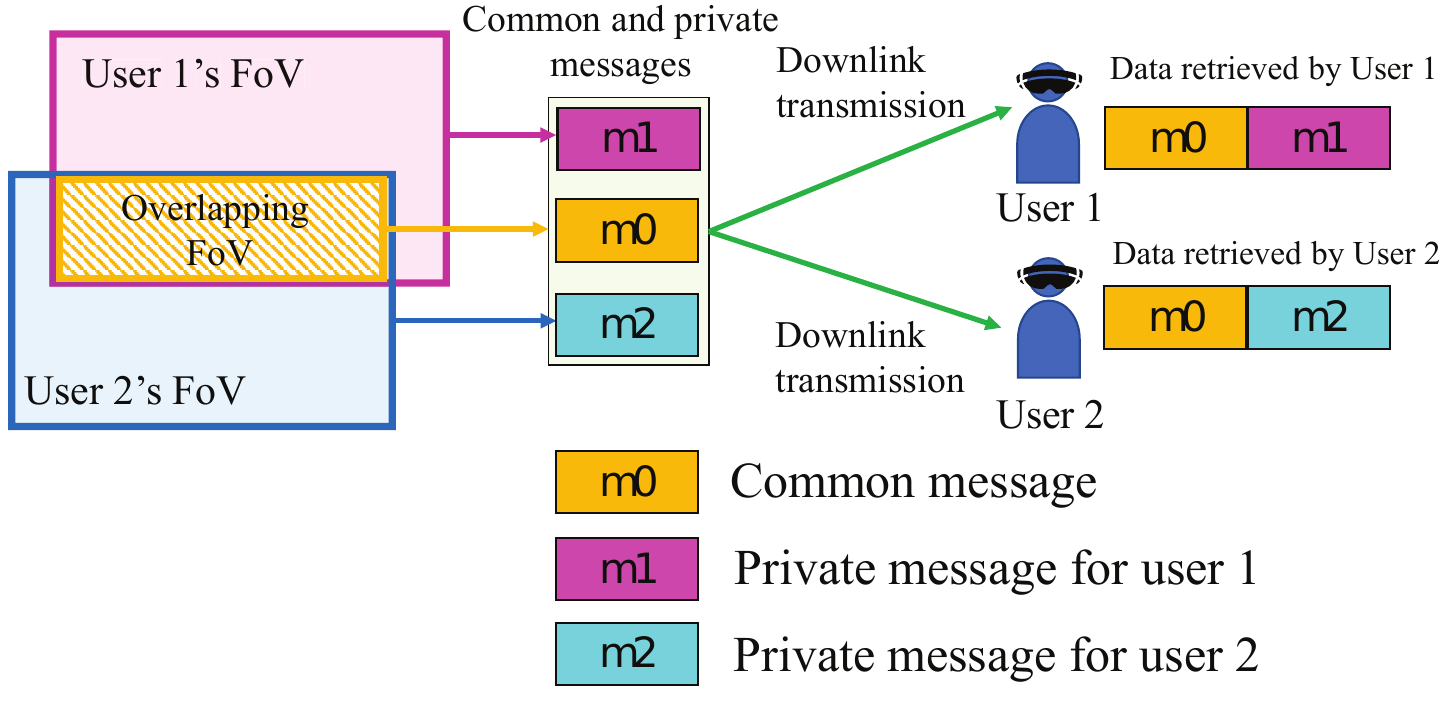}
\caption{An example of using RSMA for VR applications. Users 1 and 2 have overlapping field of view (FoV). The video tiles with overlapping FoV are sent via the common messages. The video tiles with non-overlapping FoV are sent via the private messages.}
\label{fig:VR-RSMA}
\end{figure}

RSMA manages multi-user interference by decoding a part of the interference, considering it as the common message of VR users, while treating the remaining part as noise, representing the private messages of other VR users. An example of using RSMA for VR applications is shown in Fig. \ref{fig:VR-RSMA}. 
RSMA has been the focus of recent VR studies. In~\cite{zhao2022optimization}, the authors proposed iterative algorithms to optimize beamforming vectors and rates of sub-messages in a multicast RSMA VR streaming system. In~\cite{huang2022rate}, RSMA is employed in an reconfigurable intelligent surface-aided VR streaming system. The authors proposed a deep deterministic policy gradient~(DDPG) algorithm with imitation learning to jointly optimize beamforming vectors, IRS phase shifts, rate-splitting parameters, and video tile bit rate selection. In~\cite{zhao2021adaptive}, RSMA is exploited for adaptive streaming of tiled 360-degree videos from a multi-antenna base station to multiple single antenna users. The authors utilized the concave-convex procedure (CCCP) to optimize parameters such as the rates of common and private messages in RSMA, as well as transmission beamforming vectors. In~\cite{hieu2023virtual}, the users in a 360-degree video streaming system are clustered into different groups based on their overlapped field-of-view~(FoV). Then, to effectively manage interference among the users, the authors adopted a hierarchical RSMA approach in which the messages are split and encoded into three streams: super common stream, group common stream, and private stream. All those works demonstrate the benefits of RSMA over other MA schemes, such as SDMA and PD-NOMA, in VR applications.

\subsubsection{Technology Outlook and Future Works}
\par Some potential research challenges in MA design for VR streaming systems include designing a low-complexity receiver for VR users while simultaneously preserving the VR users' privacy and considering the HMD characteristics. For example, it is important to consider that HMDs have limited battery and computational resources. HMDs should be lightweight. Additionally, they may have multiple antennas, which raises the challenge of designing RSMA for receivers with multiple antennas. 
\par For cache-enabled wireless networks, placing VR content in the cache requires a larger timescale than delivering VR content using an MA scheme. To enhance VR video streaming performance, content placement and delivery should be jointly optimized, leading to a multi-timescale optimization problem. \par In UAV-assisted cellular networks, the deployment and trajectory of UAVs should be designed according to the mobility of VR users. This design is essential to manage interference in NOMA and RSMA schemes and to prevent degradation in VR streaming performance. 
\par In IRS-aided VR streaming systems, it is important to consider the impact of imperfectly estimated CSI for both the IRS-user and BS-IRS links on the performance of the MA schemes. 
\par In VR video streaming, particularly in the case of live VR streaming, multiple users may request the same part of a 360-degree video frame due to their shared interests. This correlated data presents the opportunity to design a more efficient RSMA for VR streaming, in contrast to typical scenarios in which the data requested by users are independent and uncorrelated. However, RSMA should also perform well when the requested tiles by users do not overlap and the users watch distinct parts of the video frames. Therefore, the RSMA parameters need to be optimized, considering all possible cases in VR video streaming systems.
\par The integration of RSMA with other MA schemes, such as OFDMA, has not yet been investigated to determine whether RSMA performance can be further enhanced in VR video streaming systems. This approach provides greater flexibility, especially when dealing with multigroup multicasting problems, by pairing a group of users and serving them on a designated resource block using RSMA. Beyond OFDMA-RSMA, integration of CD-NOMA and RSMA would also be of interest to address the need for bit rate in VR.
\par In VR streaming systems, certain information, such as users' head movements, needs to be transmitted from the users to the server. Using RSMA, users can share the uplink channel to send their information simultaneously. Furthermore, federated learning can be applied within a VR streaming system to obtain a viewport prediction model for the users. In particular, federated learning enables the users to train a shared viewport prediction model, while their privacy is preserved without sharing their historical head movement data \cite{zhang2021buffer, setayesh2023PredFramework}. RSMA can be incorporated in federated learning to facilitate model aggregation without compromising model accuracy.
\par Higher frequency bands, such as mmWave and THz, can be employed in VR streaming systems to fulfill the bitrate requirements of high-resolution 360-degree videos. The use of RSMA in combination with beamforming design in these high-frequency bands represents an interesting research area. Moreover, communication in mmWave and THz bands faces high path loss and blocking issues. It would be interesting to explore how RSMA can be integrated with cooperative communication to tackle these challenges in VR video streaming systems.
\par 6G wireless systems will provide various services, including an evolved version of eMBB, known as eMBB+. The high data rate and low latency requirements for VR video streaming position it as one of the key eMBB+ services provided by 6G systems. Whether RSMA is a promising MA scheme to meet the hybrid requirements of VR users remains unknown and a potential research direction.

\subsection{Smart Radio and Reconfigurable Intelligent Surfaces}


\begin{figure*}[t]
\centering
\includegraphics[width=0.7\linewidth]{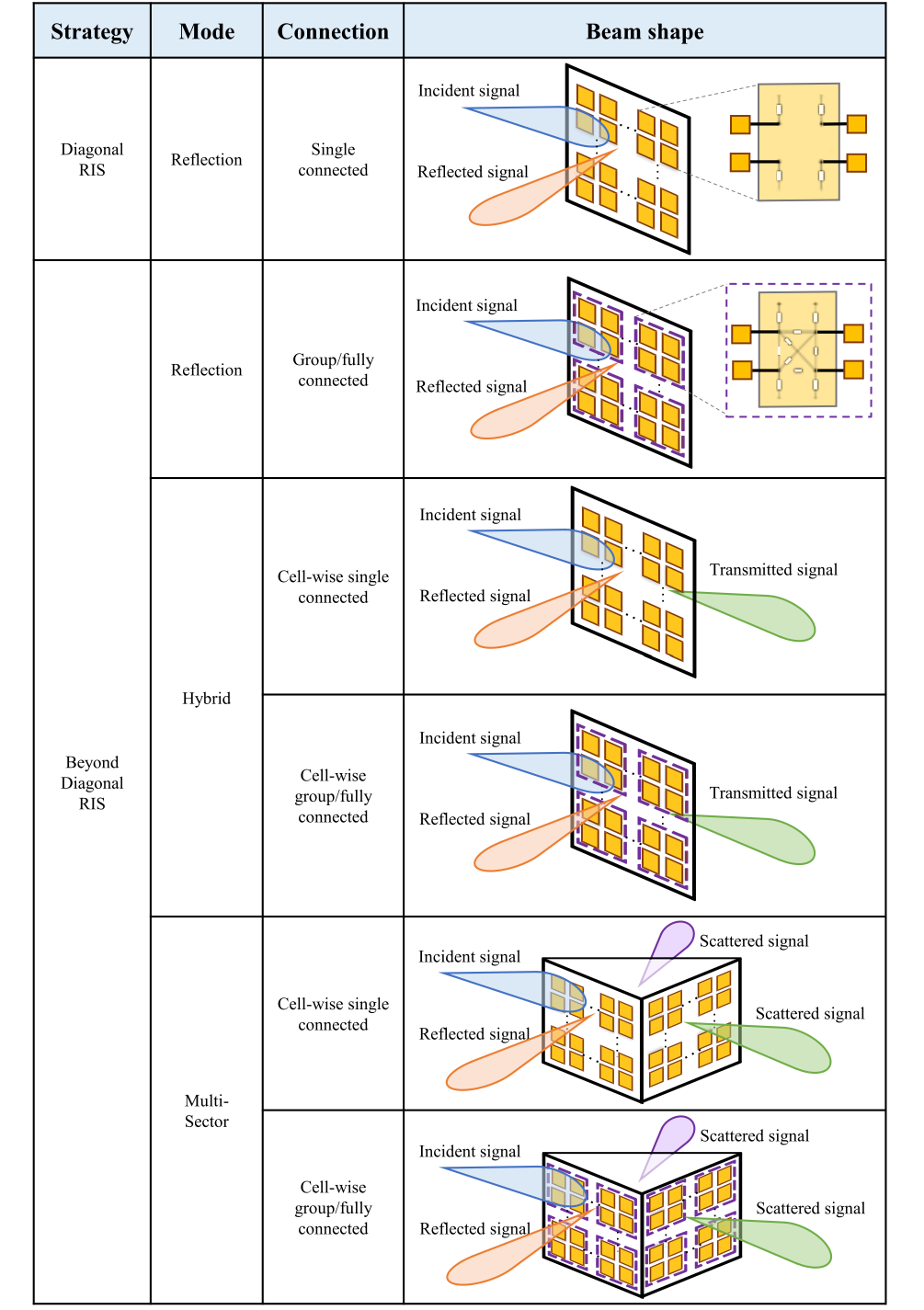}
\caption{Illustration of the family of RISs.} \label{fig:RIS}
\end{figure*}

\par Reconfigurable intelligent surface (RIS) has garnered significant recognition as a revolutionary technique for 6G. By incorporating multiple passive reconfigurable elements  with tunable amplitudes and phases, RIS is capable of effectively manipulating the direction and strength of the scattered signals, thereby exerting a transformative impact on the propagation environment.
Recent advances on RIS has witnessed its evolution from convention diagonal RIS (D-RIS)--where each element is independently controlled by a tunable impedance connected to ground, resulting in a diagonal scattering matrix--to the generalized beyond diagonal RIS (BD-RIS) \cite{hongyu2023BDRIS, hongyu2023BDRIS2}. BD-RIS facilitates interconnections among elements, leading to a beyond diagonal scattering matrix. 
Figure \ref{fig:RIS} illustrates the state-of-the-art RIS architectures as outlined in  \cite{nerini2023BDRIS}.
In general, RIS can be classified into D-RIS and BD-RIS depending on the nature of the scattering matrix. Conventional D-RIS is typically  modeled as a diagonal phase shift matrix, limiting its functionality to a reflection mode where incident signals are exclusively reflected from one side of RIS.
In contrast, BD-RISs, with beyond diagonal scattering matrices, are capable of creating group/fully-connected reconfigurable impedance networks based on the same reflection model as D-RIS.
Moreover, to achieve full-space coverage, hybrid and multi-sector modes are proposed for single/group/fully-connected BD-RIS \cite{hongyu2023BDRIS2}. The hybrid mode enables  partial reflection and transmission of incident signals to different sides of RIS, which is also known as intelligent omni-surface (IOS) or simultaneous transmitting and reflecting RIS (STAR-RIS) \cite{hongliang2022IOS}. The multi-sector mode extends the hybrid model by dividing the full space into sectors, allowing incident signals from one sector to be partially reflected and scattered to different sectors.
\par Thanks to its powerful capability for controlling wireless propagation environments and manipulating signals spatially, considerable  research efforts have been dedicated to RIS-assisted MA schemes, such as OMA, SDMA, NOMA, and RSMA.  

\subsubsection{RIS-assisted OMA}
The deployment  of RISs was initially studied and implemented in OMA, focusing on point-to-point communication with a single or multiple resource blocks \cite{WCL2020RISrelay,tcom2020RISOFDM,WCL2020RISCE,IRS2020mmWave,TCCN2020RIS,changsheng2020CERIS,beixiong2021RISCE,CL2020performanceRIS,zappone2021RIS,ICC2016prototype, infocom2018prototype,JSAC2020prototype, access2022prototype,haifan2021RISprototype,linglong2023RISprototype}. 
In this simplified context, the primary  research focus involved passive beamforming design \cite{WCL2020RISrelay,tcom2020RISOFDM,WCL2020RISCE,IRS2020mmWave},  channel modeling \cite{TCCN2020RIS},  channel estimation \cite{WCL2020RISCE,changsheng2020CERIS,beixiong2021RISCE}, theoretical limits analysis \cite{WCL2020RISrelay,CL2020performanceRIS,zappone2021RIS}, and prototype development \cite{ICC2016prototype, infocom2018prototype,JSAC2020prototype, access2022prototype,haifan2021RISprototype,linglong2023RISprototype}.
Despite the introduction of additional challenges related to channel estimation and passive beamforming design, RIS-assisted OMA (OMA) shows significant potential to cost-effectively enhance the spectrum efficiency and extend coverage without requiring additional energy consumption.

\subsubsection{RIS-assisted SDMA}
In multi-user scenarios, the most commonly considered MA is SDMA. It is worth noting that in RIS-assisted SDMA, maximizing the channel gain is not the only objective due to the introduction of multi-user interference. 
To achieve objectives, such as maximizing spectral or energy efficiency and minimizing energy consumption, transceiver design in RIS-assisted SDMA typically involves the joint optimization of passive beamforming at the RIS and active beamforming at the transmitter that require tailored optimization methods and channel estimation methods.
Numerous optimization techniques have been proposed in existing literature addressing joint beamforming design with continuous or discrete D-RIS phase shifts \cite{cunhua2022SPRIS}, or for BD-RIS \cite{nerini2023BDRIS, hongyu2023BDRIS, tianyu2023BDRIS}.
Concerning channel estimation, multi-user channel estimation in RIS-assisted SDMA brings additional research challenges due to the presence of inter-user interference.
Existing literature has focused on advanced channel estimation techniques tailored to unstructured channels, which model low-frequency rich-scattering scenarios, as well as structured channels designed for high-frequency sparse channels \cite{cunhua2022SPRIS,COMST2022RISCEsurvey}.
In terms of prototyping, most of existing research on RIS prototypes considered a single-user case. The design of RIS-aided multi-user prototypes is still in its early stages. Recent work \cite{tomluo2023RISprototype} introduced a novel blind beamforming strategy for RIS-assisted SDMA, which designs the phase shifts without relying on channel information. Real-world trails have been conducted to demonstrate the effectiveness of this strategy.
In \cite{zhou2023multi}, a new prototype that incorporates both channel estimation and beamforming design is proposed. Experimental results highlight a notable enhancement in spectrum efficiency with the proposed scheme, marking a significant advancement of RIS-assisted SDMA prototypes.

\subsubsection{RIS-assisted PD-NOMA}
In conventional PD-NOMA without RIS, the efficacy of SIC decoding relies on a substantial channel strength disparity between users.
In RIS-assisted PD-NOMA, the aid of an additional reflection/transmission link helps to enhance the channel strength disparity, thereby augmenting the benefits of SIC decoding.
However, RIS introduces challenges in the resource allocation as the optimal SIC decoding order and user grouping depends on both the active beamforming at the transmitter and the passive beamforming at the RIS. 
The interplay between NOMA and RISs have been summarized in \cite{zhiguo2022NOMARISsurvey}. 
The primary research focus in RIS-assisted NOMA is on addressing the above challenges.
Besides, RIS offers significant advantages in addressing security concerns within conventional NOMA by enhancing wireless communication covertness at the physical layer \cite{lulv2021NOMARISsecurity}. 
An interesting observation from \cite{8970580} is that in the presence of RIS, NOMA is not always a better option than OMA. Indeed the minimum power needed to achieve a target rate may be lower for TDMA than NOMA, i.e. NOMA may perform worse than TDMA. This contrasts with the non-RIS deployment where PD-NOMA is superior to TDMA.

\subsubsection{RIS-assisted RSMA}
The potential advantages of RSMA and RIS have sparked interest in their combined use in recent years \cite{daniel2022RSMARISmagazine,hongyu2022RSMARISletter}. The synergy between D-RIS and RSMA has been widely studied, and revealing substantial performance advantages over RIS-aided SDMA, NOMA, OMA in terms of spectrum efficiency \cite{hongyu2022RSMARISletter,eduard2023RSMARIS,linzhi2023RSMARIS,vincent2023RSMARISVR,9910027,9912342}, energy efficiency \cite{zhaohui2020EERSMARIS,JSAC2023EERSMARIS}, user fairness \cite{kangchun2022RSMACRSRIS,shiwen2023RSMARIS}, outage probability \cite{TVT2021RSMARISoutage,ojcom2021RSMARISoutage} etc. 
Recent works have also delved into the interplay between BD-RIS and RSMA, covering scenarios such as STAR-RIS aided RSMA \cite{tcom2022STARRISRSMA,IoTJ2023RSMASTARRIS}, RSMA with group/fully connected RIS in the reflection mode \cite{tianyu2022RSMABDRIS,TWC2023BDRISRSMA}, and RSMA with multi-sector BD-RIS \cite{hongyu2023BDRISRSMA}.
BD-RIS aided RSMA has demonstrated significant  advantages in extending coverage, further boosting both spectrum and energy efficiency.
Given the absence of an RF chain at RISs, channel estimation is a widely recognized challenge in RIS-assisted links. 
Considering the resilience of RSMA to CSI inaccuracies and user mobility as shown in many existing works,  RIS-assisted RSMA emerges as a promising new paradigm to effectively address and compensate for the channel estimation limitations inherent in RIS \cite{hongyu2022RSMARISletter}.
RIS-assisted RSMA therefore offers a mutually beneficial solution to both RSMA and RIS. 
This advantage is distinctive to RIS-assisted RSMA, as RSMA uniquely demonstrates robustness to CSI, which is not shared by other conventional MA schemes.

\subsubsection{Technology Outlook and Future Works} RIS can significantly improve the performance of various MA schemes from the following perspectives: 
\begin{itemize}
    \item \textbf{Enhanced spectrum efficiency}: By dynamically adjusting the phase and amplitude of incident signals, RISs provide narrow beams directed towards users. This versatile capability brings about various advantages in different MA schemes: 
    \begin{itemize}
        \item OMA: enhances the signal strength, thereby improve the spectrum efficiency.
        \item SDMA: effectively redirects intended signals to their respective users and  mitigates multi-user interference, thereby enhancing spectrum efficiency. 
        \item NOMA: creates an accurate beam towards the serving users and enhances the channel strength disparity between the users, and thereby enhancing the spectrum efficiency.
        \item RSMA: unifies the benefits for OMA, SDMA and NOMA, leading to higher spectrum efficiency.
    \end{itemize}

    \item \textbf{Enhanced energy efficiency}: RISs employ beamforming techniques to concentrate signals on specific users. This avoids spreading energy in multiple directions and leads to more efficient energy utilization for all MA schemes.

    \item \textbf{Improved coverage}: RISs introduce a new paradigm for enhancing coverage in wireless networks. Instead of deploying additional energy-consuming infrastructure  for relaying and forwarding signals, RISs adaptively redirect signals to navigate obstacles like trees and buildings, thereby extending coverage for all MA schemes.

    \item \textbf{Low complexity}: RISs contribute to reduce the computational and hardware complexity for various MA schemes. In terms of computational complexity, the ability of RIS to generate narrow beams and mitigate multi-user interference allows for  simplified signal processing techniques at the transceivers for all MA schemes. As demonstrated  in \cite{tianyu2023BDRIS}, the transmitter employing classical regularized zero-forcing (RZF) precoding can attain comparable performance to a scenario where the transmitter  utilizes  optimized precoders (based on fractional programming), provided that the number of RIS elements is sufficiently large.
    From the hardware complexity perspective, as RIS can enhance the spectral efficiency, it is capable of reducing the number of  antennas at the transceivers without influencing system performance. 
    Moreover, for RSMA, the integration of RIS  empowers the use of simplified schemes, i.e., 1-layer RS, to achieve performance comparable to HRS/GRS without RIS. Therefore, RIS  contributes to  reduce the receiver complexity \cite{hongyu2022CLRSIRS}.
\end{itemize}









\subsection{Internet-of-Things Networks, Massive Connectivity, and Random Access }\label{IoT_section}

\begin{figure*}[t]
\centering
\includegraphics[width=0.7\linewidth]{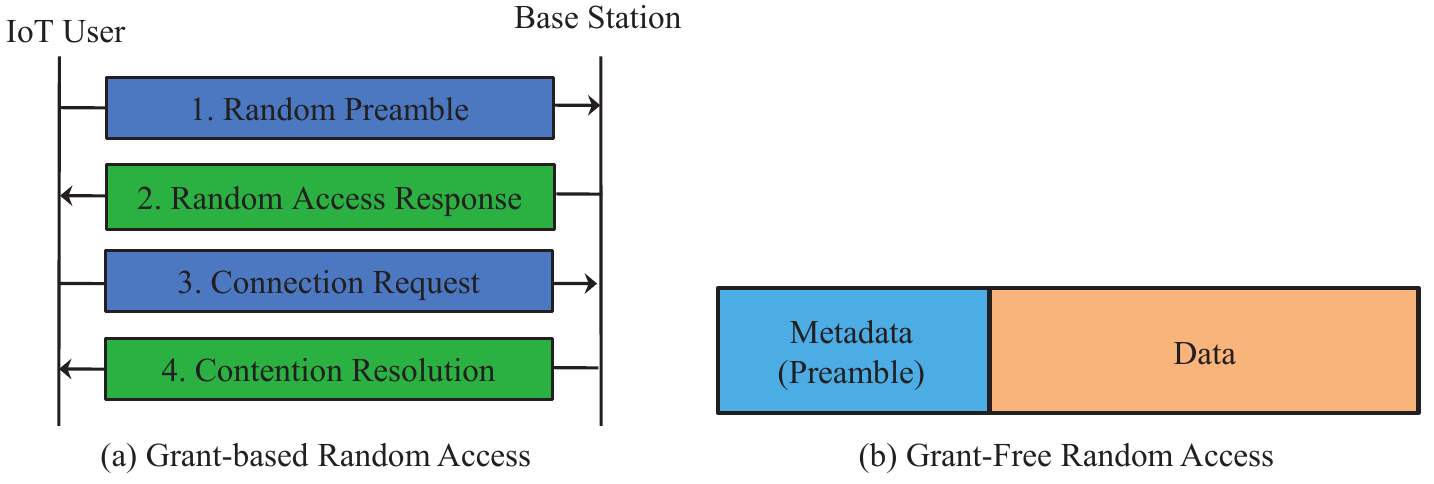}
\caption{Grant-based versus grant-free random access schemes.} \label{fig:massive connectivity}
\end{figure*}

For IoT communication network, the key is how to coordinate the transmission of small packets between the base station and a huge number of IoT devices in an reliable and low-latency manner. This is in sharp contrast to the human-type communication, which targets at high-speed transmission to/from a medium number of users. Such a difference poses new challenges for the MA design under IoT applications. Specifically, if the number of users is huge, then access collision is more likely to occur, which leads to long delay. In IoT network, it is of paramount importance to design innovative MA schemes that can greatly reduce the access delay so as to facilitate the small-packet transmission from a massive number of IoT devices. Generally speaking, in the literature, there are two solutions for MA in IoT network: the grant-based MA technique and the grant-free MA technique, as shown in Fig. \ref{fig:massive connectivity}. \par Grant-based MA technique are conventional for eMBB-like services and relies on the base station to grant users the access of the radio spectrum for data transmission. 
\par In grant-free MA, or also called grant-free random access (RA), the devices can access the channel without any prior resource requests \cite{Liu_Yu_Massive_IoT_3,9097306}. In addition, the set of active users is unknown to the receiver. Grant-free RA is suitable for supporting a massive number of devices with sporadic traffic, e.g., mMTC as in IoT, and can be divided into two different paradigms, namely sourced and unsourced RA \cite{9060999,9205230}. For sourced RA, the receiver is interested in both messages and user identities. Since the active device detection for sourced random access is a typical sparse signal processing problem, numerous compressed sensing (CS) and AI-based approaches have been employed to handle the detection and user identification problems \cite{Liu_Yu_Massive_IoT_1,8264818,8323218,8961111,9266124}. For unsourced RA, the receiver is interested in the transmitted messages only (user identity is recovered at the higher layers of the communication protocol rather than the
physical layer). 
\par In the sequel, we discuss grant-based MA, grant-free sourced random access (based on compressed sensing and AI), and grant-free unsourced random access.  

\subsubsection{Grant-based Random Access}
First, with human-type communication, the contention-based grant-based access schemes, e.g., ALOHA, are widely used. Under this scheme, there is a preamble pool consisting of lots of orthogonal preambles. When a user becomes active, it will randomly select a preamble from the poor. If an active users selects a unique preamble that is not selected by other active users, the base station will grant this user the access for data transmission. Otherwise, if some active users select the same preambles, the base station will detect their access collision and not grant them the access for data transmission. Despite of their simplicity, such schemes do work quite well in current cellular network, where the number of users is moderate. Therefore, it is not surprised that some efforts have been paid in the literature to investigate how to modify the existing grant-based access schemes according to the requirements of IoT applications. Here, the main issue to directly apply grant-based schemes in IoT network lies in delay - when there is a huge number of IoT devices to compete for the transmission grant from the base station, the possibility for collision is very high. As a result, in the literature, some works have been done to propose efficient methods for resolving the collisions. For example, in \cite{Bjornson_2017}, a strongest-user collision resolution scheme was proposed for grant-free random access in IoT network. The idea is that after a collision occurs, each colliding user will make a decision about whether it is with the strongest channel among all the colliding users. If a colliding user believes that its channel is the strongest, it will send the access request to the base station. Otherwise, it will be quiet for some time and re-compete for the access. It was shown in \cite{Bjornson_2017} that under the massive MIMO regime, the strongest-user collision resolution scheme can resolve most of the access collisions in massive connectivity.   

\subsubsection{Grant-Free Sourced Random Access based on Compressed Sensing}
Second, one may ask whether it is possible to reduce the collision probability, rather than resolving the collision, under IoT networks. The goal can be achieved via the grant-free random access scheme \cite{Liu_Yu_Massive_IoT_3}. Under the grant-based random access scheme, collision arises from the case when multiple users select the same preamble for transmission. To mitigate collision, a natural idea is that each user is allocated with a fixed preamble, while the preambles allocated to different users are different. In this case, the preamble of each user serves as the identification of this user. In other words, after receiving the preambles sent from the active IoT devices, the base station can detect which preambles are received, and then know which users are active. This scheme is called grant-free scheme because the base station just detects which users are active, instead of sending grants to permit some users to be active. However, detecting which preambles are received is quite challenging in massive IoT connectivity, because it is not possible to assign orthogonal preambles to a huge number of users. Recently, a breakthrough was made to tackle the above challenge arising from grant-based random access based on the compressed sensing technique. Specifically, although there is a huge number of IoT devices, most of them are in the sleeping model to save the energy at each time slot. Thanks to this activity sparsity, it was shown in \cite{Liu_Yu_Massive_IoT_1} that detecting the active devices, i.e., deciding which preambles are received, can be cast into a compressed sensing problem. More importantly, based on the state evolution, \cite{Liu_Yu_Massive_IoT_1} rigorously proved that in the asymptotic massive MIMO regime where the number of antennas at the base station goes to infinity, the activity detection error probability can go down to zero when the approximate message passing algorithm is used. This result theoretically justifies the effectiveness of the compressed sensing based grant-free random access scheme for IoT networks.

\subsubsection{Grant-Free Sourced Random Access based on AI}
In massive IoT connectivity setup, the number of IoT devices is huge. As a result, low-complexity multiple access design suitable for large-scale IoT network is appealing. In this sense, recently, AI-empowered multiple access schemes have attracted more and more attention in the area of massive machine-type communications.

On one hand, the performance of compressed sensing algorithms introduced in the above heavily depend on the choice of the sensing matrix, while most of the works on compressed sensing based device activity detection assume that the sensing matrix, which is determined by the pilot signals assigned to IoT devices, is given. For example, many works assume Gaussian pilots \cite{Liu_Yu_Massive_IoT_1,Fengler_Haghighatshoar_Jung_Caire_Massive_IoT,Chen_Yu_Massive_IoT} such that the approximate message passing (AMP) algorithm can be adopted to detect the active devices and analyze the detection performance based on the state evolution. Moreover, Reed-muller sequences based pilots have been utilized for device activity detection in \cite{Wang_Zhang_Hanzo_Massive_IoT}.  However, how to design the pilot signals that work the best for device activity detection is still open. This difficulty lies in the unknown relation between the sensing matrix and the detection performance. It is well-known that AI techniques are a good option to learn the complicated and unknown relations. Therefore, recently, there is a growing interest in applying AI to design the pilot signals for massive IoT connectivity. For example, it was found in \cite{Cui_Li_Zhang_Massive_IoT,Li_SPL_2019} that if the deep learning technique is used to design the pilot signals, the device detection performance can be improved compared to using Gaussian pilots and Reed-muller sequences based pilots.

On the other hand, deep learning technique can also be applied to design efficient device activity detection algorithms, when the dimension of the problem is huge due to the large number of IoT devices. In general, these deep learning based device activity detection methods can be classified into data-driven based methods and model-driven based methods. Under data-driven based methods, basic neural network architectures, e.g., fully connected neural networks, CNNs, are utilized to detect active devices \cite{Li_SPL_2019,Wu_CL_2021}. Under model-driven based methods, existing algorithms, e.g., AMP, group LASSO, etc., are utilized to train the neural networks for better activity detection performance \cite{Cui_JSAC_2021,Bai_JSAC_2022,Johnston_TWC_2022,Zhu_TWC_2021}. 

To summarize, there is a theoretical gain of the grant-free random access scheme over the grant-based random access scheme in IoT networks, because the former scheme can mitigate almost all the collisions in the massive MIMO region, but the latter scheme can merely tackle some of the collisions. But the theoretical gain is achieved with complicated device activity detection algorithms, under both the compressed sensing based scheme and the covariance based scheme. 

\subsubsection{Grant-Free Unsourced Random Access}
 For unsourced RA \cite{8006984}, the receiver is interested in the transmitted messages only (user identity is recovered at the higher layers of the communication protocol rather than the
physical layer). Moreover, all users share a common codebook, which is motivated by the practical scenario where millions of low-cost IoT devices have their codebook hardwired at the moment of production. In addition, the number of users grows with the blocklength. Only per-user error probability is of interest. The fundamental limits of the unsourced RA channel have been studied in \cite{8006984,9462926,9941000}. Various popular achievable schemes are proposed in the literature, which are based on concatenated coding structures \cite{9153051,9432925}, ALOHA \cite{8891583}, random spreading \cite{9148687,9500486}, etc.

\subsubsection{Technology Outlook and Future Works}
There are still many challenges for MA in IoT network that have not been addressed. For example, most of the works assume that all the IoT devices are perfectly synchronized, such that when they become active, they transmit their preambles at the same time. In practice, it is impossible to perfect synchronize a huge number of low-cost IoT devices. In this case, how to make the strongest-user collision resolution scheme work under the grant-based random access scheme? How to formulate the device activity detection and synchronization error estimation problem as a compressed sensing problem under the grant-based random access scheme? Whether the sample covariance matrix of the received sequence is still a sufficient statistic for device activity detection? All these questions should be carefully addressed before the MA schemes can work in IoT networks. Moreover, there exist correlation in user activity along the temporal and spatial domain. For example, if there is a fire, several sensors around the fire site tend to be active together, and each of these sensors tend to be active for some time. How to utilize these correlation as side information to improve the activity detection accuracy is also a challenge. 

\section{Roadmap to 6G Standardization}\label{Section_6G}





3GPP is the standards body that introduced UMTS/W-CDMA for 3G, LTE for 4G and NR for 5G. 3GPP also introduced GSM for 2G even before being called 3GPP.  While 6G work has not yet started in 3GPP, it is expected to quickly intensify in the 2nd half of 2025. 


At the highest conceptual level, 3GPP has delivered over its history systems based on TDMA (2G GSM), CDMA (3G UMTS) and (O)FDMA (4G LTE and 5G NR) closely following any textbook description of basic MA schemes. In this section we look ahead into what 3GPP may consider for MA for 6G by looking back first into previous generations. 

The work of 3GPP is structured in so-called Releases, the first version of the 3G specifications was introduced in Release 99 towards the end of last millennium. Release 4 was the subsequent Release after which Releases have been labeled with increasing index. For example, the first version of 4G LTE was introduced in Release 8 while the first version of 5G NR was introduced in Release 15. Release 18, the first Release of 5G-Advanced, was finalized late 2023 and Release 19 started in early 2024. The study phase for 6G is expected to launch in Release 20 in the second half of 2025. In turn, the first 6G specifications are expected in Release 21 with a completion date anticipated to be between the end of 2028 and the end of 2029 aiming at commercial 6G deployments sometime around 2030. 

MA is one of the main characteristics of any wireless system. Indeed, it is one of the first decisions made when the design of a new generation takes place as it relates closely to the underlying waveform. It is important to note that, typically, there is one main MA scheme which is supplemented by other schemes for certain specific functions as we will discuss shortly. 

\subsection{MA in 3G and 4G}
In 3G systems, the MA scheme in the downlink is based on CDMA with Hadamard or Walsh codes applied to coded modulation symbols and orthogonally separating the various channels in a given cell. Different codes can be used to transmit data of different users and multiple codes can be used to increase the data rate for a given user. The application of these codes effectively expands the transmission bandwidth by a factor called spreading factor. The spreading factor offers the ability to boost the SINR at the receiver helping to cope with multiple sources of interference, i.e., inter-cell, intra-cell (ISI). However, the spreading factor decreased over time as the target operating data rates increased. This trend increased the need to use equalizers at the receiver, which, in turn, made OFDM based transmissions more appealing as they would not require equalization at the receiver. The application of a cell-dependent scrambling based on long pseudo-noise (PN) sequences onto the aggregated transmissions of a given cell makes it appear as noise when received at neighboring cells in full frequency reuse deployments. 

For 4G, OFDMA was introduced in the downlink maintaining intra-cell orthogonality while being more resilient to ISI caused by the delay spread of the channel. In full frequency reuse deployments, transmissions from different cells are randomized through the application of a cell-dependent scrambling based on long PN sequences similar to 3G. 

In the transition from 3G to 4G, for the uplink we went from a non-orthogonal MA with CDMA in 3G to orthogonal MA with DFT-spread OFDM (DFT-S-OFDM) in 4G. Transitions are typically somewhat blurred with inroads into new schemes by way of extensions of the current scheme before an entirely new scheme is introduced. For example, in the second half of 3G days we experienced the migration from circuit switch (CS) to packet switch (PS) communications. With that migration, we also moved from continuous downlink and uplink transmissions to packetized transmissions which would need to be scheduled by the network. Therefore, Release 5 introduced scheduled downlink with much lower multiplexing capability in the code domain (lower spreading factor) and aiming more at time-domain-multiplexing (TDM’ing) transmissions to multiple users with the goal to increase the instantaneous individual user data rate with the so-called  high speed downlink packet access (HSDPA). Similarly, Release 6 introduced at the time scheduled UL with what was called enhanced uplink (eUL) or  high speed uplink packet access (HSUPA). Release 7, in turn, introduced discontinuous uplink transmissions which, for the first time, enabled uplink transmissions to be gated in time and hence favoring TDM’ing of transmissions from different users supplemented with the possibility of non-orthogonally multiplexing multiple users’ transmissions in the code domain (CDMA). 

The main driver for the switch of MA from 3G to 4G was the uplink performance for cell-edge users for which the introduction of frequency domain multiplexing (FDM’ing) brought important performance gains in terms of uplink user throughput \cite{LTE25912}. The realization of FDM in the uplink of 4G was done via the introduction of DFT-S-OFDM, which due to its single carrier properties was key to reduce the user transmissions’ peak-to-average-power-ratio (PAPR) which directly entails a coverage gain for the uplink. Also, because of its underlying block-based construction of OFDM, subframe transmissions of 1ms simply meant transmissions of 14 symbols facilitating TDM’ing of transmissions with a built-in time-domain gating function which did not exist in 3G days. 

While FDMA/TDMA was the main MA scheme for the uplink enabling orthogonal access for users with the same serving cell, it is not the only MA scheme used in 4G LTE. 

Indeed, random access is performed via the Physical Random Access Channel (PRACH) which, while orthogonalized in time and frequency with the other uplink Physical Layer (PHY) channels, PRACH transmissions from different users overlap with each other yielding a CDMA-like MA for this physical channel with possibility of recoverable or unrecoverable inter-user collisions. 

Furthermore, SDMA was envisioned to enable MU-MIMO operation from the start of 4G. In this scenario, the spatial dimensionality of the channel is exploited by enabling data transmissions from different users to overlap in time and frequency with each other relying on spatially separating them at the multi-antenna base-station receiver. As a result, while the data portion of the uplink transmission is merely spatial-division-multiplexed (SDM’d) across users, their reference signal transmissions are orthogonalized by virtue of different cyclic shifts of the underlying Zadoff-Chu sequences used to modulate the demodulation reference signals (DMRS). In turn, that orthogonalization enables the possibility to perform channel estimation for each user’s transmission off cleaner DMRS samples without inter-user interference. 

Moreover, uplink control channel transmissions typically bear few bits of information which attempting to orthogonalize solely in the frequency domain would very rapidly run out of dimensions in the frequency domain (resource limitation). In order to alleviate that limitation, the possibility to orthogonally multiplex within the same time/frequency resources via different cyclic shifts of underlying Zadoff-Chu sequences (similar to DMRS) or time-domain orthogonal covers applied to each uplink symbol was also introduced to increase the capacify of the uplink control channel. Release 10 introduced the possibility for SU-MIMO in the uplink of LTE still resorting to DFT-S-OFDM for each of the transmitted layers. 

Towards the latter part of LTE, multi-user superposition transmission (MUST) was standardized for the downlink of LTE in Release 14 after the conclusion of the corresponding study item in the preceding release. The project attempted to jointly optimize multi-user operation from both transmitter and receiver’s perspective to improve the MU system capacity even if the transmission/precoding was non-orthogonal. The outcome of the study is captured in the technical report (TR) for MUST in \cite{MUST36859}. Three MUST categories were identified during the study: 1) MUST Category 1: Superposition transmission with adaptive power ratio on component constellations and non-Gray-mapped composite constellation; 2) MUST Category 2: Superposition transmission with adaptive power ratio on component constellations and Gray-mapped composite constellation; MUST Category 3: Superposition transmission with label-bit assignment on Gray-mapped composite constellation. 

\subsection{MA in 5G and 6G}
With the exception of MUST, all the schemes mentioned for 4G LTE were also adopted for 5G NR. Indeed, for the uplink data channel, the earlier mentioned DFT-S-OFDM is only used for single layer transmissions, while multi-layer transmissions always resort to OFDMA to improve link efficiency at the cost of an increased PAPR. The DMRS of the uplink data channel for different layers of the same user and from different users in the same cell are orthogonalized in time/frequency domain by corresponding orthogonal covers. 

5G NR while introduced in Release 15, had a project in Release 16 to study NOMA \cite{NOMA_SID} as a follow up to the discussions during the 5G NR study of Release 14 \cite{NR38912}. The TR for NOMA is a good reference \cite{NOMA38812} to review the schemes that were considered during the study with some of them presented in this paper. The Release 16 NOMA study investigated the application of NOMA techniques to mMTC, URLLC, and eMBB communications. 

The uplink NOMA transmitter was generalized so that it would include all the different flavors being considered. Also, different receivers were considered tailored to the different schemes and incurring various levels of complexity. 	

The outcome of the Release 16 study was deemed to be non-conclusive and a follow up work item was not approved at the time. Part of the reason was the proliferation of schemes during the study without a clear winner amongst them which made it impossible to define the scope of a normative project for only one scheme. Perhaps the scope of the project was too broad aiming at NOMA for mMTC, URLLC and eMBB use cases. Note that the mMTC use case of IMT-2020 is actually served by LTE CatM (eMTC) and NBIOT originally developed in Release 13 and evolved thereafter. 

Indeed, it was discussed that NOMA could be enabled in NR without specification impact provided that there was adequate receiver processing. The possibility to expand the number of orthogonal DMRS ports was discussed which, in combination with multi-user detectors at the receiver, would enable the possibility for processing fully overlapping data transmissions from different users. Those transmissions from multiple users could be resulting from dynamic scheduling from the network (with individual grants to each of the users) or from overlapping configured grants (pre-configured via L2 or L3 allocations).   

In the absence of a normative projects for NOMA specification, a minor improvement to the Random Access procedure which used to take 4-steps was approved with the goal to shorten it to 2-steps \cite{2stepRACH_SID}. It is, however, likely that the investigations that took part during Release 16 will be revisited  when NOMA is considered again in 3GPP in the future. 

Clearly, there is no universal MA (yet) that optimally suits all applications and use cases. The MA scheme aiming at optimizing individual user’s high throughput operation (eMBB-like use case) is not the same as the one aiming at maximizing multiplexing capability for low-rate applications of many, many users (mMTC-like use case). To some extent, that trade-off is already seen by looking into the nuanced MA scheme used today for random access, uplink control channel, uplink data channel, etc. 

\par 6G will see a resurgence of MA discussions. Though traditionally there is one main MA scheme supplemented by other schemes for certain specific functions, it is unclear whether this approach will still hold for 6G with the proliferation of schemes to cope with eMBB and mMTC but also and mainly to enable new use cases fueled by emerging intelligent and multi-functional applications. Aside the ones already discussed as part of 5G, more recent MA schemes have made their way toward standardization. This is the case of RSMA whose "universal" capability has been found attractive by various industries and recently been proposed for the first time for 6G for various use cases such as 1) to boost the performance of unicast transmissions beyond what has been achieved traditionally in 4G and 5G by SDMA-multi-user MIMO \cite{Viavi}, and 2) to enable non-orthogonal unicast and multicast transmissions where multicast traffic are superimposed over unicast traffic instead of being served in orthogonal resources as in 4G and 5G \cite{BBC}. It will be interesting to see in the coming years whether 3GPP will favour a single unified and general MA scheme as main MA scheme instead of a combination of multiple MA schemes, each optimized for specific conditions. 

\subsection{AI/ML in 5G and 6G}
A Study Item for AI/ML applied to the NR Air Interface was agreed for Release 18 \cite{AIML_SID}. The TR for this study is available at \cite{AIML38843} and constitutes a good reference to review what is being studied (project completion planned for December 2023). While AI/ML first appeared in 3GPP for the purpose of users’ data collection, this project is the first time in 3GPP where the direct application of AI/ML is sought for some fundamental air interface problems. In order to avoid too abstract, high level conceptual discussions and to give some shape to the project, a number of pilot use cases was identified to be analyzed for this project. Namely, CSI enhancements, beam management (BM) and positioning accuracy enhancements. Each of those three use cases was further refined into two sub-use cases. Namely, CSI (spatial-frequency) compression and CSI (temporal) prediction for CSI enhancement, Spatial beam prediction and Temporal beam prediction for BM, direct AI/ML positioning and AI/ML assisted positioning for Positioning accuracy enhancements. All sub use cases but CSI compression rely on single-sided models at the user terminal or the network, while CSI compression resorts to two-sided AI/ML models with one part of the model running at the user terminal side and the other at the network side. For the CSI compression sub use case, the user terminal side performs the encoding or compression of the CSI information, and the network side performs the decoder or decompression to obtain the terminal’s CSI. 

It is expected that normative work to address the specification impact of these use cases will be carried out as part of Release 19. Moreover, further use cases, for example Mobility enhancements, may be additionally investigated. 

As indicated by the ITU-R WP 5D framework document for IMT-2030 \cite{ITUR_IMT2030}, Integrated AI and communication has been identified as one of the usage scenarios for IMT-2030. As a result, it is expected that 6G will be AI/ML native. 

While nobody knows precisely what that will imply from 3GPP perspective, the foundational work on AI/ML for NR air interface that is underway will for sure play a very important role on how native 6G will be specified \cite{QC_6G}\cite{ERI_6G}\cite{NOK_6G}\cite{HW_6G}\cite{SS_6G}. 

The application of AI/ML techniques to MA may be investigated for 6G with the aim at figuring out the optimal MA scheme for the given application and use case. Indeed, channel access policies that fluently transition from contention- to schedule-based depending on the use case and environment are deemed desirable \cite{NOK_6G}. 

Similar to the Rel-18 study where single-sided and two-sided (or cross-node) models are being considered depending on the use case, different AI/ML functions for 6G will require one or another approach. In particular,  how the AI/ML engines running at the terminals and the network will figure out the optimal MA scheme is subject of research and will for sure open up new opportunities for improved system performance and user experience. Adaptability of the models through further offline or even online training will also play a key role in the end-to-end performance optimization. 

As a result, the current investigations on join training of two-sided models, as well as, their interoperability and performance testing will be very relevant for the future native AI work in 6G.

\section{Conclusions}

Wireless networks experience a paradigm shift with the integration of multi-functionality and the rise of AI and ML. 6G and beyond will be AI-native and multi-functionality-native, spurring the need to re-think multiple access techniques to make use of the available time, frequency, space, power, signal dimensions to serve users, devices, machines, services, training nodes in the most efficient way and cope with those new network advances. The paper highlighted that much work is left for researchers in this area. Aside many other schemes, RSMA has emerged in the past few years by providing a unified and conceptually simple understanding of SDMA, PD-NOMA, physical layer multicasting, and uniquely shrinks rather grows the knowledge tree of MA schemes based on space,
power, signal dimensions. Consequently RSMA has found numerous applications for 6G. Moving next, UMA should further shrink the knowledge tree of MA schemes by unifying RSMA with all other dimensions, such as code domain MAs, and ultimately provide a unified and conceptually simple understanding of the current and future morass of MA schemes. Such UMA does not exist yet. It is hoped that the techniques and outlook presented in this article will help inspiring future research in this exciting and important area and pave the way for designing and implementing efficient multiple access techniques in future wireless systems.

\bibliographystyle{IEEEtran}
\bibliography{Ref}

\end{document}